\documentclass{article}

\usepackage{arxiv}

\usepackage[utf8]{inputenc} 
\usepackage[T1]{fontenc}    
\usepackage{hyperref}       
\usepackage{url}            
\usepackage{booktabs}       
\usepackage{amsfonts}       
\usepackage{nicefrac}       
\usepackage{microtype}      
\usepackage{lipsum}
\usepackage{amsmath}
\usepackage{graphicx}
\usepackage{adjustbox}
\usepackage{multirow}
\usepackage{multicol}
\usepackage[table]{xcolor}
\usepackage{subfig}
\usepackage[section]{placeins}
\usepackage{subfiles}
\usepackage[capitalise]{cleveref}       

\usepackage{fancyvrb}

\RecustomVerbatimCommand{\VerbatimInput}{VerbatimInput}%
{fontsize=\footnotesize,
 frame=lines,  
 framesep=2em, 
 rulecolor=\color{Gray},
 label=\fbox{\color{Black}data.txt},
 labelposition=topline,
 commandchars=\|\(\), 
 commentchar=*        
}

\graphicspath{ {./images/} }

\makeatletter
\newcommand*{\rom}[1]{\expandafter\@slowromancap\romannumeral #1@}
\makeatother

\usepackage{enumitem} 

\begin{document}

\vspace*{0.35in}

\begin{flushleft}
{\Large
\textbf\newline{Preoperative brain tumor imaging: models and software for segmentation and standardized reporting}
}
\newline
\\
David Bouget\textsuperscript{1,*},
André Pedersen\textsuperscript{1,2,3},
Asgeir Store Jakola\textsuperscript{4,5},
Vasileios Kavouridis\textsuperscript{6},
Kyrre Eeg Emblem\textsuperscript{7},
Roelant S. Eijgelaar\textsuperscript{8,9},
Ivar Kommers\textsuperscript{8,9},
Hilko Ardon\textsuperscript{10},
Frederik Barkhof\textsuperscript{11,12},
Lorenzo Bello\textsuperscript{13},
Mitchel S. Berger\textsuperscript{14},
Marco Conti Nibali\textsuperscript{13},
Julia Furtner\textsuperscript{15},
Shawn Hervey-Jumper\textsuperscript{14},
Albert J.S. Idema\textsuperscript{16},
Barbara Kiesel\textsuperscript{17},
Alfred Kloet\textsuperscript{18},
Emmanuel Mandonnet\textsuperscript{19},
Domenique M.J. Müller\textsuperscript{8,9},
Pierre A. Robe\textsuperscript{20},
Marco Rossi\textsuperscript{13},
Tommaso Sciortino\textsuperscript{13},
Wimar Van den Brink\textsuperscript{21},
Michiel Wagemakers\textsuperscript{22},
Georg Widhalm\textsuperscript{17},
Marnix G. Witte\textsuperscript{23},
Aeilko H. Zwinderman\textsuperscript{24},
Philip C. De Witt Hamer\textsuperscript{8,9},
Ole Solheim\textsuperscript{6,25},
Ingerid Reinertsen\textsuperscript{1,26}
\\
\bigskip
\bf{1} Department of Health Research, SINTEF Digital, NO-7465 Trondheim, Norway
\\
\bf{2} Department of Clinical and Molecular Medicine, Norwegian University of Science and Technology, NO-7491 Trondheim, Norway
\\
\bf{3} Clinic of Surgery, St. Olavs hospital, Trondheim University Hospital, NO-7030 Trondheim, Norway
\\
\bf{4} Department of Neurosurgery, Sahlgrenska University Hospital, 41345 Gothenburg, Sweden
\\
\bf{5} Department of Clinical Neuroscience, Institute of Neuroscience and Physiology, Sahlgrenska Academy, University of Gothenburg, 40350 Gothenburg, Sweden
\\
\bf{6} Department of Neurosurgery, St. Olavs hospital, Trondheim University Hospital, NO-7030 Trondheim, Norway
\\
\bf{7} Department of Physics and Computational Radiology, Division of Radiology and Nuclear Medicine, Oslo University Hospital, 0450 Oslo, Norway
\\
\bf{8} Department of Neurosurgery, Amsterdam University Medical Centers, Vrije Universiteit, 1081 HV Amsterdam, The Netherlands
\\
\bf{9} Cancer Center Amsterdam, Brain Tumor Center, Amsterdam University Medical Centers, 1081 HV Amsterdam, The Netherlands
\\
\bf{10} Department of Neurosurgery, Twee Steden Hospital, 5042 AD Tilburg, The Netherlands
\\
\bf{11} Department of Radiology and Nuclear Medicine, Amsterdam University Medical Centers, Vrije Universiteit, 1081 HV Amsterdam, The Netherlands
\\
\bf{12} Institutes of Neurology and Healthcare Engineering, University College London, London WC1E 6BT, UK
\\
\bf{13} Neurosurgical Oncology Unit, Department of Oncology and Hemato-Oncology, Humanitas Research Hospital, Università Degli Studi di Milano, 20122 Milano, Italy
\\
\bf{14} Department of Neurological Surgery, University of California San Francisco, San Francisco, CA 94143, USA
\\
\bf{15} Department of Biomedical Imaging and Image-Guided Therapy, Medical University Vienna, 1090 Wien, Austria
\\
\bf{16} Department of Neurosurgery, Northwest Clinics, 1815 JD Alkmaar, The Netherlands
\\
\bf{17} Department of Neurosurgery, Medical University Vienna, 1090 Wien, Austria
\\
\bf{18} Department of Neurosurgery, Haaglanden Medical Center, 2515 VA The Hague, The Netherlands
\\
\bf{19} Department of Neurological Surgery, Hôpital Lariboisière, 75010 Paris, France
\\
\bf{20} Department of Neurology and Neurosurgery, University Medical Center Utrecht, 3584 CX Utrecht, The Netherlands
\\
\bf{21} Department of Neurosurgery, Isala, 8025 AB Zwolle, The Netherlands
\\
\bf{22} Department of Neurosurgery, University Medical Center Groningen, University of Groningen, 9713 GZ Groningen, The Netherlands
\\
\bf{23} Department of Radiation Oncology, The Netherlands Cancer Institute, 1066 CX Amsterdam, The Netherlands
\\
\bf{24} Department of Clinical Epidemiology and Biostatistics, Amsterdam University Medical Centers, University of Amsterdam, 1105 AZ Amsterdam, The Netherlands
\\
\bf{25} Department of Neuromedicine and Movement Science, Norwegian University of Science and Technology, NO-7491 Trondheim, Norway
\\
\bf{26} Department of Circulation and Medical Imaging, Norwegian University of Science and Technology, NO-7491 Trondheim, Norway
\\
\bigskip
* david.bouget@sintef.no

\end{flushleft}

\begin{abstract}
For patients suffering from brain tumor, prognosis estimation and treatment decisions are made by a multidisciplinary team based on a set of preoperative MR scans. Currently, the lack of standardized and automatic methods for tumor detection and generation of clinical reports, incorporating a wide range of tumor characteristics, represents a major hurdle.
In this study, we investigate the most occurring brain tumor types: glioblastomas, lower grade gliomas, meningiomas, and metastases, through four cohorts of up to $4\,000$ patients. Tumor segmentation models were trained using the AGU-Net architecture with different preprocessing steps and protocols. Segmentation performances were assessed in-depth using a wide-range of voxel and patient-wise metrics covering volume, distance, and probabilistic aspects. Finally, two software solutions have been developed, enabling an easy use of the trained models and standardized generation of clinical reports: Raidionics and Raidionics-Slicer.
Segmentation performances were quite homogeneous across the four different brain tumor types, with an average true positive Dice ranging between 80\% and 90\%, patient-wise recall between 88\% and 98\%, and patient-wise precision around 95\%. In conjunction to Dice, the identified most relevant other metrics were the relative absolute volume difference, the variation of information, and the Hausdorff, Mahalanobis, and object average symmetric surface distances. With our Raidionics software, running on a desktop computer with CPU support, tumor segmentation can be performed in $16$ to $54$ seconds depending on the dimensions of the MRI volume. For the generation of a standardized clinical report, including the tumor segmentation and features computation, $5$ to $15$ minutes are necessary.
All trained models have been made open-access together with the source code for both software solutions and validation metrics computation.
In the future, a method to convert results from a set of metrics into a final single score would be highly desirable for easier ranking across trained models. In addition, an automatic classification of the brain tumor type would be necessary to replace manual user input. Finally, the inclusion of post-operative segmentation in both software solutions will be key for generating complete post-operative standardized clinical reports.
\end{abstract}


\section{Introduction}
\label{intro}
Prognosis in patients with brain tumors is heterogeneous with survival rates varying from weeks to several years depending on the tumor grade and type, and for which most patients will experience progressive neurological and cognitive deficit~\cite{day2016neurocognitive}. Brain tumors can be classified as either primary or secondary. In the former, tumors originate from the brain itself or its supporting tissues whereas in the latter cancer cells have spread from tumors located elsewhere in the body to reach the brain (i.e., brain metastasis). According to the World Health Organization classification of tumors~\cite{louis20212021}, primary brain tumors are graded by histopathological and genetic analyses and can be regrouped in $100$ different subtypes with frequent to relatively rare occurrences. Amongst the most frequent subtypes, tumors arising from the brain's supportive cell population(i.e., glial tissue) are referred to as gliomas. The more aggressive entities are labelled as high-grade gliomas (HGGs) and are graded between $3$ and $4$, while the less aggressive entities are referred to as diffuse lower grade gliomas (LGGs) and are graded between $2$ and $3$. Tumors arising from the meninges, which form the external membranous covering the brain, are referred to as meningiomas. Aside from the aforementioned large categories, other and less frequent tumor types exist (e.g., in pituitary, sellar, or pineal regions). Each tumor category has a distinct biology, prognosis, and treatment~\cite{deangelis2001brain,fisher2007epidemiology}. The most common primary malignant brain tumor type in adults is high-grade glioma which remains among the most difficult cancers to treat with a limited 5-year overall survival~\cite{lapointe2018primary}.

For patients affected by brain tumors, prognosis estimation and treatment decisions are made by a multidisciplinary team (including neurosurgeons, oncologists, and radiologists), and based on a set of preoperative MR scans. A high accuracy in the preoperative diagnostics phase is of utmost importance for patient outcomes.
Judgements concerning the complexity or radicality of surgery, or the risks of postoperative complications hinge on data gleaned from MR scans. Additionally, tumor-specific characteristics such as volume and location, or cortical structures profile can to a large degree be collected~\cite{kickingereder2016radiomic}.
Retrospectively, such measurements can be gathered from the analysis of surgical cohorts, multicenter trials, or registries in order to devise patient outcome prediction models~\cite{sawaya1998neurosurgical,mathiesen2011two,zinn2013extent}.
Reliable measurements and reporting of tumor characteristics are therefore instrumental in patient care. Standard reporting and data systems (RADSs) have been established for several solid tumors such as prostate cancer~\cite{weinreb2016pi} and lung cancer~\cite{dyer2020implications}. Very few attempts have been made for brain cancer in general~\cite{ellingson2015consensus} or glioblastomas~\cite{kommers2021glioblastoma}. The main goal of RADSs is to provide rules for imaging techniques, terminology of reports, definitions of tumor features, and treatment response to reduce practice variation and obtain reproducible tumor classification. A broad implementation can facilitate collaborations and stimulate evaluation for development and improvement of RADSs.

Currently, the lack of standardized and automatic methods for tumor detection in brain MR scans represents a major hurdle towards the generation of clinical reports incorporating a wide range of tumor characteristics. Manual tumor delineation or assessment by radiologists is time-consuming and subject to intra and inter-rater variations that are difficult to characterize~\cite{binaghi2016collection} and therefore rarely done in clinical practice. As a result, informative tumor features (e.g., location or volume) are often estimated from the images solely based on crude measuring techniques (e.g., eyeballing)~\cite{berntsen2020volumetric}. 

\subsection{Related work}
From the fast-growing development in the field of deep learning, convolutional neural networks have demonstrated impressive performance in various segmentation tasks and benchmark challenges, with the added-value of being fully automatic and deterministic~\cite{minaee2021image}. Regarding brain tumor segmentation, performances have specifically been assessed on the Brain Tumor Segmentation challenge (BraTS) dataset~\cite{menze2014multimodal,bakas2017advancing}. Occurring every year since 2012, the challenge focuses on gliomas (i.e., HGGs and LGGs) and has reached a notable cohort size with a total of $2\,040$ patients included in the $2021$ edition, and multiple MR sequences included for each patient (i.e., T1c, T1w, T2, FLAIR). Segmentation performance has been assessed using the Dice similarity coefficient and the $95$th percentile Hausdorff distance (HD95) as metrics~\cite{baid2021rsna}. The current state-of-the-art is an extension of the nnU-Net architecture~\cite{isensee2018nnu} with an asymmetrical number of filters between the encoding and decoding paths, substitution of all batch normalization layers by group normalization, and addition of axial attention~\cite{luu2021extending}. An average Dice score of $85$\% together with a $17.70$\,mm HD95 were obtained for the enhancing tumor segmentation task in glioblastomas.
The segmentation of other brain tumor types has been sparsely investigated in the literature in comparison, possibly due to a lack of open-access annotated data, as illustrated by recent reviews or studies investigating brain tumor segmentation in general~\cite{tiwari2020brain, pereira2016brain}.
Grovik et al. used a multicentric and multi-sequence dataset of $165$ metastatic patients to train a segmentation model with the DeepLabV3 architecture~\cite{grovik2020deep, grovik2021handling}. The best segmentation results were around $79$\% Dice score with $3.6$ false positive detections per patient on average. Other prior studies have focused on using variations of the DeepMedic architecture~\cite{kamnitsas2016deepmedic}, using contrast-enhanced T1-weighted MRI volumes as input, to train their segmentation models~\cite{liu2017deep, charron2018automatic}. Datasets were of a similar magnitude with around $200$ patients. However, in both cases the test sets were limited to up to $20$ patients, making it difficult to assess the generalization ability of the trained models in the absence of cross-validation studies. Obtained average Dice scores over the contrast-enhancing tumor were approximating $75$\%, with almost $8$ false positive detections per patient.
From a recent review on the use of machine learning applied to different meningioma-related tasks using MRI scans~\cite{neromyliotis2022machine}, more than $30$ previous studies have investigated automatic diagnosis or grading but only a handful focused on the segmentation task. In addition, datasets' magnitude used for segmentation purposes has been consistently smaller than for the other tasks, with barely up to $126$ patients in the reported studies. Laukamp et al. reported the best Dice scores using well-known 3D neural network architectures such as DeepMedic and BioMedIA, though at the expense of heavy preprocessing techniques the likes of atlas registration~\cite{laukamp2019fully, laukamp2020automated}. In a previous study, we achieved equally promising performance using an attention-based U-Net architecture, reaching an average Dice score of up to $88$\% on contrast-enhanced T1-weighted MRI volumes~\cite{bouget2021meningioma}. In addition, the cross-validation studies performed over up to $600$ patients with a wide range of tumor sizes, coming from the hospital and the outpatient clinic, exhibited a proper ability to generalize from the trained models.

To summarize, with the exception of the BraTS challenge, there is a dearth of high-quality MRI datasets for brain tumor segmentation. Furthermore, open-access pretrained models and inference code are scarce and can be cumbersome to operate, hence hindering the generation of private datasets for brain tumor segmentation tasks. On the other hand, open-source tools are being developed to assist in image labeling and generation of AI models for clinical evaluation, such as MONAI label~\cite{the_monai_consortium_2020_4323059}. Yet, they do not integrate nor provide access to the latest and highest performing brain tumor segmentation models from the literature.
From a validation standpoint, the focus has been on reporting Dice scores and often Hausdorff distances, while many other meaningful and possibly more relevant metrics exist and could be investigated to better highlight strengths and weaknesses of the different segmentation methods~\cite{reinke2021common,taha2015metrics}.

The literature on RADSs for brain tumors is equally scarce with only few attempts for preoperative glioblastoma surgery~\cite{kommers2021glioblastoma} or post-treatment investigation~\cite{weinberg2018management}. In the former, automatic segmentation and computation of relevant tumor features was provided, and an excellent agreement has been shown between characteristics computed over the manual and automatic segmentations. In the latter, the interpretation of the post-treatment MR scans was provided using a structured set of rules, but deprived of any automatic tumor segmentation or image analysis support.

\subsection{Contributions}
While research is exceedingly ahead for glioma segmentation under the aegis of the BraTS challenge community, the segmentation of meningiomas and metastases is trailing behind. In addition, validation studies in the literature have too often been dominated by Dice score reporting and a broader inspection is essential to ensure clinical relevance. Finally, the outcome of this research is often not readily available, especially for the intended end-users who are clinicians without programming experience.
As such, the contributions of our study are: (i) the training of robust segmentation models for glioblastomas, lower grade gliomas, meningiomas, and metastases assessed using a panel of more than $20$ different metrics to better highlight performance, (ii) the development of two software solutions enabling easy use of the trained models and tumor features computation: Raidionics and Raidionics-Slicer, and (iii) open-access models and source code for the software and validation metrics computation.

\section{Data}
\label{sec:dataset}
\begin{figure}[!ht]
\centering
\includegraphics[scale=1.40]{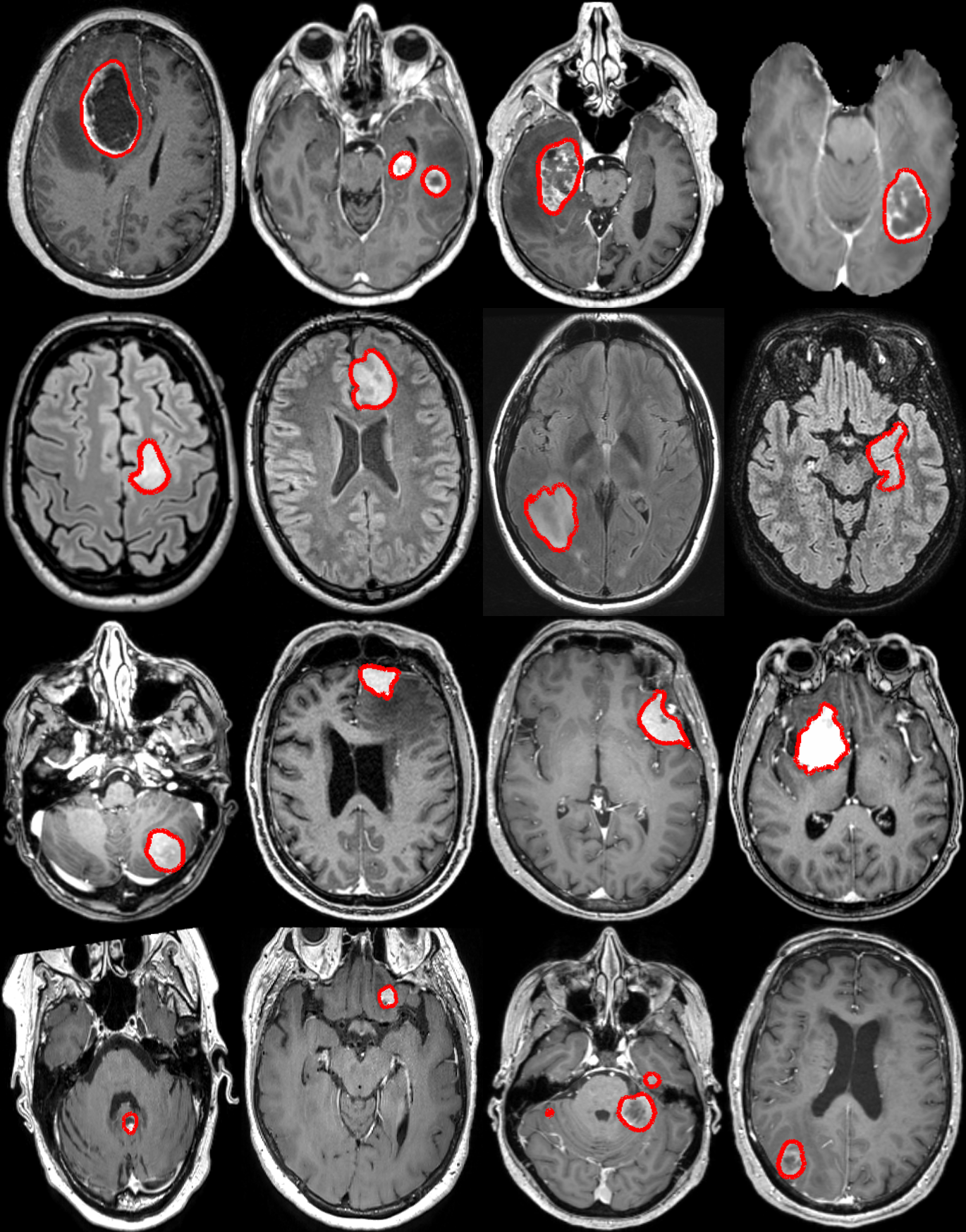}
\caption{Examples of brain tumors from the raw MRI volumes collected in this study. Each row illustrates a tumor type: glioblastoma, lower grade glioma, meningioma, metastasis (from top to bottom). The manual annotation contours are overlaid in red.}
\label{fig:datasets-examples}
\end{figure}

For this study, four different datasets have been assembled, one for each main tumor type considered: glioblastoma, lower grade glioma, meningioma, and metastasis. The tumor type was assessed at time of surgery, when applicable, following the currently applicable guidelines (i.e., either WHO 2007 or WHO 2016). Tumors were manually segmented in 3D by trained raters using as support either a region growing algorithm~\cite{huber2017reliability} or a grow cut algorithm~\cite{vezhnevets2005growcut}, and subsequent manual editing. Trained raters were supervised by neuroradiologists and neurosurgeons. On contrast-enhanced T1-weighted scans, the tumor was defined as gadolinium-enhancing tissue including non-enhancing enclosed necrosis or cysts. On FLAIR scans, the tumor was defined as the hyperintense region. 
The four datasets are introduced in-depth in the subsequent sections. An overall summary of the data available is reported in \cref{tab:all-datasets-description}, and some visual examples are provided in \cref{fig:datasets-examples}.

\begin{table}[ht]
\centering
\caption{Overview of the datasets gathered for the four brain tumor types considered. Only one MRI sequence is available for each patient, and T1c corresponds to Gd-enhanced T1-weighted MR scans.}
\begin{tabular}{c|ccccc}
Tumor type & Sequence type & \# patients & \# sources & Volume average (ml) & Volume range (ml)\tabularnewline
\hline
Glioblastoma & T1c & $2134$ & $15$ & $34.37\pm28.83$ & $[0.01, 243.39] $\tabularnewline
Lower grade glioma & FLAIR & $659$ & $4$ & $51.71\pm78.60$ & $[0.14, 478.83]$ \tabularnewline
Meningioma & T1c & $719$ & $2$ & $19.40\pm28.62$ & $[0.07, 209.38]$\tabularnewline
Metastasis & T1c & $396$ & $2$ & $17.53\pm17.97$ & $[0.01, 114.77]$ \tabularnewline
\end{tabular}
\label{tab:all-datasets-description}
\end{table}

\subsection{Glioblastomas}
The glioblastoma dataset is made of a total of $2\,134$ Gd-enhanced T1-weighted MRI volumes originating from fourteen different hospitals, and one public challenge.

The first $1\,841$ patients have been collected from fourteen different hospitals worldwide: $38$ patients from the Northwest Clinics, Alkmaar, Netherlands (ALK); $97$ patients from the Amsterdam University Medical Centers, location VU medical center, Netherlands (AMS); $86$ patients from the University Medical Center Groningen, Netherlands (GRO); $103$ patients from the Medical Center Haaglanden, the Hague, Netherlands (HAG); $75$ patients from the Humanitas Research Hospital, Milano, Italy (MIL); $74$ patients from the Hôpital Lariboisière, Paris, France (PAR); $134$ patients from the University of California San Francisco Medical Center, U.S. (SFR); $49$ patients from the Medical Center Slotervaart, Amsterdam, Netherlands (SLO); $153$ patients from the St Elisabeth Hospital, Tilburg, Netherlands (TIL); $171$ patients from the University Medical Center Utrecht, Netherlands (UTR); $83$ patients from the Medical University Vienna, Austria (VIE); $72$ patients from the Isala hospital, Zwolle, Netherlands (ZWO); $456$ patients from the St. Olavs hospital, Trondheim University Hospital, Norway (STO); and $249$ patients from the Sahlgrenska University Hospital, Gothenburg, Sweden. An in-depth description of most cohorts can be found in a recent study~\cite{kommers2021glioblastoma}.
The remaining $293$ patients correspond to the training set of the BraTS challenge (edition $2020$), but have already undergone preprocessing transformations such as skull-stripping.

Overall, MRI volume dimensions are covering $[159; 896]\times[86; 896]\times[17; 512]$\,voxels, and the voxel size ranges $[0.26; 1.25]\times[0.26; 2.00]\times[0.47; 7.50]$\,mm$^{3}$. An average MRI volume is $[303\times323\times193]$ pixels with a spacing of $[0.86\times0.84\times1.24]$\,mm$^3$.

\subsection{Lower grade gliomas}
The lower grade glioma dataset is made of a total of $659$ FLAIR MRI volumes, with mostly grade 2 diffuse gliomas, coming from four different hospitals: $330$ patients from the Brigham and Womens Hospital, Boston, USA; $165$ patients from the St. Olavs hospital, Trondheim University Hospital, Norway; $154$ patients from the Sahlgrenska University Hospital, Gothenburg, Sweden; and $10$ from the University Hospital of North Norway, Norway.

Overall, MRI volume dimensions are covering $[192; 576]\times[240; 640]\times[16; 400]$\,voxels, and the voxel size ranges $[0.34; 1.17]\times[0.34; 1.17]\times[0.50; 8.0]$\,mm$^{3}$. An average MRI volume is $[349\times363\times85]$ pixels with a spacing of $[0.72\times0.72\times4.21]$\,mm$^3$.

\subsection{Meningiomas}
The meningioma dataset is made of $719$ Gd-enhanced T1-weighted MRI volumes, mostly built around a dataset previously introduced~\cite{bouget2021fast}, showcasing patients either followed at the outpatient clinic or recommended for surgery at the St. Olavs hospital, Trondheim University Hospital, Norway.

Overall, MRI volume dimensions are covering $[192; 512]\times[224; 512]\times[11; 290]$\,voxels, and the voxel size ranges $[0.41; 1.05]\times[0.41; 1.05]\times[0.60; 7.00]$\,mm$^{3}$. An average MRI volume is $[343\times350\times147]$ pixels with a spacing of $[0.78\times0.78\times1.67]$\,mm$^3$.

\subsection{Metastases}
The metastasis dataset is made of a total of $396$ Gd-enhanced T1-weighted MRI volumes, collected from two different hospitals: $329$ patients from the St. Olavs hospital, Trondheim University Hospital, Norway; and $67$ patients from Oslo University Hospital, Oslo, Norway.

Overall, MRI volume dimensions are covering $[128; 560]\times[114; 560]\times[19; 561]$\,voxels, and the voxel size ranges $[0.43; 1.33]\times[0.43; 1.80]\times[0.45; 7.0]$\,mm$^{3}$. An average MRI volume is $[301\times370\times289]$ pixels with a spacing of $[0.85\times0.76\times1.08]$\,mm$^3$.

\section{Methods}
\label{sec:methods}
First, the process for automatic brain tumor segmentation including data preprocessing, neural network architecture, and training design is introduced in \cref{subsec:segm}. Second, the tumor characteristics extraction process, using the generated tumor segmentation as input, is summarized in \cref{subsec:tumor-features}. Finally, a description of the two developed software solutions for performing segmentation and standardized reporting is given in \cref{subsec:software}

\subsection{Tumor segmentation}
\label{subsec:segm}
The architecture selected to train segmentation models for each brain tumor type is AGU-Net, which has shown to perform well on glioblastoma and meningioma segmentation~\cite{bouget2021glioblastoma,bouget2021meningioma}. In the following, the different training blocks are presented with some inner variations specified by roman numbers inside brackets. A global overview is provided in \cref{tab:training-strat-summary} summarizing used variants.

\begin{table}[!h]
\centering
\caption{Summary of the model training strategy followed for each tumor type.}
\adjustbox{max width=\textwidth}{
\begin{tabular}{l|lll}
Tumor type & Preprocessing & Strategy & Protocol\tabularnewline
\hline
glioblastoma & (ii) skull-stripping & (i) from-scratch & (i) leave-one-out \tabularnewline
lower grade glioma & (i) tight clipping & (i) from-scratch & (ii) 5-fold \tabularnewline
Meningioma & (i) tight clipping & (i) from-scratch & (ii) 5-fold \tabularnewline
Metastasis & (ii) skull-stripping & (ii) transfer-learning & (ii) 5-fold \tabularnewline
\end{tabular}
}
\label{tab:training-strat-summary}
\end{table}

\paragraph{Architecture:} Single-stage approach leveraging multi-scale input and deep supervision to preserve details, coupled to a single attention module. The loss function used was the class-averaged Dice loss, excluding the background. The final architecture was as described in the original article with $5$ levels and $[16, 32, 128, 256, 256]$ as convolution blocks.

\paragraph{Preprocessing:} The following preprocessing steps were used:
\begin{enumerate}
    \item resampling to an isotropic spacing of $1\,\text{mm}^3$ using spline interpolation of order 1 from NiBabel~\footnote{https://github.com/nipy/nibabel}.
    \item (i) tight clipping around the patient's head, excluding the void background, or (ii) skull-stripping using a custom brain segmentation model.
    \item volume resizing to $128\times128\times144\,\text{voxels}$ using spline interpolation of order 1.
    \item intensity normalization to the range $[0, 1]$.
\end{enumerate}

\paragraph{Training strategy:} Models were trained using the Adam optimizer over a batch size of $32$ samples with accumulated gradients (actual batch size $2$), stopped after $30$ consecutive epochs without validation loss improvement, following either: (i) training from scratch with $1e^{-3}$ initial learning rate, or transfer learning with an initial learning rate of $1e^{-4}$ fine-tuning over the best glioblastoma model.

For the data augmentation strategy, the following transforms were applied to each input sample with a probability of $50$\%: horizontal and vertical flipping, random rotation in the range $[-20^{\circ}, 20^{\circ}]$, and translation up to 10\% of the axis dimension.

\paragraph{Training protocol:} Given the magnitude difference within our four datasets, two different protocols were considered: (i) a three-way split at the hospital level whereby MRI volumes from one hospital constituted the validation fold; MRI volumes from a second hospital constituted the test fold; and the remaining MRI volumes constituted the training fold. As such, each hospital was used in turn as the test set in order to properly assess the ability of the different models to generalize. Or (ii) a 5-fold cross-validation with random two-way split over MRI volumes whereby four folds are used in turn as training set and the remaining one as validation set, without the existence of a proper separate test set.

\subsection{Preoperative clinical reporting}
\label{subsec:tumor-features}
For the generation of standardized preoperative clinical reports in a reproducible fashion, the computation of tumor characteristics was performed after alignment to a standard reference space. As described in-depth in our previous study~\cite{kommers2021glioblastoma}, the reference space was constituted by the symmetric Montreal Neurological Institute ICBM2009a atlas (MNI)~\cite{fonov2009unbiased}. The atlas space not possessing any brain average as FLAIR sequence, the T1 atlas file was used for all tumor types.

For each tumor type, the collection of features includes: volume, laterality, multifocality, cortical structure location profile, and subcortical structure location profile.
Specifically tailored for glioblastomas, resectability features are therefore not available for the other brain tumor types.

\subsection{Proposed software}
\label{subsec:software}
In order to make our models and tumor features easily available to the community, we have developed two software solutions. The first one is a stand-alone software called Raidionics, and the second one is a plugin to 3D Slicer given its predominant and widespread use in the field~\cite{fedorov20123d}.
Both solutions provide access to a similar back-end including inference and processing code. However, the GUI and intended user interactions differ. The trained models are stored in a separate online location and are downloaded on the user's computer at runtime. Models can be improved over time and a change will be automatically detected, resulting in the replacement of outdated models on the user's machine.

\begin{figure}[!ht]
\centering
\includegraphics[scale=0.42]{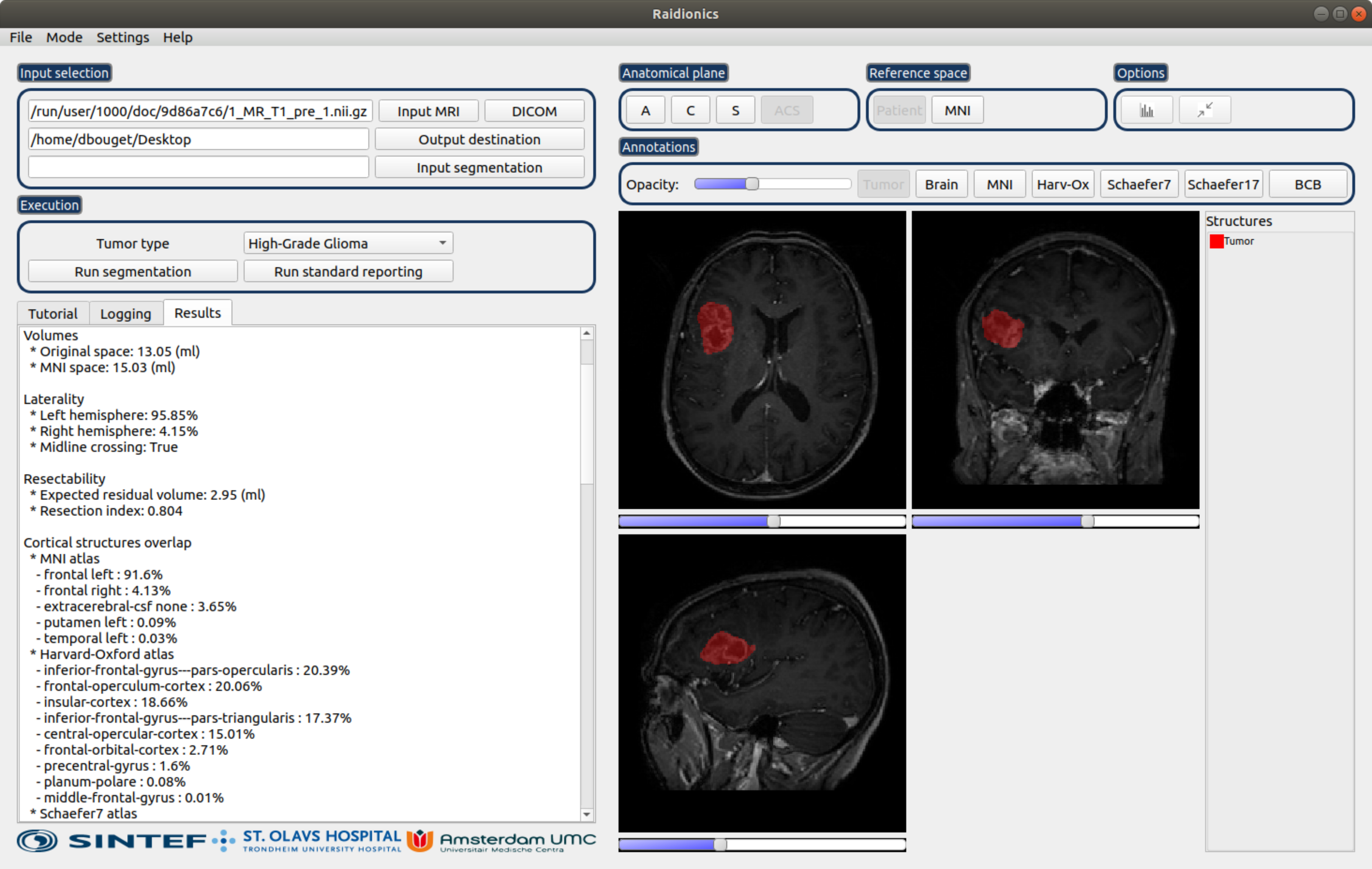}
\caption{Illustration of the Raidionics software after generating the standardized report for a patient suffering from a glioblastoma. The left side presents the tumor characteristics belonging to the report, whereas the right side offers a simplistic viewer.}
\label{fig:neurorads-illu}
\end{figure}

\subsubsection{Stand-alone solution: Raidionics}
\label{subsubsec:neurorads-description}
The software proposes two modes: (i) single-use where only one patient is to be processed and results can be visually assessed in the 2D viewer, and (ii) batch-mode whereby a collection of patients can be processed sequentially without any viewing possibility. In each mode, the option is left to the user to solely perform tumor segmentation, or to compute the whole set of tumor characteristics and generate the standardized report.
For each patient, the software expects an MRI scan as input (i.e., Gd-enhanced T1-weighted or FLAIR sequence) and the tumor type must be manually selected.  Additionally, a pre-existing tumor segmentation mask can be provided to bypass the automatic segmentation, if collecting the tumor characteristics is the main interest and manual annotations have been performed beforehand.
The total set of processed files saved on disk includes the standardized reports, brain and tumor segmentation masks in both patient and MNI space, cortical and subcortical structures masks in both patient and MNI space, and the registration files to navigate from patient to MNI space. To complement the reporting and give the possibility for follow-up statistical studies, the complete set of computed features is also provided in comma separated value format (i.e., .csv).

The software has been developed in Python 3.6.9, using PySide2 v5.15.2 for the graphical user interface, and only uses the Central Processing Unit (CPU) for the various computations. The software has been tested and is compatible with Windows ($\geq$~10), macOS ($\geq$~Catalina 10.15), and Ubuntu Linux ($\geq$~18.04). 
An illustration of the software is provided in \cref{fig:neurorads-illu}. Cross-platform installers and source code are freely available at \url{https://github.com/dbouget/Raidionics}.

\subsubsection{3D Slicer plugin: Raidionics-Slicer}
\label{subsubsec:slicer-plugin-description}
The 3D Slicer plugin has been developed using the DeepInfer plugin as baseline~\cite{mehrtash2017deepinfer}, and is mostly intended for tumor segmentation purposes. Through a slider, the possibility is provided to manually alter the probability threshold cutoff in order to refine the proposed binary mask. Further manual editing can be performed thereafter using the existing 3D Slicer functionalities.
The back-end processing code has been bundled into a Docker image for convenience, and therefore administrator rights are required for the end-user to perform the installation locally. The same inputs, behaviour, and outputs can be expected as for the stand-alone software.

\begin{figure}[!ht]
\centering
\includegraphics[scale=0.325]{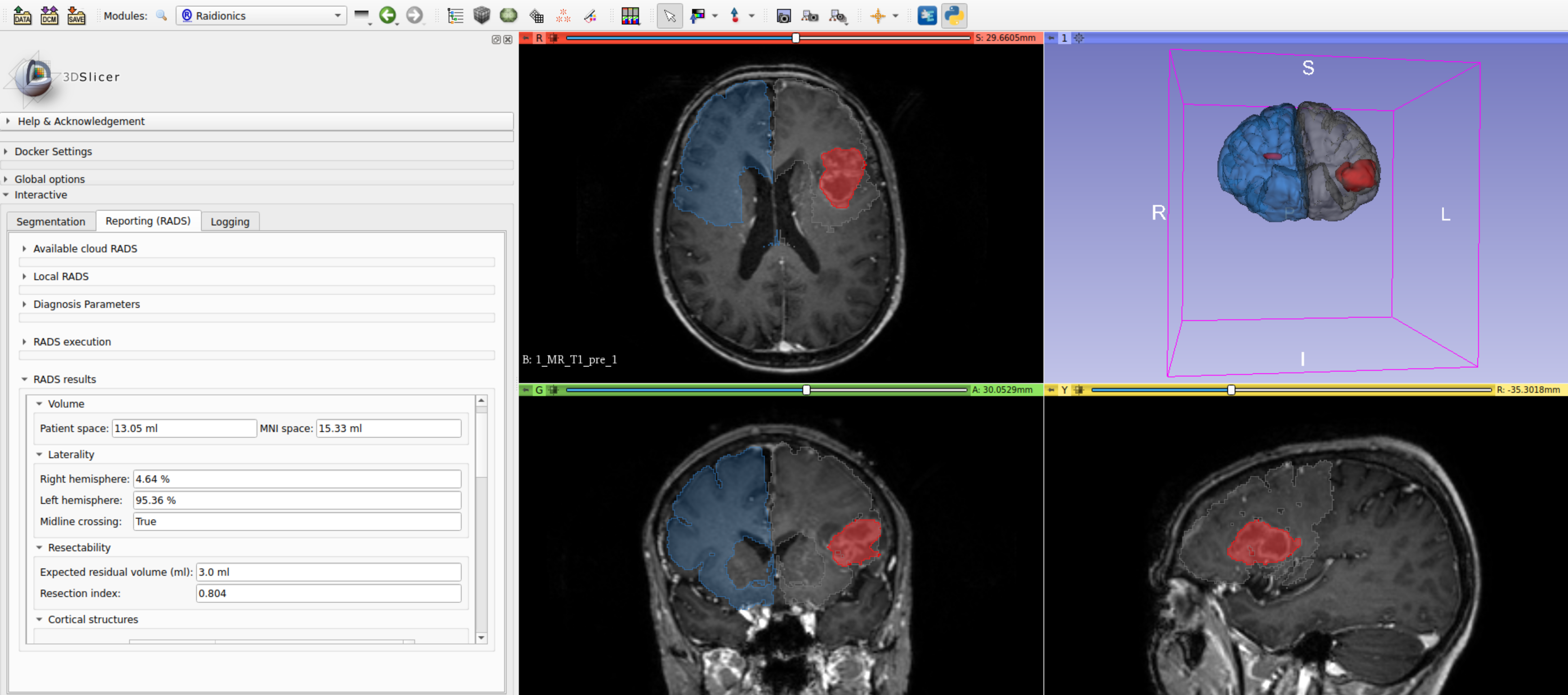}
\caption{Illustration of the Raidionics-Slicer plugin after generating the standardized report for a patient suffering from a glioblastoma.}
\label{fig:deepsintef-illu}
\end{figure}

The GitHub repository for the 3D Slicer plugin can be found at \url{https://github.com/dbouget/Raidionics-Slicer}, and an illustration is provided in \cref{fig:deepsintef-illu}.

\section{Validation studies}
\label{sec:validation}
In the validation studies, only the automatic segmentation performances are assessed. The clinical validity and relevance of the extracted tumor features has been addressed thoroughly in a previous  study~\cite{kommers2021glioblastoma}.
To better grasp the different aspects of the segmentation performance, a wider set of metrics is studied as described in \cref{subsec:metrics-description}. For the voxel-wise segmentation task, only two classes are considered as the whole tumor extent (including contrast-enhancing regions, cysts, and necrosis) is the target: non-tumor tissue or tumor tissue. In that sense, a positive voxel is a voxel exhibiting tumor tissue, whereas a negative voxel is a voxel exhibiting background or normal tissue.

\subsection{Metrics}
\label{subsec:metrics-description}
Following a review on metrics for evaluating 3D medical image segmentation~\cite{taha2015metrics}, a broad spectrum of $25$ metrics was selected, computed either voxel-wise or instance-wise, and grouped according to the following categories: overlap-based, volume-based, information theory-based, probabilistic, and spatial distance-based.

\paragraph{Voxel-wise:} For quantifying semantic segmentation performance, we have selected the following metrics computed directly and indiscriminately over all voxels of a given patient MRI volume:
\begin{enumerate}
    \item Overlap-based: (i) True Positive Rate (TPR), also called recall or sensitivity, is the probability that an actual positive voxel will test positive; (ii) True Negative Rate (TNR), also called specificity, is the probability that an actual negative voxel will test negative; (iii) False Positive Rate (FPR), is the probability that a false alarm will be raised (i.e., a negative voxel will test positive); (iv) False Negative Rate (FNR), also called missed rate, is the probability that a true positive voxel will test negative; (v) Positive Predictive Value (PPV), also referred to as precision, is the ratio of truly positive voxels over all voxels which tested positive; (vi) Dice score (Dice), also called the overlap index and gauging the similarity of two samples, is the most commonly used metric in validating medical volume segmentation~\cite{dice1945measures}; (vii) Dice True Positive score (Dice-TP) is similar to the Dice score, but is only computed over the true positive predictions (i.e., when the model found the tumor); (viii) Intersection Over Union (IoU), also called the Jaccard index, measures the volume similarity as the size of the intersection divided by the size of the union of two samples~\cite{jaccard1912distribution}; (ix) Global Consistency Error (GCE), defined as the error measure averaged over all voxels~\cite{martin2001database}.
    \item Volume-based: (i) Volumetric Similarity (VS), as the absolute volume difference divided by the sum of the compared volumes~\cite{cardenes2009multidimensional}; (ii) Relative Absolute Volume Difference (RAVD), as the relative absolute volume difference between the joint binary objects in the two images. This is a percentage value in the range $[-1.0, \infty)$ for which a $0$ denotes an ideal score.
    \item Information theory-based: (i) Normalized Mutual Information (MI), normalization of the mutual information score to scale the results between $0$ (no mutual information) and $1$ (perfect correlation)~\cite{russakoff2004image}; (ii) Variation Of Information (VOI), measuring the amount of information lost or gained when changing from one variable to the other, in this case to compare clustering partitions~\cite{meilua2003comparing}.
    \item Probabilistic: (i) Cohen's Kappa Score (CKS), measuring the agreement between two samples~\cite{cohen1960coefficient}. The metric ranges between $-1.0$ and $1.0$ whereby the maximum value means complete agreement, and zero or lower means chance agreement; (ii) Area Under the Curve (AUC), first presented as the measure of accuracy in the diagnostic radiology~\cite{bradley1997use}, further adjusted for the validation of machine learning algorithms; (iii) Volume Correlation (VC), as the linear correlation in binary object volume, measured through the Pearson product-moment correlation coefficient where the coefficient ranges $[-1., 1.]$; (iv) Matthews Correlation Coefficient (MCC), as a measure of the quality of binary and multiclass classifications, taking into account true and false positives and negatives and generally regarded as a balanced measure~\cite{baldi2000assessing}. The metric ranges between $-1.0$ and $1.0$ whereby $1.0$ represents a perfect prediction, $0.0$ an average random prediction, and $-1.0$ an inverse prediction; (v) Probabilistic Distance (PBD), as a measure of the distance between fuzzy segmentations~\cite{gerig2001valmet}.
    \item Spatial-distance-based: (i) $95$th percentile Hausdorff distance (HD$95$), measuring the boundary delineation quality (i.e., contours). The $95\%$ version is used to make measurements more robust to small outliers~\cite{huttenlocher1993comparing}; (ii) the Mahalanobis distance (MHD), measuring the correlation of all points and calculated according to the variant described for the validation of image segmentation~\cite{mclachlan1999mahalanobis}; (iii) Average Symmetric Surface Distance (ASSD), as the average symmetric surface distance between the binary objects in two images.
\end{enumerate}

\paragraph{Instance-wise:} For quantifying instance detection performance, we chose the following metrics, reported in a patient-wise fashion (PW) or in an object-wise fashion (OW). In the latter, and in case of multifocal tumors, each focus is considered as a separate tumor. The detection threshold has been set to $0.1$\% Dice to determine whether an automatic segmentation is eligible to be considered as a true detection or a false positive.
\begin{enumerate}
    \item Overlap-based: (i) Recall, as the ratio in \% of tumors properly identified; (ii) Precision, as the ratio in \% of tumors incorrectly detected; (iii) F1-score (F1), measuring information retrieval as a trade-off between the recall and precision~\cite{chinchor1993muc}; (iv) False Positives Per Patient (FPPP), as the average number of incorrect detections per patient.
    \item Probabilistic: (i) Adjusted Rand Index (ARI), as a similarity measure between two clusters by considering all pairs of samples and counting pairs that are assigned in the same or different clusters between the model prediction and the ground truth~\cite{hubert1985comparing}. The metric ranges from $-1.0$ to $1.0$, whereby random segmentation has an ARI close to $0.0$ and $1.0$ stands for perfect match.
    \item Spatial-distance-based: (i) Object Average Symmetric Surface Distance (OASSD), as the average symmetric surface distance (ASSD) between the binary objects in two volumes.
\end{enumerate}

\subsection{Measurements}
Pooled estimates, computed from each fold's results, are reported for each measurement~\cite{killeen2005alternative}. Overall, measurements are reported as mean and standard deviation (indicated by $\pm$) in the tables.

\paragraph{Voxel-wise:} For semantic segmentation performance, the Dice score is computed between the ground truth volume and a binary representation of the probability map generated by a trained model. The binary representation is computed for ten different equally-spaced probability thresholds (PT) in the range $]0, 1]$.

\paragraph{Instance-wise:} For instance detection performance, a connected components approach coupled to a pairing strategy was employed to associate ground truth and detected tumor parts. A minimum size threshold of $50$\,voxels was set and objects below that limit were discarded. A detection was deemed true positive for any Dice score strictly higher than $0$\%.

\subsection{Experiments}
To validate the trained models, the following set of experiments was conducted:
\begin{enumerate}[label=(\roman*)]
    \item \textbf{Overall performance study:} k-fold cross-validation studies for the different tumor types for assessing segmentation performance. For easy interpretation, only Dice scores together with patient-wise and object-wise recall, precision, and F1-score values are reported.
    \item \textbf{Metrics analysis:} in-depth performance comparison using the additional metrics, and confusion matrix computation between the metrics to identify redundancy in their use.
    \item \textbf{Representative models selection:} identification of one final segmentation model for each tumor type, which will be made available for use in our software solutions.
    \item \textbf{Speed study:} computation of the pure inference speed and the total elapsed time required to generate predictions for a new patient, obtained with CPU support and reported in seconds. The operations required to prepare the data to be sent through the network, to initialize the environment, to load the trained model, and to reconstruct the probability map in the referential space of the original volume are accounted for. The experiment was repeated ten consecutive times over the same MRI volume for each model, using a representative sample of each dataset in terms of dimension and spacing.
\end{enumerate}

\section{Results}
\label{sec:results}

\subsection{Implementation details}
Results were obtained using a computer with the following specifications: Intel Core Processor (Broadwell, no TSX, IBRS) CPU with $16$ cores, $64$GB of RAM, Tesla V100S ($32$GB) dedicated GPU, and a regular hard-drive. Training and inference processes were implemented in Python 3.6 using \texttt{TensorFlow} v1.13.1, and the data augmentation was performed using the Imgaug Python library~\cite{imgaug}. The metrics were for the most part computed manually using the equations described in the supplementary material, or alternatively using the sklearn v0.24.2~\cite{scikit-learn} and medpy v0.4.0~\cite{oskar_maier_2019_2565940} Python libraries. The source code used for computing the metrics and performing the validation studies is made publicly available at~\url{https://github.com/dbouget/validation_metrics_computation}.

\subsection{Overall performance study}
\label{subsec:overall-perf-results}

\begin{figure}[!ht]
\centering
\includegraphics[scale=0.75]{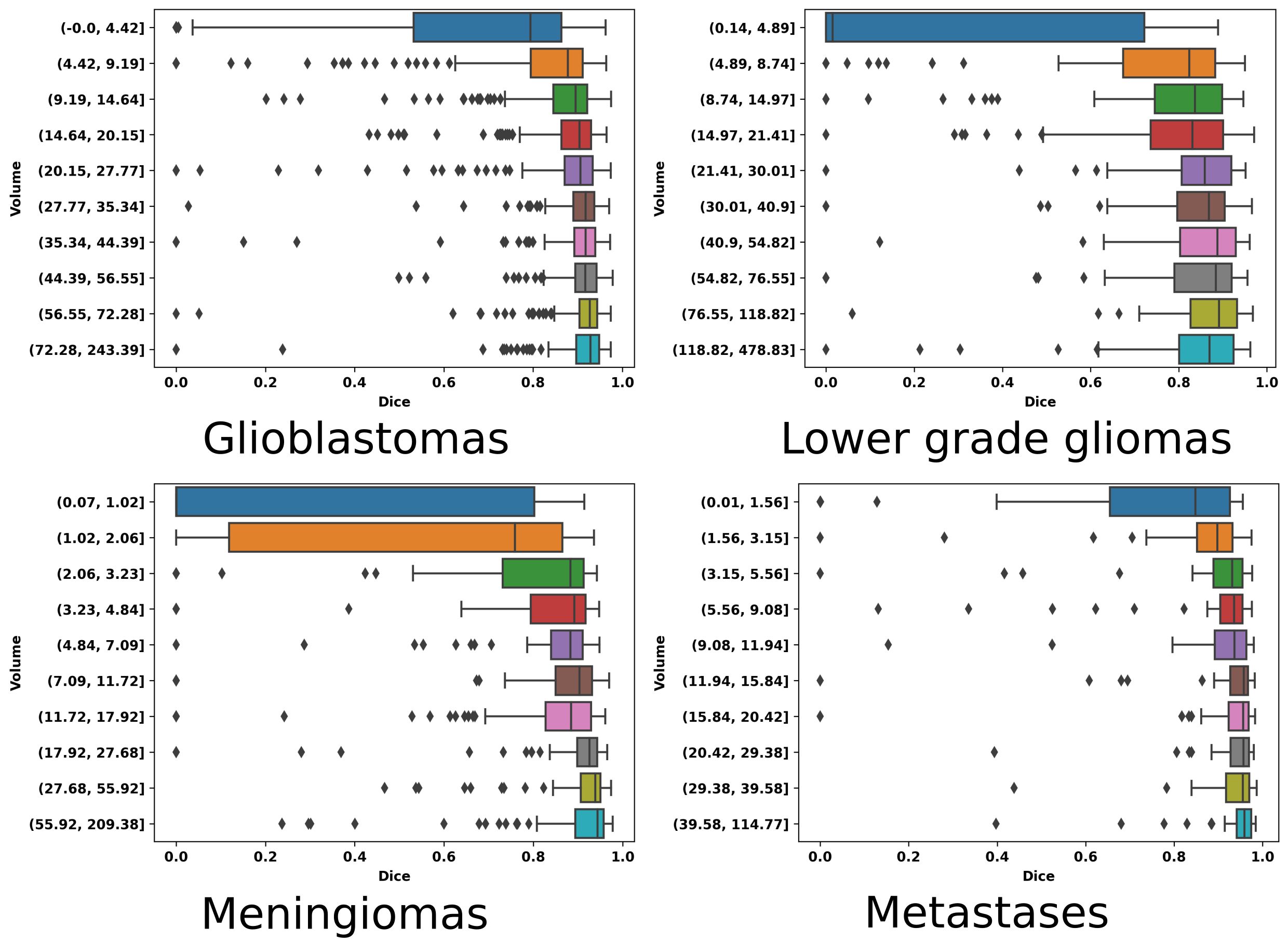}
\caption{Volume-wise (equally binned) Dice performance as boxplots for each of the four tumor types.}
\label{fig:volumewise-boxplots}
\end{figure}

\begin{table}[!hb]
\caption{Segmentation performance summary for each tumor type.}
\adjustbox{max width=\textwidth}{
\begin{tabular}{l|cc|ccc|ccc}
 & \multicolumn{2}{c|}{Voxel-wise} & \multicolumn{3}{c|}{Patient-wise} & \multicolumn{3}{c}{Object-wise} \tabularnewline
Tumor type & Dice & Dice-TP & F1-score & Recall & Precision & F1-score & Recall & Precision\tabularnewline
\hline
Glioblastoma & $85.69\pm16.97$ & $87.36\pm12.17$ & $97.40\pm01.01$ & $98.08\pm01.29$ & $96.76\pm01.43$ & $89.61\pm04.11$ & $85.78\pm07.95$ & $94.19\pm02.71$\tabularnewline
LGG & $75.39\pm25.95$ & $81.24\pm16.01$ & $93.60\pm01.74$ & $92.86\pm03.19$ & $94.42\pm01.07$ & $81.58\pm02.25$ & $75.58\pm02.41$ & $88.70\pm03.16$\tabularnewline
Meningioma & $75.00\pm30.52$ & $84.81\pm15.07$ & $90.67\pm01.42$ & $88.46\pm02.12$ & $93.25\pm04.76$ & $83.85\pm03.60$ & $80.93\pm04.34$ & $87.77\pm08.30$\tabularnewline
Metastasis  & $87.73\pm18.94$ & $90.02\pm12.80$ & $97.54\pm00.76$ & $97.46\pm01.38$ & $97.63\pm00.77$ & $88.71\pm01.34$ & $82.80\pm02.38$ & $95.60\pm01.45$\tabularnewline
\end{tabular}
}
\label{tab:all-seg-results}
\end{table}

\noindent An overall summary of brain tumor segmentation performance for all four tumor subtypes is presented in \cref{tab:all-seg-results}. Meningiomas and lower grade gliomas appear more difficult to segment given average Dice scores of $75$\%, compared to average Dice scores of $85$\% for glioblastomas and metastases. 
A similar trend, yet with a slightly smaller gap, can be noted for the Dice-TP scores ranging between $81$\% and $90$\% with a standard deviation around $15$\%, indicating the quality and relative stability of the trained models. From a patient-wise perspective, those results demonstrate the difficulty of achieving good recall while keeping the precision steadily above $95$\%.
Even though a direct comparison to the literature is impossible since different datasets have been used, obtained performance is on-par if not better than previously reported performances where Dice scores have been ranging from $75$\% to $85$\%.

Regarding the lower grade glioma tumor subtype, the diffuse nature of the tumors and less pronounced gradients over image intensities are possible explanations for the lower segmentation performance. For the meningioma category, the reason for the lower Dice-score and recall values can be attributed to the larger number of small tumors ($<2$\,ml) compared to other subtypes. In addition, outliers have been identified in this dataset whereby a small extent of the tumors were either partly enhancing because of calcification, or non-enhancing due to intraosseous growth.
For all tumor types, Dice-score distributions are reported against tumor volumes in \cref{fig:volumewise-boxplots} for ten equally-sized bins. For meningiomas, four bins are necessary to group tumors with a volume up to $4$\,ml while only one bin is necessary for the glioblastomas, indicating a volume distribution imbalance between the two types. The diamond-shaped points outside the boxes represent cases where the segmentation model did not perform well (cf. Figures S1, S2, S3, and S4, Sup. Mat.).

\begin{figure}[!ht]
\centering
\includegraphics[scale=1.15]{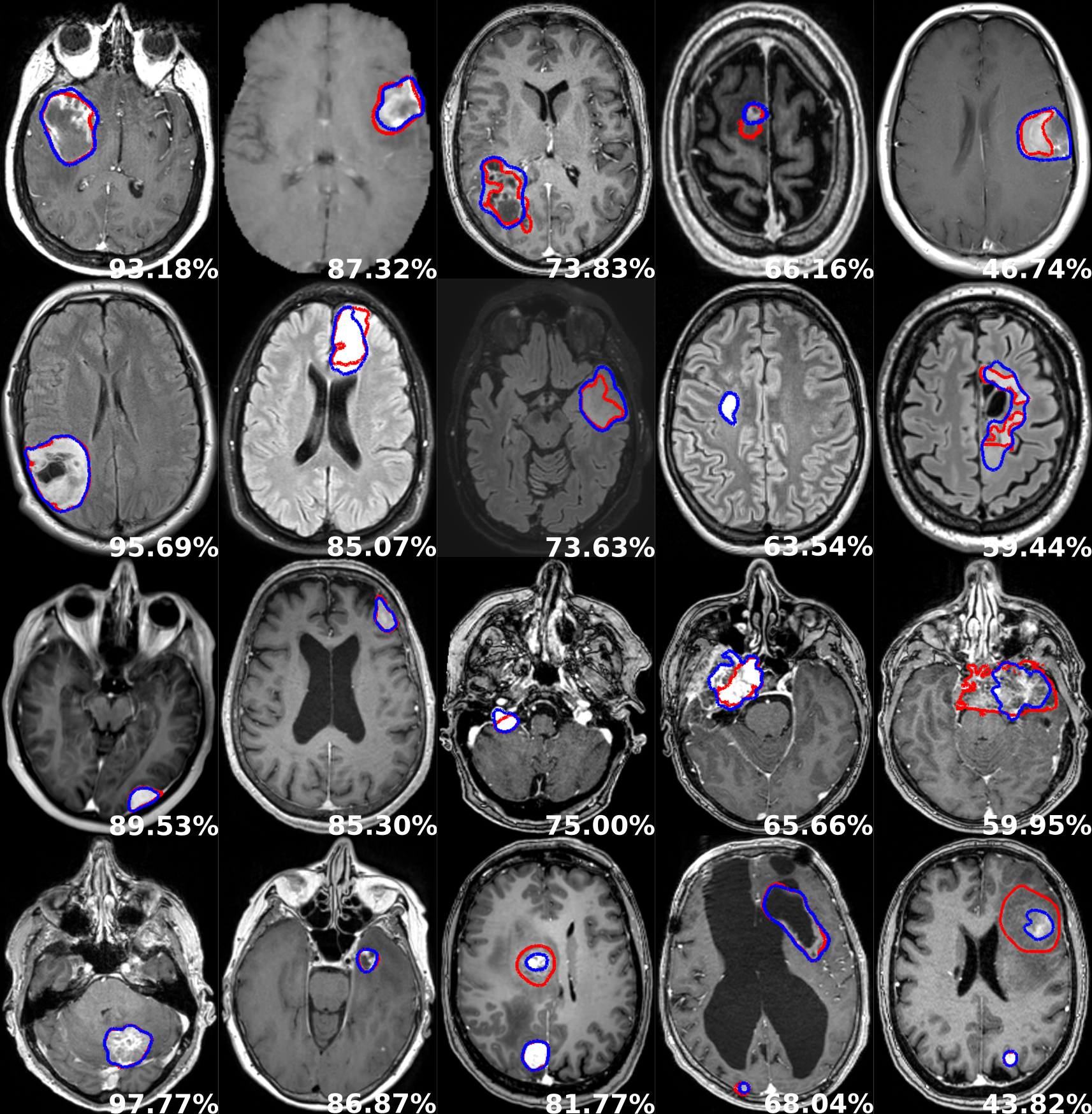}
\caption{Examples of segmentation performances. One row illustrates one tumor type: glioblastoma, lower grade glioma, meningioma, metastasis (from top to bottom), and each column depicts a different patient. The manual delineation is shown in red, the automatic segmentation in blue, and the patient-wise Dice score in white.}
\label{fig:overall-seg-results}
\end{figure}

While tumor volumes and outlier MR scans are reasons for the discrepancy in Dice and recall values across the board, precision is rather unaffected and more stable. The nature of the convolutional neural network architecture and training strategy used can explain those results. By leveraging volumes covering the full brain, global relationships can be learned by the trained model hence reducing the confusion between tumor regions and other contrast-enhancing structures such as blood vessels. Given GPU memory limitation, the preprocessed MR scans have undergone a significant downsampling, and as such small tumors are reduced to very few voxels, impacting mainly recall performance.

\noindent Finally, an average decrease of $\sim10$\% can be noticed between patient-wise and object-wise detection metrics, whereby satellite tumors are on average an order of magnitude smaller than the main tumor, and are hence more prone to be omitted or poorly segmented by our models. Segmentation performance is illustrated in \cref{fig:overall-seg-results}. Each row corresponds to one tumor type and each column depict a different patient.

\subsection{Metrics analysis}
\label{subsec:indepth-metrics-results}
Side-by-side voxel-wise performances regarding the overlap-based metrics are reported in \cref{tab:indepth-seg-results-overlapbased} and \cref{tab:indepth-seg-results-overlapvolumebased}. Unsurprisingly, given the good precision performance and the absence of patients without a tumor, both TNR and its opposite FPR scores are almost perfect for all tumor types. Similarly, the TPR and its opposite FNR metrics are scoring similarly to Dice. Within each tumor category, the overlap-based metrics are following the same trend whereby a higher average Dice score would correlate with a higher score for any other metrics and vice versa (e.g.,IoU). An exception can be made regarding the behaviour of the GCE metric, scoring on average higher for glioblastomas than for meningiomas and as such not following the same pattern as Dice. Upon careful visual inspection, the GCE metric seems to be extremely sensitive to outliers, either coming from the image quality or manual ground truth correctness (cf. top row in Figures S1-S4, Sup. Mat.). Given the non-normalized state of the GCE metric, and its absence of any upper bound, an extremely poor agreement between manual ground truth and automatic segmentation will result in a score orders of magnitude higher than its average expression over a given dataset.
Regarding the two volume-based metrics, featured rightmost in the second table, an antagonistic pattern towards Dice can be observed. The VS metric has the same cross-type trend as Dice with similar yet slightly greater scores. On the other hand, while the RAVD metric scores best over the metastasis group similar to Dice, its worst average value is obtained for the glioblastoma group, hence potentially exhibiting the same frailty towards outliers as for the GCE metric.

\begin{table}[!h]
\centering
\caption{Voxel-wise overlap-based metrics performance summary for each tumor type.}
\adjustbox{max width=\textwidth}{
\begin{tabular}{l|ccccc}
Tumor type & TPR & TNR & FPR & FNR & PPV\tabularnewline
\hline
Glioblastoma & $87.88\pm17.64$ & $99.96\pm00.06$ & $00.04\pm00.06$ & $12.12\pm17.64$ & $87.35\pm13.29$ \tabularnewline
LGG & $77.91\pm27.89$ & $99.90\pm00.16$ & $00.09\pm00.16$ & $22.08\pm27.89$ & $82.16\pm17.01$ \tabularnewline
Meningioma & $77.44\pm32.48$ & $99.97\pm00.04$ & $00.02\pm00.04$ & $22.56\pm32.48$ & $84.77\pm15.69$ \tabularnewline
Metastasis & $88.45\pm20.82$ & $99.98\pm00.03$ & $00.01\pm00.03$ & $11.54\pm20.82$ & $89.43\pm16.78$ \tabularnewline
\end{tabular}
}
\label{tab:indepth-seg-results-overlapbased}
\end{table}

\begin{table}[!h]
\caption{Voxel-wise performance summary for each tumor type for overlap-based and volume-based metrics.}
\adjustbox{max width=\textwidth}{
\begin{tabular}{l|cccc|cc}
& \multicolumn{4}{c|}{Overlap-based} & \multicolumn{2}{c}{Volume-based}\tabularnewline
Tumor type & Dice & Dice-TP & IoU & GCE (1e4) & VS & RAVD\tabularnewline
\hline
Glioblastoma & $85.69\pm16.97$ & $87.36\pm12.17$ & $77.59\pm17.99$ & $12.34\pm12.57$ & $90.43\pm16.94$& $13.98\pm171.2$  \tabularnewline
LGG & $75.39\pm25.95$ & $81.24\pm16.01$ & $65.72\pm25.32$ & $34.15\pm46.34$ & $82.20\pm26.44$ & $07.88\pm60.14$ \tabularnewline
Meningioma & $75.00\pm30.52$ & $84.81\pm15.07$ & $67.13\pm29.39$ & $09.04\pm17.53$ & $80.21\pm31.08$ & $07.87\pm61.31$ \tabularnewline
Metastasis & $87.73\pm18.94$ & $90.02\pm12.80$ & $81.56\pm20.42$ & $04.55\pm07.62$ & $91.37\pm18.61$ & $02.11\pm55.35$ \tabularnewline
\end{tabular}
}
\label{tab:indepth-seg-results-overlapvolumebased}
\end{table}

Next off, voxel-wise performance for information theory-based and probabilistic metric are regrouped in \cref{tab:indepth-seg-results-infoproba-metrics}. The MI and VOI metrics, both based on information theory, are exhibiting an inverse behaviour in line with observations about the relationship between Dice and GCE metrics. The normalized mutual information metric ranges from $0.668$ to $0.829$ for Dice scores between $75$\% and $87$\%, showcasing stability but also correlation. On the contrary, the VOI metric expresses a behaviour concurrent to GCE whereby worst performance is obtained for the lower grade gliomas and then glioblastomas categories, while it performs best over metastases where Dice also scores the highest.
Alike the aforementioned metric groups exhibiting inner discrepancies, three of the five probabilistic metrics follow a similar trend scoring high alongside Dice, with an average gap of $0.1$ corresponding to a $\sim10$\% Dice score difference. Meanwhile, the PBD metric has a behaviour of its own scoring an order of magnitude worse for the meningioma category than for the three other subtypes. The metric is not normalized and an extremely poor agreement between the manual ground truth and automatic segmentation would result in an extremely large score, similar to the GCE metric, hence reporting the median score in addition might be of interest (cf. second row in Figures S1-S4, Sup. Mat.).

\begin{table}[!h]
\caption{Voxel-wise performance summary for each tumor type for information theory-based and probabilistic metrics.}
\adjustbox{max width=\textwidth}{
\begin{tabular}{l|cc|ccccc}
& \multicolumn{2}{c|}{Information theory-based} & \multicolumn{5}{c}{Probabilistic}\tabularnewline
Tumor type & MI & VOI & CKS & AUC & VC & MCC & PBD \tabularnewline
\hline
Glioblastoma & $0.787\pm0.168$ & $0.011\pm0.009$ & $0.856\pm0.169$ & $0.939\pm0.088$ & $0.978\pm0.089$ & $0.875\pm0.122$ & $0.840\pm24.02$\tabularnewline
LGG & $0.668\pm0.246$ & $0.026\pm0.030$ & $0.753\pm0.259$ & $0.889\pm0.139$ & $0.961\pm0.119$ & $0.812\pm0.167$ & $0.573\pm04.82$\tabularnewline
Meningioma & $0.691\pm0.291$ & $0.008\pm0.013$ & $0.749\pm0.305$ & $0.887\pm0.162$ & $0.954\pm0.149$ & $0.841\pm0.171$ & $5.358\pm103.4$\tabularnewline
Metastasis & $0.829\pm0.191$ & $0.004\pm0.006$ & $0.877\pm0.189$ & $0.942\pm0.104$ & $0.978\pm0.100$ & $0.901\pm0.127$ & $0.152\pm0.623$\tabularnewline
\end{tabular}
}
\label{tab:indepth-seg-results-infoproba-metrics}
\end{table}

\noindent Finally, the voxel-wise distance-based metrics are reported in \cref{tab:indepth-seg-results-distance}. Similar cross-type trends can also be noted whereby the best HD95 of $4.97$\,mm is obtained for the glioblastoma category and the worst HD95 of $10$\,mm for meningiomas, heavily correlated to Dice performance. Our average HD95 results appear lower than previously reported results in the literature, however a strong statement can hardly be made as the tumors featured can vary highly in terms of volume and number of satellites which might reflect greatly on metrics' average scores. The other two spatial distance-based metrics display a similar behaviour to HD95, whereby tumor types can be ranked as follows based on best to worse performance: glioblastoma, metastasis, lower grade glioma, and meningioma.

\begin{table}[!h]
\centering
\caption{Voxel-wise performance summary for each tumor type for spatial distance-based metrics.}
\adjustbox{max width=\textwidth}{
\begin{tabular}{l|ccc}
Tumor type & HD95 & MHD & ASSD \tabularnewline
\hline
Glioblastoma & $04.97\pm09.06$ & $00.41\pm03.69$ & $01.46\pm03.22$ \tabularnewline
LGG & $08.37\pm13.31$ & $00.53\pm03.27$ &$02.19\pm05.06$ \tabularnewline
Meningioma & $10.11\pm21.82$ & $00.72\pm03.57$ & $02.77\pm07.91$ \tabularnewline
Metastasis & $07.54\pm20.61$ & $00.54\pm04.56$ & $01.73\pm05.89$ \tabularnewline
\end{tabular}
}
\label{tab:indepth-seg-results-distance}
\end{table}

\noindent Regarding instance-wise metrics, grouped in \cref{tab:indepth-seg-results-instancewise}, the close OASSD average values between glioblastomas and meningiomas represents the most surprising outcome given the $5$\% difference in F1-score. Unsurprisingly, the lower grade glioma category achieves the highest average OASSD with $2.6$\,mm together with the lowest F1-score. As one might expect, the amount of FPPP correlates greatly with the average precision values obtained. Ultimately, the ARI metric generates scores extremely similar to voxel-wise Dice and correlates highly with the F1-score whereby the glioblastoma and metastasis categories obtain almost $0.1$ more than for the meningioma and lower grade glioma subtypes.

\begin{table}[!h]
\caption{Instance-wise performance for each tumor type.}
\adjustbox{max width=\textwidth}{
\begin{tabular}{l|cccccc}
Tumor type & F1-score & Recall & Precision & FPPP & ARI & OASSD \tabularnewline
\hline
Glioblastoma & $89.61\pm04.11$ & $85.78\pm07.95$ & $94.19\pm02.71$ & $0.078\pm0.037$ & $0.856\pm0.169$ & $01.45\pm02.82$ \tabularnewline
LGG & $81.58\pm02.25$ & $75.57\pm02.40$ & $88.67\pm03.16$ & $0.129\pm0.041$ & $0.751\pm0.259$ & $02.60\pm06.10$\tabularnewline
Meningioma & $83.85\pm03.60$ & $80.93\pm04.34$ & $87.77\pm08.30$ & $0.151\pm0.128$ & $0.749\pm0.305$ & $01.62\pm04.09$\tabularnewline
Metastasis & $88.71\pm01.34$ & $82.79\pm02.38$ & $95.60\pm01.45$ & $0.061\pm0.020$ & $0.877\pm0.189$ & $0.672\pm0.869$ \tabularnewline
\end{tabular}
}
\label{tab:indepth-seg-results-instancewise}
\end{table}

\noindent For completeness, the correlation between the different metrics computed in this study has been assessed, and the results over the glioblastoma category are shown in \cref{fig:confusion-matrix-glioblastoma-main} (cf. other correlation matrices in Tables S2, S4, S6, and S8, Sup. Mat.). Some metrics have been excluded given inherent correlation from their computation, such as FPR and FNR being the opposite of TNR and TPR. Similarly, metrics having computation in a voxel-wise, patient-wise, or instance-wise fashion were not considered in the matrix (i.e., recall, precision, and F1-score). 
Overall, the conclusions identified by analyzing the raw average results are further confirmed whereby a majority of voxel-wise metrics correlate with one another and thus do not bring any additional information to Dice. However, relevant insight can be obtained from the RAVD and GCE/VOI metrics given their low correlation to Dice and their higher sensitivity towards outliers, enabling to quantify the ability to generalize of the model or potentially the quality of the data and manual ground truth (cf. third row in Figures S1-S4, Sup. Mat.). The correlation between HD95 and MHD appears also quite low for spatial distance-based metrics, indicating a potential usefulness. Finally, in the instance-wise category, the OASSD is a stand-alone metric offering to properly assess model performance over the  detection of satellite tumors.
To conclude, a final pool of metrics to consider for benchmarking purposes and capturing all aspects of the segmentation performances are: Dice, RAVD, VOI, HD95, MHD, and OASSD. Given the task, reporting patient-wise and instance-wise recall, precision, and F1-score is always of interest because of an innate comprehension of their meaning, easy to interpret for clinicians or other end-users.

\begin{table}[h]
\caption{Metrics correlation matrix for glioblastoma segmentation. The color intensity of each cell represents the strength of the correlation, where blue denotes direct correlation and red denotes inverse correlation.}
\label{fig:confusion-matrix-glioblastoma-main}
\adjustbox{max width=\textwidth}{
\begin{tabular}{lrrrrrr|rr|rr|rrrrr|rrr|rr}
\toprule
 & \multicolumn{6}{c|}{Overlap} & \multicolumn{2}{c|}{Volume} & \multicolumn{2}{c|}{Information theory} & \multicolumn{5}{c|}{Probabilistic} & \multicolumn{3}{c|}{Spatial distance} & \multicolumn{2}{c}{Instance-wise}\tabularnewline
{} & Dice & TPR & TNR & PPV & IoU & GCE & VS & RAVD & MI & VOI & CKS & AUC & VC & MCC & PBD & HD95 & MHD & ASSD & ARI & OASSD\tabularnewline
Dice & \cellcolor{blue!100}1.0 & \cellcolor{blue!70}0.7 & \cellcolor{blue!28}0.29 & \cellcolor{blue!61}0.62 & \cellcolor{blue!98}0.98 & \cellcolor{red!22}-0.22 & \cellcolor{blue!94}0.94 & \cellcolor{red!34}-0.35 & \cellcolor{blue!98}0.99 & \cellcolor{red!23}-0.23 & \cellcolor{blue!99}1.0 & \cellcolor{blue!70}0.71 & \cellcolor{blue!78}0.78 & \cellcolor{blue!99}1.0 & \cellcolor{red!34}-0.34 & \cellcolor{red!54}-0.55 & \cellcolor{red!43}-0.43 & \cellcolor{red!70}-0.71 & \cellcolor{blue!99}1.0 & \cellcolor{red!30}-0.3\tabularnewline
TPR & \cellcolor{blue!70}0.7 & \cellcolor{blue!100}1.0 & \cellcolor{red!16}-0.17 & \cellcolor{red!6}-0.07 & \cellcolor{blue!70}0.71 & \cellcolor{red!8}-0.08 & \cellcolor{blue!62}0.62 & \cellcolor{blue!9}0.1 & \cellcolor{blue!70}0.7 & \cellcolor{red!8}-0.08 & \cellcolor{blue!70}0.7 & \cellcolor{blue!99}1.0 & \cellcolor{blue!51}0.51 & \cellcolor{blue!70}0.71 & \cellcolor{red!26}-0.26 & \cellcolor{red!37}-0.38 & \cellcolor{red!33}-0.34 & \cellcolor{red!47}-0.47 & \cellcolor{blue!70}0.7 & \cellcolor{red!20}-0.2\tabularnewline
TNR & \cellcolor{blue!28}0.29 & \cellcolor{red!16}-0.17 & \cellcolor{blue!100}1.0 & \cellcolor{blue!57}0.58 & \cellcolor{blue!28}0.28 & \cellcolor{red!75}-0.76 & \cellcolor{blue!28}0.29 & \cellcolor{red!36}-0.36 & \cellcolor{blue!33}0.33 & \cellcolor{red!75}-0.76 & \cellcolor{blue!28}0.29 & \cellcolor{red!16}-0.17 & \cellcolor{blue!22}0.23 & \cellcolor{blue!28}0.29 & \cellcolor{red!4}-0.04 & \cellcolor{red!16}-0.16 & \cellcolor{red!4}-0.04 & \cellcolor{red!27}-0.27 & \cellcolor{blue!29}0.29 & \cellcolor{red!22}-0.22\tabularnewline
PPV & \cellcolor{blue!61}0.62 & \cellcolor{red!6}-0.07 & \cellcolor{blue!57}0.58 & \cellcolor{blue!100}1.0 & \cellcolor{blue!64}0.64 & \cellcolor{red!24}-0.24 & \cellcolor{blue!54}0.55 & \cellcolor{red!49}-0.49 & \cellcolor{blue!64}0.64 & \cellcolor{red!24}-0.25 & \cellcolor{blue!61}0.62 & \cellcolor{red!6}-0.07 & \cellcolor{blue!47}0.47 & \cellcolor{blue!62}0.63 & \cellcolor{red!15}-0.16 & \cellcolor{red!37}-0.38 & \cellcolor{red!21}-0.21 & \cellcolor{red!47}-0.47 & \cellcolor{blue!61}0.62 & \cellcolor{red!21}-0.22\tabularnewline
IoU & \cellcolor{blue!98}0.98 & \cellcolor{blue!70}0.71 & \cellcolor{blue!28}0.28 & \cellcolor{blue!64}0.64 & \cellcolor{blue!100}1.0 & \cellcolor{red!23}-0.24 & \cellcolor{blue!89}0.9 & \cellcolor{red!28}-0.29 & \cellcolor{blue!99}0.99 & \cellcolor{red!23}-0.24 & \cellcolor{blue!98}0.98 & \cellcolor{blue!70}0.71 & \cellcolor{blue!71}0.71 & \cellcolor{blue!98}0.99 & \cellcolor{red!27}-0.28 & \cellcolor{red!55}-0.55 & \cellcolor{red!37}-0.37 & \cellcolor{red!69}-0.7 & \cellcolor{blue!98}0.98 & \cellcolor{red!30}-0.31\tabularnewline
GCE & \cellcolor{red!22}-0.22 & \cellcolor{red!8}-0.08 & \cellcolor{red!75}-0.76 & \cellcolor{red!24}-0.24 & \cellcolor{red!23}-0.24 & \cellcolor{blue!100}1.0 & \cellcolor{red!19}-0.19 & \cellcolor{blue!12}0.13 & \cellcolor{red!30}-0.3 & \cellcolor{blue!99}1.0 & \cellcolor{red!22}-0.23 & \cellcolor{red!8}-0.09 & \cellcolor{red!13}-0.14 & \cellcolor{red!22}-0.23 & \cellcolor{blue!2}0.02 & \cellcolor{blue!17}0.18 & \cellcolor{blue!2}0.03 & \cellcolor{blue!28}0.29 & \cellcolor{red!22}-0.23 & \cellcolor{blue!27}0.28\tabularnewline
\hline
VS & \cellcolor{blue!94}0.94 & \cellcolor{blue!62}0.62 & \cellcolor{blue!28}0.29 & \cellcolor{blue!54}0.55 & \cellcolor{blue!89}0.9 & \cellcolor{red!19}-0.19 & \cellcolor{blue!100}1.0 & \cellcolor{red!37}-0.37 & \cellcolor{blue!90}0.9 & \cellcolor{red!20}-0.2 & \cellcolor{blue!94}0.94 & \cellcolor{blue!62}0.62 & \cellcolor{blue!75}0.76 & \cellcolor{blue!92}0.92 & \cellcolor{red!35}-0.36 & \cellcolor{red!48}-0.48 & \cellcolor{red!43}-0.43 & \cellcolor{red!65}-0.65 & \cellcolor{blue!94}0.94 & \cellcolor{red!26}-0.26\tabularnewline
RAVD & \cellcolor{red!34}-0.35 & \cellcolor{blue!9}0.1 & \cellcolor{red!36}-0.36 & \cellcolor{red!49}-0.49 & \cellcolor{red!28}-0.29 & \cellcolor{blue!12}0.13 & \cellcolor{red!37}-0.37 & \cellcolor{blue!100}1.0 & \cellcolor{red!31}-0.31 & \cellcolor{blue!14}0.15 & \cellcolor{red!34}-0.35 & \cellcolor{blue!9}0.1 & \cellcolor{red!38}-0.39 & \cellcolor{red!34}-0.34 & \cellcolor{blue!17}0.18 & \cellcolor{blue!19}0.19 & \cellcolor{blue!14}0.14 & \cellcolor{blue!28}0.28 & \cellcolor{red!34}-0.35 & \cellcolor{blue!14}0.15\tabularnewline
\hline
MI & \cellcolor{blue!98}0.99 & \cellcolor{blue!70}0.7 & \cellcolor{blue!33}0.33 & \cellcolor{blue!64}0.64 & \cellcolor{blue!99}0.99 & \cellcolor{red!30}-0.3 & \cellcolor{blue!90}0.9 & \cellcolor{red!31}-0.31 & \cellcolor{blue!100}1.0 & \cellcolor{red!30}-0.31 & \cellcolor{blue!98}0.99 & \cellcolor{blue!70}0.7 & \cellcolor{blue!73}0.74 & \cellcolor{blue!99}0.99 & \cellcolor{red!30}-0.31 & \cellcolor{red!56}-0.56 & \cellcolor{red!39}-0.4 & \cellcolor{red!71}-0.71 & \cellcolor{blue!98}0.99 & \cellcolor{red!32}-0.32\tabularnewline
VOI & \cellcolor{red!23}-0.23 & \cellcolor{red!8}-0.08 & \cellcolor{red!75}-0.76 & \cellcolor{red!24}-0.25 & \cellcolor{red!23}-0.24 & \cellcolor{blue!99}1.0 & \cellcolor{red!20}-0.2 & \cellcolor{blue!14}0.15 & \cellcolor{red!30}-0.31 & \cellcolor{blue!100}1.0 & \cellcolor{red!23}-0.23 & \cellcolor{red!8}-0.08 & \cellcolor{red!15}-0.15 & \cellcolor{red!23}-0.24 & \cellcolor{blue!2}0.03 & \cellcolor{blue!18}0.18 & \cellcolor{blue!3}0.03 & \cellcolor{blue!29}0.3 & \cellcolor{red!23}-0.24 & \cellcolor{blue!28}0.28\tabularnewline
\hline
CKS & \cellcolor{blue!99}1.0 & \cellcolor{blue!70}0.7 & \cellcolor{blue!28}0.29 & \cellcolor{blue!61}0.62 & \cellcolor{blue!98}0.98 & \cellcolor{red!22}-0.23 & \cellcolor{blue!94}0.94 & \cellcolor{red!34}-0.35 & \cellcolor{blue!98}0.99 & \cellcolor{red!23}-0.23 & \cellcolor{blue!100}1.0 & \cellcolor{blue!70}0.71 & \cellcolor{blue!78}0.78 & \cellcolor{blue!99}1.0 & \cellcolor{red!34}-0.34 & \cellcolor{red!54}-0.55 & \cellcolor{red!43}-0.43 & \cellcolor{red!70}-0.71 & \cellcolor{blue!99}1.0 & \cellcolor{red!30}-0.3\tabularnewline
AUC & \cellcolor{blue!70}0.71 & \cellcolor{blue!99}1.0 & \cellcolor{red!16}-0.17 & \cellcolor{red!6}-0.07 & \cellcolor{blue!70}0.71 & \cellcolor{red!8}-0.09 & \cellcolor{blue!62}0.62 & \cellcolor{blue!9}0.1 & \cellcolor{blue!70}0.7 & \cellcolor{red!8}-0.08 & \cellcolor{blue!70}0.71 & \cellcolor{blue!100}1.0 & \cellcolor{blue!51}0.51 & \cellcolor{blue!71}0.71 & \cellcolor{red!26}-0.27 & \cellcolor{red!38}-0.38 & \cellcolor{red!33}-0.34 & \cellcolor{red!47}-0.47 & \cellcolor{blue!70}0.71 & \cellcolor{red!20}-0.2\tabularnewline
VC & \cellcolor{blue!78}0.78 & \cellcolor{blue!51}0.51 & \cellcolor{blue!22}0.23 & \cellcolor{blue!47}0.47 & \cellcolor{blue!71}0.71 & \cellcolor{red!13}-0.14 & \cellcolor{blue!75}0.76 & \cellcolor{red!38}-0.39 & \cellcolor{blue!73}0.74 & \cellcolor{red!15}-0.15 & \cellcolor{blue!78}0.78 & \cellcolor{blue!51}0.51 & \cellcolor{blue!100}1.0 & \cellcolor{blue!77}0.78 & \cellcolor{red!49}-0.49 & \cellcolor{red!51}-0.51 & \cellcolor{red!58}-0.58 & \cellcolor{red!71}-0.71 & \cellcolor{blue!78}0.78 & \cellcolor{red!22}-0.22\tabularnewline
MCC & \cellcolor{blue!99}1.0 & \cellcolor{blue!70}0.71 & \cellcolor{blue!28}0.29 & \cellcolor{blue!62}0.63 & \cellcolor{blue!98}0.99 & \cellcolor{red!22}-0.23 & \cellcolor{blue!92}0.92 & \cellcolor{red!34}-0.34 & \cellcolor{blue!99}0.99 & \cellcolor{red!23}-0.24 & \cellcolor{blue!99}1.0 & \cellcolor{blue!71}0.71 & \cellcolor{blue!77}0.78 & \cellcolor{blue!100}1.0 & \cellcolor{red!35}-0.36 & \cellcolor{red!55}-0.55 & \cellcolor{red!44}-0.44 & \cellcolor{red!71}-0.71 & \cellcolor{blue!99}1.0 & \cellcolor{red!30}-0.31\tabularnewline
PBD & \cellcolor{red!34}-0.34 & \cellcolor{red!26}-0.26 & \cellcolor{red!4}-0.04 & \cellcolor{red!15}-0.16 & \cellcolor{red!27}-0.28 & \cellcolor{blue!2}0.02 & \cellcolor{red!35}-0.36 & \cellcolor{blue!17}0.18 & \cellcolor{red!30}-0.31 & \cellcolor{blue!2}0.03 & \cellcolor{red!34}-0.34 & \cellcolor{red!26}-0.27 & \cellcolor{red!49}-0.49 & \cellcolor{red!35}-0.36 & \cellcolor{blue!100}1.0 & \cellcolor{blue!15}0.16 & \cellcolor{blue!97}0.97 & \cellcolor{blue!28}0.29 & \cellcolor{red!34}-0.34 & \cellcolor{blue!5}0.05\tabularnewline
\hline
HD95 & \cellcolor{red!54}-0.55 & \cellcolor{red!37}-0.38 & \cellcolor{red!16}-0.16 & \cellcolor{red!37}-0.38 & \cellcolor{red!55}-0.55 & \cellcolor{blue!17}0.18 & \cellcolor{red!48}-0.48 & \cellcolor{blue!19}0.19 & \cellcolor{red!56}-0.56 & \cellcolor{blue!18}0.18 & \cellcolor{red!54}-0.55 & \cellcolor{red!38}-0.38 & \cellcolor{red!51}-0.51 & \cellcolor{red!55}-0.55 & \cellcolor{blue!15}0.16 & \cellcolor{blue!100}1.0 & \cellcolor{blue!25}0.25 & \cellcolor{blue!89}0.89 & \cellcolor{red!55}-0.55 & \cellcolor{blue!13}0.14\tabularnewline
MHD & \cellcolor{red!43}-0.43 & \cellcolor{red!33}-0.34 & \cellcolor{red!4}-0.04 & \cellcolor{red!21}-0.21 & \cellcolor{red!37}-0.37 & \cellcolor{blue!2}0.03 & \cellcolor{red!43}-0.43 & \cellcolor{blue!14}0.14 & \cellcolor{red!39}-0.4 & \cellcolor{blue!3}0.03 & \cellcolor{red!43}-0.43 & \cellcolor{red!33}-0.34 & \cellcolor{red!58}-0.58 & \cellcolor{red!44}-0.44 & \cellcolor{blue!97}0.97 & \cellcolor{blue!25}0.25 & \cellcolor{blue!100}1.0 & \cellcolor{blue!39}0.4 & \cellcolor{red!43}-0.43 & \cellcolor{blue!5}0.06\tabularnewline
ASSD & \cellcolor{red!70}-0.71 & \cellcolor{red!47}-0.47 & \cellcolor{red!27}-0.27 & \cellcolor{red!47}-0.47 & \cellcolor{red!69}-0.7 & \cellcolor{blue!28}0.29 & \cellcolor{red!65}-0.65 & \cellcolor{blue!28}0.28 & \cellcolor{red!71}-0.71 & \cellcolor{blue!29}0.3 & \cellcolor{red!70}-0.71 & \cellcolor{red!47}-0.47 & \cellcolor{red!71}-0.71 & \cellcolor{red!71}-0.71 & \cellcolor{blue!28}0.29 & \cellcolor{blue!89}0.89 & \cellcolor{blue!39}0.4 & \cellcolor{blue!100}1.0 & \cellcolor{red!71}-0.71 & \cellcolor{blue!19}0.2\tabularnewline
\hline
ARI & \cellcolor{blue!99}1.0 & \cellcolor{blue!70}0.7 & \cellcolor{blue!29}0.29 & \cellcolor{blue!61}0.62 & \cellcolor{blue!98}0.98 & \cellcolor{red!22}-0.23 & \cellcolor{blue!94}0.94 & \cellcolor{red!34}-0.35 & \cellcolor{blue!98}0.99 & \cellcolor{red!23}-0.24 & \cellcolor{blue!99}1.0 & \cellcolor{blue!70}0.71 & \cellcolor{blue!78}0.78 & \cellcolor{blue!99}1.0 & \cellcolor{red!34}-0.34 & \cellcolor{red!55}-0.55 & \cellcolor{red!43}-0.43 & \cellcolor{red!71}-0.71 & \cellcolor{blue!100}1.0 & \cellcolor{red!30}-0.3\tabularnewline
OASSD & \cellcolor{red!30}-0.3 & \cellcolor{red!20}-0.2 & \cellcolor{red!22}-0.22 & \cellcolor{red!21}-0.22 & \cellcolor{red!30}-0.31 & \cellcolor{blue!27}0.28 & \cellcolor{red!26}-0.26 & \cellcolor{blue!14}0.15 & \cellcolor{red!32}-0.32 & \cellcolor{blue!28}0.28 & \cellcolor{red!30}-0.3 & \cellcolor{red!20}-0.2 & \cellcolor{red!22}-0.22 & \cellcolor{red!30}-0.31 & \cellcolor{blue!5}0.05 & \cellcolor{blue!13}0.14 & \cellcolor{blue!5}0.06 & \cellcolor{blue!19}0.2 & \cellcolor{red!30}-0.3 & \cellcolor{blue!100}1.0\tabularnewline
\bottomrule
\end{tabular}
}
\end{table}

\subsection{Representative models selection}
Only one model can be provided in the software solutions for each tumor type, and the best model selection was done empirically according to the following criterion: size of the validation or test set, average Dice score and patient-wise F1-score performances.
The exhaustive list of chosen models is the following: the model trained for fold $0$ was selected for the glioblastomas, the model trained for fold $3$ was selected for the lower grade gliomas, for the meningiomas the model trained for fold $2$ was selected, and finally for the metastases the model trained for fold $2$ was selected.

\subsection{Speed study}
A comparison in processing speed regarding pure tumor segmentation and complete generation of standardized reports is provided in \cref{fig:speed-study-main}, when using the Raidionics software with CPU support. The high-end computer is the computer used for training the models, whereas the mid-end computer is a Windows laptop with an Intel Core Processor ($i7@2.20$GHz), and $16$GB of RAM.

For the smallest MRI volumes on average, $17$\,seconds are needed to perform tumor segmentation whereas $4.5$\,minutes are required to generate the complete standardized report with the high-end computer. Unsurprisingly, the larger the MRI volume the more time required to perform the different processing operations (cf. Section S3, Sup. Mat.). For the largest MRI volumes overall, $59$\,seconds are needed to perform tumor segmentation whereas $15$\,minutes are required to generate the complete standardized report.
When using the mid-end laptop, overall runtime is increased by $1.5$\,times for the different MRI volume sizes. On average, $9$\,minutes are necessary to generate the standardized report for MRI volumes of reasonable quality.

\begin{table}[!ht]
\centering
\caption{Segmentation (Segm.) and standardized reporting (SR) execution speeds for each tumor subtype, using our Raidionics software.}
\label{fig:speed-study-main}
\adjustbox{max width=\textwidth}{
\begin{tabular}{l|c|cc|cc}
& & \multicolumn{2}{c|}{High-end computer (Desktop) } & \multicolumn{2}{c}{Mid-end computer (Laptop)}\tabularnewline
& Dimensions (voxels) & Segm. (s) & SR (m) & Segm. (s) & SR (m)\tabularnewline
\hline
LGG & $394\times394\times80$ & $16.69\pm0.426$ & $04.50\pm0.09$ & $28.69\pm0.577$ & $07.32\pm0.07$ \tabularnewline
Meningioma & $256\times256\times170$ & $17.21\pm0.425$ & $05.48\pm0.12$ & $31.41\pm0.862$ & $09.09\pm0.32$ \tabularnewline
Glioblastoma & $320\times320\times220$ & $21.99\pm0.177$ & $05.89\pm0.03$ & $33.65\pm1.429$ & $09.06\pm0.24$ \tabularnewline
Metastasis & $560\times560\times561$\ & $59.06\pm1.454$ & $15.35\pm0.41$ & $98.54\pm2.171$ & $24.06\pm0.93$ \tabularnewline
\end{tabular}
}
\end{table}

\section{Discussion}
\label{sec:discussion}
In this study, we have investigated the segmentation of a range of common main brain tumor types in 3D preoperative MR scans using a variant of the Attention U-Net architecture. We have conducted experiments to assess the performances of each trained model using close to $30$ metrics, and developed two software solutions for end-users to freely benefit from our segmentation models and standardized clinical reports. The main contributions are the high performances of the models, on-par with performances reported in the literature for the glioblastomas, with illustrated robustness and ability to generalize thanks to the multiple and widespread data sources. In addition, the two proposed open-access and open-source software solutions include our best models, together with a RADS for computing tumor characteristics. This is the first open RADS solution which supports all major brain tumor types. The software is user-friendly, requiring only a few clicks and no programming to use, making it easily accessible for clinicians. The overall limitations are those already known for deep learning approaches whereby a higher amount of patients or data sources would improve the ability to generalize, boost segmentation performances, and increase the immunity toward rare tumor expressions. The employed architecture also struggles with smaller tumors given the large downsampling to feed the entire 3D MR scan in the network, hence the need for a better design combining local and global features either through multiple steps or ensembling.

The architecture and training strategy used in this study were identical to our previously published work considering that the intent was not to directly make advances on the segmentation task. Nevertheless, the stability and robustness to train efficient models had been documented, alongside performance comparison to another well-known architecture (e.g., nnU-Net~\cite{isensee2018nnu}), thus not precluding its use to train models for other brain tumor types. Aside from evident outliers in the datasets, where either tumors with partial or missing contrast uptake or suboptimal manual annotations were identified, the major pitfall from using the AGU-Net architecture lies in its struggle to segment equally satisfactorily small tumor pieces with a volume below $2$\,ml.
Overall, the glioblastoma model is expected to be the most robust and able to generalize since patient data from $15$ different sources was used. For other models trained on data from much fewer hospitals, with an expected limited variability in MR scan quality, their robustness is likely to be inferior.
While larger datasets is often correlated with improved segmentation performance, the metastasis model is the best performing with the lowest amount of patients included. The relative easiness of the task from a clear demarcation of the tumor from surrounding normal tissue in contrast-enhanced T1-weighted volumes, and the potentially low variance in tumor characteristics with patient data coming from two hospitals only, can explain the results.
Additionally, the metastasis model has been trained by transfer-learning using as input the second best performing glioblastoma model where the most data was used, which may have been the compelling factor. Lower-grade gliomas represent the most difficult type to manually segment since tumors are diffuse and infiltrating with an average volume in FLAIR sequences a lot higher than in T1 sequences for the other tumor types, and as such overall worse performances were expected.

The in-depth assessment of a larger pool of metrics allowed us to identify redundancy and uniqueness, and proved that the Dice score is overall quite robust and indicative of expected performance. However, the sole use of Dice score cannot cover all aspects of a model performance, and spatial distance-based metrics (e.g., HD95 and MHD) are suggested to be used in conjunction as providing values uncorrelated to Dice. In addition, some metrics were identified to be more sensitive to outliers and are as such powerful to either assess the ability to generalize of a model across data acquired on different scanners from multiple sources, or quickly identify potential issues in a large body of data. Finally, and depending on the nature of the patients included in one's study and amount of satellite tumors, specific object-wise metrics are imperative to use (e.g., OASSD). Only a combination of various metrics computed either voxel-wise, patient-wise, or instance-wise can give the full picture of a model's performance. Unfortunately, interpreting and comparing sets of metrics can prove challenging and as such further investigations regarding their merging into a unique informative and coherent score are fundamental (e.g., Roza~\cite{melek2021roza}).
Furthermore, an inadequacy lies in the nature of the different metrics whereby some can be computed across all segmentations generated by a trained model, whereas others are exclusively eligible on true positive cases, i.e., when the model has correctly segmented some extent of the tumor. For models generating perfect patient-wise recall, all metrics will be eligible for every segmentation. However, in this field of research and as of today, no trained model can fulfill this requirement due to the substantially large inter-patient variability.
Ideally, the identification of relevant metrics, bringing unique information for interpreting the results, should not be confined to the validation studies. More metrics should be considered to be a part of the loss function computation during training of neural network architectures. Attempts have been made towards using the Hausdorff distance as loss function, but a direct minimization is challenging from an optimization viewpoint. For example, approximation of Hausdorff distance based on distance transforms, on morphological operations, or with circular and spherical kernels showed potential for medical image segmentation~\cite{karimi2019reducing}. In general, a careful mix between losses (e.g., Dice, cross-entropy, and HD95) is challenging to achieve and adaptive strategies might be required to avoid reaching a local minimum where overall segmentation performance may suffer~\cite{heydari2019softadapt}.

As a current trend in the community, inference code and trained segmentation models are often at best available on GitHub repositories. As a consequence, only engineers, or people with some extent of knowledge in machine learning and programming, can benefit from such research advances. Besides, the research focus is heavily angled towards gliomas, due to the BraTS challenge influence, whereby segmentation models are expected to yield superior performance than for meningiomas and metastases.
By developing and giving free and unrestricted access to our two proposed software solutions, we hope to facilitate more research on all brain tumor types. Willing research institutes have the opportunity to generate private annotated datasets at a faster pace than through fully manual labour by exploiting our trained models. Having made all source code available on GitHub, as customarily done, we made the effort to further make stand-alone solutions with easy-to-use GUIs. Hopefully, clinicians and other non-programming end-users should feel more comfortable manipulating such tools, available across the three major operating systems and necessitating only a computer with average hardware specifications.
For the generation of standardized clinical reports, the computation of tumor characteristics relies heavily on the quality of the automatic segmentation, occasional mishaps are expected as models are not perfect and can omit the tumor. Therefore, manual inputs will be required sporadically to correct the tumor segmentation. Over time, new and better models will be generated and made available seamlessly into the two software through regular updates.

In the future, an approach incorporating a set of metrics and converting them into one final score would be highly desirable (e.g., Roza). Not only would it help to automatically select the best model from a k-fold validation study from one unique score, but a proper assessment and ranking across multiple methods would be enabled.
With all preoperative brain tumor types available for segmentation and reporting in our software, a key missing component is the automatic tumor type classification to supplement manual user input. Concurrently, the variety and amount of tumor characteristics to compute should be extended, considering more type-specific features similar to the resection index for glioblastomas.
Alternatively, bringing a similar focus on post-operative segmentation of residual tumor is of great interest to both assess the quality of the surgery and refine the estimated patient outcome. The generation of a complete post-operative standardized clinical report would also be permitted with new features such as the extent of resection.
Otherwise, intensifying the gathering of patient data from more widespread hospital centers and a larger array of MRI scanners is always of importance. The inclusion of more than one MR sequence per patient as segmentation input has the potential to boost overall performance, but at the same time might reduce models' potency as not always routinely available across all centers worldwide.

\section{Conclusion}
\label{sec:conclusion}
Efficient and robust segmentation models have been trained on pre-operative MR scans for the four main brain tumor types: glioblastoma, lower grade glioma, meningioma, and metastasis. In-depth performance assessment allowed to identify the most relevant metrics from a large panel, computed either voxel-wise, patient-wise, or instance-wise. Trained models and standardized reporting have been made publicly available and packaged into a stand-alone software and a 3D Slicer plugin to enable effortless widespread use. 

\subsection*{Disclosures}
The authors declare that the research was conducted in the absence of any commercial or financial relationships that could be construed as a potential conflict of interest.\\
Informed consent was obtained from all individual participants included in the study.

\subsection* {Acknowledgments}
Data were processed in digital labs at HUNT Cloud, Norwegian University of Science and Technology, Trondheim, Norway.

\subsection*{Author Contributions}
Funding acquisition, I.R., O.S., P.C.D.W.H., K.E.E., and A.S.J.;
Data curation, A.S.J., K.E.E., V.K., I.K., D.B., H.A., F.B., L.B., M.S.B., M.C.N., J.F., S.H.-J., A.J.S.I., B.K., A.K., E.M., D.M.J.M., P.A.R., M.R., T.S., W.A.v.d.B., M.W., G.W., O.S. and P.C.D.W.H.;
Conceptualization, D.B., A.P., I.R., O.S. and P.C.D.W.H.;
Methodology, D.B.;
Software, D.B. and A.P.;
Validation, D.B. and A.P.;
Visualization, D.B.;
Supervision, I.R., O.S. and P.C.D.W.H.;
Project administration, I.R., O.S., and P.C.D.W.H.;
Writing—original draft, D.B., A.P., I.R., O.S., A.S.J., K.E.E., and P.C.D.W.H.;
Writing—review and editing, H.A., F.B., L.B., M.S.B., M.C.N., J.F., S.H.-J., A.J.S.I., B.K., A.K., E.M., D.M.J.M., P.A.R., M.R., T.S., W.A.v.d.B., M.W., G.W., M.G.W. and A.H.Z.

\subsection*{Funding}
This work was funded by the Norwegian National Advisory Unit for Ultrasound and Image-Guided Therapy (usigt.org); South-Eastern Norway Regional Health Authority; Contract grant numbers: 2016102 and 2013069; Contract grant sponsor: Research Council of Norway; Contract grant number: 261984; Contract grant sponsor: Norwegian Cancer Society; Contract grant numbers: 6817564 and 3434180; Contract grant sponsor: European Research Council under the European Union's Horizon 2020 Program; Contract grant number: 758657-ImPRESS; an unrestricted grant of Stichting Hanarth fonds, “Machine learning for better neurosurgical decisions in patients with glioblastoma”; a grant for public-private partnerships (Amsterdam UMC PPP-grant) sponsored by the Dutch government (Ministry of Economic Affairs) through the Rijksdienst voor Ondernemend Nederland (RVO) and Topsector Life Sciences and Health (LSH), “Picturing predictions for patients with brain tumors”; a grant from the Innovative Medical Devices Initiative program, project number 10-10400-96-14003; The Netherlands Organisation for Scientific Research (NWO), 2020.027; a grant from the Dutch Cancer Society, VU2014-7113; the Anita Veldman foundation, CCA2018-2-17.

\bibliographystyle{unsrt}  
\bibliography{references}

\begin{thebibliography}{10}

\bibitem{day2016neurocognitive}
Julia Day, David~C Gillespie, Alasdair~G Rooney, Helen~J Bulbeck, Karolis
  Zienius, Florien Boele, and Robin Grant.
\newblock Neurocognitive deficits and neurocognitive rehabilitation in adult
  brain tumors.
\newblock {\em Current treatment options in neurology}, 18(5):1--16, 2016.

\bibitem{louis20212021}
David~N Louis, Arie Perry, Pieter Wesseling, Daniel~J Brat, Ian~A Cree,
  Dominique Figarella-Branger, Cynthia Hawkins, HK~Ng, Stefan~M Pfister, Guido
  Reifenberger, et~al.
\newblock The 2021 who classification of tumors of the central nervous system:
  a summary.
\newblock {\em Neuro-oncology}, 23(8):1231--1251, 2021.

\bibitem{deangelis2001brain}
Lisa~M DeAngelis.
\newblock Brain tumors.
\newblock {\em New England journal of medicine}, 344(2):114--123, 2001.

\bibitem{fisher2007epidemiology}
James~L Fisher, Judith~A Schwartzbaum, Margaret Wrensch, and Joseph~L Wiemels.
\newblock Epidemiology of brain tumors.
\newblock {\em Neurologic clinics}, 25(4):867--890, 2007.

\bibitem{lapointe2018primary}
Sarah Lapointe, Arie Perry, and Nicholas~A Butowski.
\newblock Primary brain tumours in adults.
\newblock {\em The Lancet}, 392(10145):432--446, 2018.

\bibitem{kickingereder2016radiomic}
Philipp Kickingereder, Sina Burth, Antje Wick, Michael G{\"o}tz, Oliver Eidel,
  Heinz-Peter Schlemmer, Klaus~H Maier-Hein, Wolfgang Wick, Martin Bendszus,
  Alexander Radbruch, et~al.
\newblock Radiomic profiling of glioblastoma: identifying an imaging predictor
  of patient survival with improved performance over established clinical and
  radiologic risk models.
\newblock {\em Radiology}, 280(3):880--889, 2016.

\bibitem{sawaya1998neurosurgical}
Raymond Sawaya, Maarouf Hammoud, Derek Schoppa, Kenneth~R Hess, Shu~Z Wu,
  Wei-Ming Shi, and David~M WiIdrick.
\newblock Neurosurgical outcomes in a modern series of 400 craniotomies for
  treatment of parenchymal tumors.
\newblock {\em Neurosurgery}, 42(5):1044--1055, 1998.

\bibitem{mathiesen2011two}
Tiit Mathiesen, Inti Peredo, and Stefan L{\"o}nn.
\newblock Two-year survival of low-grade and high-grade glioma patients using
  data from the swedish cancer registry.
\newblock {\em Acta neurochirurgica}, 153(3):467--471, 2011.

\bibitem{zinn2013extent}
Pascal~O Zinn, Rivka~R Colen, Ekkehard~M Kasper, and Jan-Karl Burkhardt.
\newblock Extent of resection and radiotherapy in gbm: A 1973 to 2007
  surveillance, epidemiology and end results analysis of 21,783 patients.
\newblock {\em International journal of oncology}, 42(3):929--934, 2013.

\bibitem{weinreb2016pi}
Jeffrey~C Weinreb, Jelle~O Barentsz, Peter~L Choyke, Francois Cornud, Masoom~A
  Haider, Katarzyna~J Macura, Daniel Margolis, Mitchell~D Schnall, Faina
  Shtern, Clare~M Tempany, et~al.
\newblock Pi-rads prostate imaging--reporting and data system: 2015, version 2.
\newblock {\em European urology}, 69(1):16--40, 2016.

\bibitem{dyer2020implications}
Spencer~C Dyer, Brian~J Bartholmai, and Chi~Wan Koo.
\newblock Implications of the updated lung ct screening reporting and data
  system (lung-rads version 1.1) for lung cancer screening.
\newblock {\em Journal of Thoracic Disease}, 12(11):6966, 2020.

\bibitem{ellingson2015consensus}
Benjamin~M Ellingson, Martin Bendszus, Jerrold Boxerman, Daniel Barboriak,
  Bradley~J Erickson, Marion Smits, Sarah~J Nelson, Elizabeth Gerstner, Brian
  Alexander, Gregory Goldmacher, et~al.
\newblock Consensus recommendations for a standardized brain tumor imaging
  protocol in clinical trials.
\newblock {\em Neuro-oncology}, 17(9):1188--1198, 2015.

\bibitem{kommers2021glioblastoma}
Ivar Kommers, David Bouget, Andr{\'e} Pedersen, Roelant~S Eijgelaar, Hilko
  Ardon, Frederik Barkhof, Lorenzo Bello, Mitchel~S Berger, Marco Conti~Nibali,
  Julia Furtner, et~al.
\newblock Glioblastoma surgery imaging—reporting and data system:
  Standardized reporting of tumor volume, location, and resectability based on
  automated segmentations.
\newblock {\em Cancers}, 13(12):2854, 2021.

\bibitem{binaghi2016collection}
Elisabetta Binaghi, Valentina Pedoia, and Sergio Balbi.
\newblock Collection and fuzzy estimation of truth labels in glial tumour
  segmentation studies.
\newblock {\em Computer Methods in Biomechanics and Biomedical Engineering:
  Imaging \& Visualization}, 4(3-4):214--228, 2016.

\bibitem{berntsen2020volumetric}
Erik~Magnus Berntsen, Anne~Line Stensj{\o}en, Maren~Staurset Langlo,
  Solveig~Quam Simonsen, P{\aa}l Christensen, Viggo~Andreas Moholdt, and Ole
  Solheim.
\newblock Volumetric segmentation of glioblastoma progression compared to
  bidimensional products and clinical radiological reports.
\newblock {\em Acta Neurochirurgica}, 162(2):379--387, 2020.

\bibitem{minaee2021image}
Shervin Minaee, Yuri~Y Boykov, Fatih Porikli, Antonio~J Plaza, Nasser
  Kehtarnavaz, and Demetri Terzopoulos.
\newblock Image segmentation using deep learning: A survey.
\newblock {\em IEEE transactions on pattern analysis and machine intelligence},
  2021.

\bibitem{menze2014multimodal}
Bjoern~H Menze, Andras Jakab, Stefan Bauer, Jayashree Kalpathy-Cramer, Keyvan
  Farahani, Justin Kirby, Yuliya Burren, Nicole Porz, Johannes Slotboom, Roland
  Wiest, et~al.
\newblock The multimodal brain tumor image segmentation benchmark (brats).
\newblock {\em IEEE transactions on medical imaging}, 34(10):1993--2024, 2014.

\bibitem{bakas2017advancing}
Spyridon Bakas, Hamed Akbari, Aristeidis Sotiras, Michel Bilello, Martin
  Rozycki, Justin~S Kirby, John~B Freymann, Keyvan Farahani, and Christos
  Davatzikos.
\newblock Advancing the cancer genome atlas glioma mri collections with expert
  segmentation labels and radiomic features.
\newblock {\em Scientific data}, 4(1):1--13, 2017.

\bibitem{baid2021rsna}
Ujjwal Baid, Satyam Ghodasara, Suyash Mohan, Michel Bilello, Evan Calabrese,
  Errol Colak, Keyvan Farahani, Jayashree Kalpathy-Cramer, Felipe~C Kitamura,
  Sarthak Pati, et~al.
\newblock The rsna-asnr-miccai brats 2021 benchmark on brain tumor segmentation
  and radiogenomic classification.
\newblock {\em arXiv preprint arXiv:2107.02314}, 2021.

\bibitem{isensee2018nnu}
Fabian Isensee, Jens Petersen, Andre Klein, David Zimmerer, Paul~F Jaeger,
  Simon Kohl, Jakob Wasserthal, Gregor Koehler, Tobias Norajitra, Sebastian
  Wirkert, et~al.
\newblock nnu-net: Self-adapting framework for u-net-based medical image
  segmentation.
\newblock {\em arXiv preprint arXiv:1809.10486}, 2018.

\bibitem{luu2021extending}
Huan~Minh Luu and Sung-Hong Park.
\newblock Extending nn-unet for brain tumor segmentation.
\newblock {\em arXiv preprint arXiv:2112.04653}, 2021.

\bibitem{tiwari2020brain}
Arti Tiwari, Shilpa Srivastava, and Millie Pant.
\newblock Brain tumor segmentation and classification from magnetic resonance
  images: Review of selected methods from 2014 to 2019.
\newblock {\em Pattern Recognition Letters}, 131:244--260, 2020.

\bibitem{pereira2016brain}
S{\'e}rgio Pereira, Adriano Pinto, Victor Alves, and Carlos~A Silva.
\newblock Brain tumor segmentation using convolutional neural networks in mri
  images.
\newblock {\em IEEE transactions on medical imaging}, 35(5):1240--1251, 2016.

\bibitem{grovik2020deep}
Endre Gr{\o}vik, Darvin Yi, Michael Iv, Elizabeth Tong, Daniel Rubin, and Greg
  Zaharchuk.
\newblock Deep learning enables automatic detection and segmentation of brain
  metastases on multisequence mri.
\newblock {\em Journal of Magnetic Resonance Imaging}, 51(1):175--182, 2020.

\bibitem{grovik2021handling}
Endre Gr{\o}vik, Darvin Yi, Michael Iv, Elizabeth Tong, Line~Brennhaug Nilsen,
  Anna Latysheva, Cathrine Saxhaug, Kari~Dolven Jacobsen, {\AA}slaug Helland,
  Kyrre~Eeg Emblem, et~al.
\newblock Handling missing mri sequences in deep learning segmentation of brain
  metastases: a multicenter study.
\newblock {\em NPJ digital medicine}, 4(1):1--7, 2021.

\bibitem{kamnitsas2016deepmedic}
Konstantinos Kamnitsas, Enzo Ferrante, Sarah Parisot, Christian Ledig, Aditya~V
  Nori, Antonio Criminisi, Daniel Rueckert, and Ben Glocker.
\newblock Deepmedic for brain tumor segmentation.
\newblock In {\em International workshop on Brainlesion: Glioma, multiple
  sclerosis, stroke and traumatic brain injuries}, pages 138--149. Springer,
  2016.

\bibitem{liu2017deep}
Yan Liu, Strahinja Stojadinovic, Brian Hrycushko, Zabi Wardak, Steven Lau,
  Weiguo Lu, Yulong Yan, Steve~B Jiang, Xin Zhen, Robert Timmerman, et~al.
\newblock A deep convolutional neural network-based automatic delineation
  strategy for multiple brain metastases stereotactic radiosurgery.
\newblock {\em PloS one}, 12(10):e0185844, 2017.

\bibitem{charron2018automatic}
Odelin Charron, Alex Lallement, Delphine Jarnet, Vincent Noblet, Jean-Baptiste
  Clavier, and Philippe Meyer.
\newblock Automatic detection and segmentation of brain metastases on
  multimodal mr images with a deep convolutional neural network.
\newblock {\em Computers in biology and medicine}, 95:43--54, 2018.

\bibitem{neromyliotis2022machine}
Eleftherios Neromyliotis, Theodosis Kalamatianos, Athanasios Paschalis,
  Spyridon Komaitis, Konstantinos~N Fountas, Eftychia~Z Kapsalaki, George
  Stranjalis, and Ioannis Tsougos.
\newblock Machine learning in meningioma mri: past to present. a narrative
  review.
\newblock {\em Journal of Magnetic Resonance Imaging}, 55(1):48--60, 2022.

\bibitem{laukamp2019fully}
Kai~Roman Laukamp, Frank Thiele, Georgy Shakirin, David Zopfs, Andrea
  Faymonville, Marco Timmer, David Maintz, Michael Perkuhn, and Jan Borggrefe.
\newblock Fully automated detection and segmentation of meningiomas using deep
  learning on routine multiparametric mri.
\newblock {\em European radiology}, 29(1):124--132, 2019.

\bibitem{laukamp2020automated}
Kai~Roman Laukamp, Lenhard Pennig, Frank Thiele, Robert Reimer, Lukas
  G{\"o}rtz, Georgy Shakirin, David Zopfs, Marco Timmer, Michael Perkuhn, and
  Jan Borggrefe.
\newblock Automated meningioma segmentation in multiparametric mri.
\newblock {\em Clinical Neuroradiology}, pages 1--10, 2020.

\bibitem{bouget2021meningioma}
David Bouget, Andr{\'e} Pedersen, Sayied Abdol~Mohieb Hosainey, Ole Solheim,
  and Ingerid Reinertsen.
\newblock Meningioma segmentation in t1-weighted mri leveraging global context
  and attention mechanisms.
\newblock {\em arXiv preprint arXiv:2101.07715}, 2021.

\bibitem{the_monai_consortium_2020_4323059}
The~MONAI Consortium.
\newblock Project monai, December 2020.

\bibitem{reinke2021common}
Annika Reinke, Matthias Eisenmann, Minu~D Tizabi, Carole~H Sudre, Tim
  R{\"a}dsch, Michela Antonelli, Tal Arbel, Spyridon Bakas, M~Jorge Cardoso,
  Veronika Cheplygina, et~al.
\newblock Common limitations of image processing metrics: A picture story.
\newblock {\em arXiv preprint arXiv:2104.05642}, 2021.

\bibitem{taha2015metrics}
Abdel~Aziz Taha and Allan Hanbury.
\newblock Metrics for evaluating 3d medical image segmentation: analysis,
  selection, and tool.
\newblock {\em BMC medical imaging}, 15(1):29, 2015.

\bibitem{weinberg2018management}
Brent~D Weinberg, Ashwani Gore, Hui-Kuo~G Shu, Jeffrey~J Olson, Richard Duszak,
  Alfredo~D Voloschin, and Michael~J Hoch.
\newblock Management-based structured reporting of posttreatment glioma
  response with the brain tumor reporting and data system.
\newblock {\em Journal of the American College of Radiology}, 15(5):767--771,
  2018.

\bibitem{huber2017reliability}
T~Huber, G~Alber, S~Bette, T~Boeckh-Behrens, J~Gempt, F~Ringel, E~Alberts,
  C~Zimmer, and JS~Bauer.
\newblock Reliability of semi-automated segmentations in glioblastoma.
\newblock {\em Clinical neuroradiology}, 27(2):153--161, 2017.

\bibitem{vezhnevets2005growcut}
Vladimir Vezhnevets and Vadim Konouchine.
\newblock Growcut: Interactive multi-label nd image segmentation by cellular
  automata.
\newblock In {\em proc. of Graphicon}, volume~1, pages 150--156. Citeseer,
  2005.

\bibitem{bouget2021fast}
David Bouget, Andr{\'e} Pedersen, Sayied Abdol~Mohieb Hosainey, Johanna Vanel,
  Ole Solheim, and Ingerid Reinertsen.
\newblock Fast meningioma segmentation in t1-weighted magnetic resonance
  imaging volumes using a lightweight 3d deep learning architecture.
\newblock {\em Journal of Medical Imaging}, 8(2):024002, 2021.

\bibitem{bouget2021glioblastoma}
David Bouget, Roelant~S Eijgelaar, Andr{\'e} Pedersen, Ivar Kommers, Hilko
  Ardon, Frederik Barkhof, Lorenzo Bello, Mitchel~S Berger, Marco~Conti Nibali,
  Julia Furtner, et~al.
\newblock Glioblastoma surgery imaging--reporting and data system: Validation
  and performance of the automated segmentation task.
\newblock {\em Cancers}, 13(18):4674, 2021.

\bibitem{fonov2009unbiased}
Vladimir~S Fonov, Alan~C Evans, Robert~C McKinstry, C~Robert Almli, and
  DL~Collins.
\newblock Unbiased nonlinear average age-appropriate brain templates from birth
  to adulthood.
\newblock {\em NeuroImage}, (47):S102, 2009.

\bibitem{fedorov20123d}
Andriy Fedorov, Reinhard Beichel, Jayashree Kalpathy-Cramer, Julien Finet,
  Jean-Christophe Fillion-Robin, Sonia Pujol, Christian Bauer, Dominique
  Jennings, Fiona Fennessy, Milan Sonka, et~al.
\newblock 3d slicer as an image computing platform for the quantitative imaging
  network.
\newblock {\em Magnetic resonance imaging}, 30(9):1323--1341, 2012.

\bibitem{mehrtash2017deepinfer}
Alireza Mehrtash, Mehran Pesteie, Jorden Hetherington, Peter~A Behringer, Tina
  Kapur, William~M Wells~III, Robert Rohling, Andriy Fedorov, and Purang
  Abolmaesumi.
\newblock Deepinfer: Open-source deep learning deployment toolkit for
  image-guided therapy.
\newblock In {\em Medical Imaging 2017: Image-Guided Procedures, Robotic
  Interventions, and Modeling}, volume 10135, pages 410--416. SPIE, 2017.

\bibitem{dice1945measures}
Lee~R Dice.
\newblock Measures of the amount of ecologic association between species.
\newblock {\em Ecology}, 26(3):297--302, 1945.

\bibitem{jaccard1912distribution}
Paul Jaccard.
\newblock The distribution of the flora in the alpine zone. 1.
\newblock {\em New phytologist}, 11(2):37--50, 1912.

\bibitem{martin2001database}
David Martin, Charless Fowlkes, Doron Tal, and Jitendra Malik.
\newblock A database of human segmented natural images and its application to
  evaluating segmentation algorithms and measuring ecological statistics.
\newblock In {\em Proceedings Eighth IEEE International Conference on Computer
  Vision. ICCV 2001}, volume~2, pages 416--423. IEEE, 2001.

\bibitem{cardenes2009multidimensional}
Rub{\'e}n C{\'a}rdenes, Rodrigo de~Luis-Garcia, and Meritxell Bach-Cuadra.
\newblock A multidimensional segmentation evaluation for medical image data.
\newblock {\em Computer methods and programs in biomedicine}, 96(2):108--124,
  2009.

\bibitem{russakoff2004image}
Daniel~B Russakoff, Carlo Tomasi, Torsten Rohlfing, and Calvin~R Maurer.
\newblock Image similarity using mutual information of regions.
\newblock In {\em European Conference on Computer Vision}, pages 596--607.
  Springer, 2004.

\bibitem{meilua2003comparing}
Marina Meil{\u{a}}.
\newblock Comparing clusterings by the variation of information.
\newblock In {\em Learning theory and kernel machines}, pages 173--187.
  Springer, 2003.

\bibitem{cohen1960coefficient}
Jacob Cohen.
\newblock A coefficient of agreement for nominal scales.
\newblock {\em Educational and psychological measurement}, 20(1):37--46, 1960.

\bibitem{bradley1997use}
Andrew~P Bradley.
\newblock The use of the area under the roc curve in the evaluation of machine
  learning algorithms.
\newblock {\em Pattern recognition}, 30(7):1145--1159, 1997.

\bibitem{baldi2000assessing}
Pierre Baldi, S{\o}ren Brunak, Yves Chauvin, Claus~AF Andersen, and Henrik
  Nielsen.
\newblock Assessing the accuracy of prediction algorithms for classification:
  an overview.
\newblock {\em Bioinformatics}, 16(5):412--424, 2000.

\bibitem{gerig2001valmet}
Guido Gerig, Matthieu Jomier, and Miranda Chakos.
\newblock Valmet: A new validation tool for assessing and improving 3d object
  segmentation.
\newblock In {\em International conference on medical image computing and
  computer-assisted intervention}, pages 516--523. Springer, 2001.

\bibitem{huttenlocher1993comparing}
Daniel~P Huttenlocher, Gregory~A. Klanderman, and William~J Rucklidge.
\newblock Comparing images using the hausdorff distance.
\newblock {\em IEEE Transactions on pattern analysis and machine intelligence},
  15(9):850--863, 1993.

\bibitem{mclachlan1999mahalanobis}
Goeffrey~J McLachlan.
\newblock Mahalanobis distance.
\newblock {\em Resonance}, 4(6):20--26, 1999.

\bibitem{chinchor1993muc}
Nancy Chinchor and Beth~M Sundheim.
\newblock Muc-5 evaluation metrics.
\newblock In {\em Fifth Message Understanding Conference (MUC-5): Proceedings
  of a Conference Held in Baltimore, Maryland, August 25-27, 1993}, 1993.

\bibitem{hubert1985comparing}
Lawrence Hubert and Phipps Arabie.
\newblock Comparing partitions.
\newblock {\em Journal of classification}, 2(1):193--218, 1985.

\bibitem{killeen2005alternative}
Peter~R Killeen.
\newblock An alternative to null-hypothesis significance tests.
\newblock {\em Psychological science}, 16(5):345--353, 2005.

\bibitem{imgaug}
Alexander~B. Jung, Kentaro Wada, Jon Crall, Satoshi Tanaka, Jake Graving,
  Christoph Reinders, Sarthak Yadav, Joy Banerjee, Gábor Vecsei, Adam Kraft,
  Zheng Rui, Jirka Borovec, Christian Vallentin, Semen Zhydenko, Kilian
  Pfeiffer, Ben Cook, Ismael Fernández, François-Michel De~Rainville,
  Chi-Hung Weng, Abner Ayala-Acevedo, Raphael Meudec, Matias Laporte, et~al.
\newblock {imgaug}.
\newblock \url{https://github.com/aleju/imgaug}, 2020.
\newblock Online; accessed 01-Feb-2020.

\bibitem{scikit-learn}
F.~Pedregosa, G.~Varoquaux, A.~Gramfort, V.~Michel, B.~Thirion, O.~Grisel,
  M.~Blondel, P.~Prettenhofer, R.~Weiss, V.~Dubourg, J.~Vanderplas, A.~Passos,
  D.~Cournapeau, M.~Brucher, M.~Perrot, and E.~Duchesnay.
\newblock Scikit-learn: Machine learning in {P}ython.
\newblock {\em Journal of Machine Learning Research}, 12:2825--2830, 2011.

\bibitem{oskar_maier_2019_2565940}
Oskar Maier, Alex Rothberg, Pradeep~Reddy Raamana, Rémi Bèges, Fabian
  Isensee, Michael Ahern, mamrehn, VincentXWD, and Jay Joshi.
\newblock loli/medpy: Medpy 0.4.0, February 2019.

\bibitem{melek2021roza}
Mesut Melek and Negin Melek.
\newblock Roza: a new and comprehensive metric for evaluating classification
  systems.
\newblock {\em Computer Methods in Biomechanics and Biomedical Engineering},
  pages 1--13, 2021.

\bibitem{karimi2019reducing}
Davood Karimi and Septimiu~E Salcudean.
\newblock Reducing the hausdorff distance in medical image segmentation with
  convolutional neural networks.
\newblock {\em IEEE Transactions on medical imaging}, 39(2):499--513, 2019.

\bibitem{heydari2019softadapt}
A~Ali Heydari, Craig~A Thompson, and Asif Mehmood.
\newblock Softadapt: Techniques for adaptive loss weighting of neural networks
  with multi-part loss functions.
\newblock {\em arXiv preprint arXiv:1912.12355}, 2019.

\end{thebibliography}

\end{document}


\section{Supplementary material}

\subsection{Metrics}
Given a ground truth segmentation (denoted as g) and an automatic segmentation coming from a trained model (denoted as d), the following four voxel-wise cardinalities can be defined, given a total amount of voxels X in a given MRI volume (of dimensions $H\times W\times D$).

\begin{equation}
    TP = \sum_{x=1}^{|X|}(g(x) == 1) \And (d(x) == 1)
\end{equation}

\begin{equation}
    TN = \sum_{x=1}^{|X|}(g(x) == 0) \And (d(x) == 0)
\end{equation}

\begin{equation}
    FP = \sum_{x=1}^{|X|}(g(x) == 0) \And (d(x) == 1)
\end{equation}

\begin{equation}
    FN = \sum_{x=1}^{|X|}(g(x) == 1) \And (d(x) == 0)
\end{equation}

Using those cardinalities as base elements, the other voxel-wise metrics can be computed according to the following equations.

\paragraph{True Positive Rate (TPR):}
\begin{equation}
    TPR = \frac{TP}{TP+FN}
\end{equation}

\paragraph{True Negative Rate (TNR):}
\begin{equation}
    TNR = \frac{TN}{TN+FP}
\end{equation}

\paragraph{False Positive Rate (FPR):}
\begin{equation}
    FPR = \frac{FP}{FP+TN}
\end{equation}

\paragraph{False Negative Rate (FNR):}
\begin{equation}
    FNR = \frac{FN}{FN+TP}
\end{equation}

\paragraph{Positive Predictive Value (PPV):}
\begin{equation}
    PPV = \frac{TP}{TP+FP}
\end{equation}

\paragraph{Dice:}
\begin{equation}
    Dice = \frac{2\cdot TP}{2\cdot TP+FP+FN}
\end{equation}

\paragraph{Jaccard:}
\begin{equation}
    J = \frac{TP}{TP+FP+FN}
\end{equation}

\paragraph{Intersection over Union (IoU):}
\begin{equation}
    IoU = \frac{\sum_{x=1}^{|X|}g(x) \cap d(x)}{\sum_{x=1}^{|X|}(g(x) \cup d(x))}
\end{equation}

\paragraph{Area Under the Curve (AUC):}
\begin{equation}
    AUC = 1 - \frac{FPR + FNR}{2}
\end{equation}

\paragraph{Global Consistency Error (GCE):}
\begin{equation}
    GCE = \frac{1}{X} min\bigg\{\frac{FN(FN + 2TP)}{TP+FN} + \frac{FP(FP+2TN)}{TN+FP}, \frac{FP(FP+2TP)}{TP+FP} + \frac{FN(FN+2TN)}{TN+FN}\bigg\}
\end{equation}

\paragraph{Mathews Correlation Coefficient (MCC):}
\begin{equation}
    MCC = \frac{(TP\cdot TN) - (FP\cdot FN)}{\sqrt{(TP+FP)\cdot (TP+FN)\cdot (TN+FP)\cdot (TN+FN)}}
\end{equation}

\paragraph{Adjusted Rand Index (ARI):}
\begin{equation}
\begin{aligned}
    a = 0.5\cdot ((TP\cdot(TP-1)) + (FP\cdot(FP-1)) + (TN\cdot(TN-1)) + (FN\cdot(FN-1))) \\
    b = 0.5\cdot (((TP + FN)^{2} + (TN + FP)^{2}) - (TP^{2} + TN^{2} + FP^{2} + FN^{2})) \\
    c = 0.5\cdot (((TP + FP)^{2} + (TN + FN)^{2}) - (TP^{2} + TN^{2} + FP^{2} + FN^{2})) \\
    d = \frac{(H\cdot W\cdot D)\cdot (H\cdot W\cdot D - 1)}{2\cdot (a + b + c)} \\
    ARI = \frac{2\cdot (a\cdot b - b\cdot c)}{c^{2} + b^{2} + 2\cdot a\cdot b + (a+d)\cdot (c+b)}
\end{aligned}
\end{equation}

\paragraph{Variation Of Information (VOI):}
\begin{equation}
\begin{aligned}
    h_{1} = - \Bigg(\frac{FN + TP}{X}\cdot log_{2}\bigg(\frac{FN + TP}{X}\bigg) + \frac{1 - (FN + TP)}{X} \cdot log_{2}\bigg(\frac{1 - (FN + TP)}{X}\bigg)\Bigg) \\
    h_{2} = - \Bigg(\frac{FP + TP}{X}\cdot log_{2}\bigg(\frac{FP + TP}{X}\bigg) + \frac{1 - (FP + TP)}{X} \cdot log_{2}\bigg(\frac{1 - (FP + TP)}{X}\bigg)\Bigg) \\
    p_{00} = 1\; if\; TN == 0\;else\;\frac{TN}{X} \\
    p_{01} = 1\; if\; FN == 0\; else\; \frac{FN}{X} \\
    p_{10} = 1\; if\; FP == 0\; else\; \frac{FP}{X} \\
    p_{11} = 1\; if\; TP == 0\; else\; \frac{TP}{X} \\
    h_{12} = - \Bigg(log_{2}(p_{00})\cdot \frac{TN}{X} + log_{2}(p_{01})\cdot \frac{FN}{X} + log_{2}(p_{10})\cdot \frac{FP}{X} + log_{2}(p_{11})\cdot \frac{TP}{X}\Bigg)\\
    MI = h_{1} + h_{2} - h_{12} \\
    VOI = h_{1} + h_{2} - (2 \cdot MI)
\end{aligned}
\end{equation}

\paragraph{Probabilistic Distance (PBD):}
\begin{equation}
    PBD = \frac{\sum_{x=1}^{|X|}|g(x) - d(x)|}{2 \cdot \sum_{x=1}^{|X|}(g(x) \cdot d(x))}
\end{equation}



\newpage
\FloatBarrier

\subsection{Metrics analysis}
For each of the four brain tumor subtypes, the following elements are consistently reported: the fold-wise overall performances, the metrics correlation matrix, and a figure showing outlier cases according to different metrics.

\subsection{Glioblastoma segmentation}

Overall segmentation performances for the glioblastoma tumor type are reported fold-wise in \cref{tab:suppl-glioblastoma-foldwise}. The average Dice score is the lowest on fold $9$ which is unsurprising given the size of the test set, the largest of all. Otherwise, the average Dice score is relatively stable across all folds between $85$\% and $88$\%.

\begin{figure}[!ht]
\centering
\includegraphics[scale=1.40]{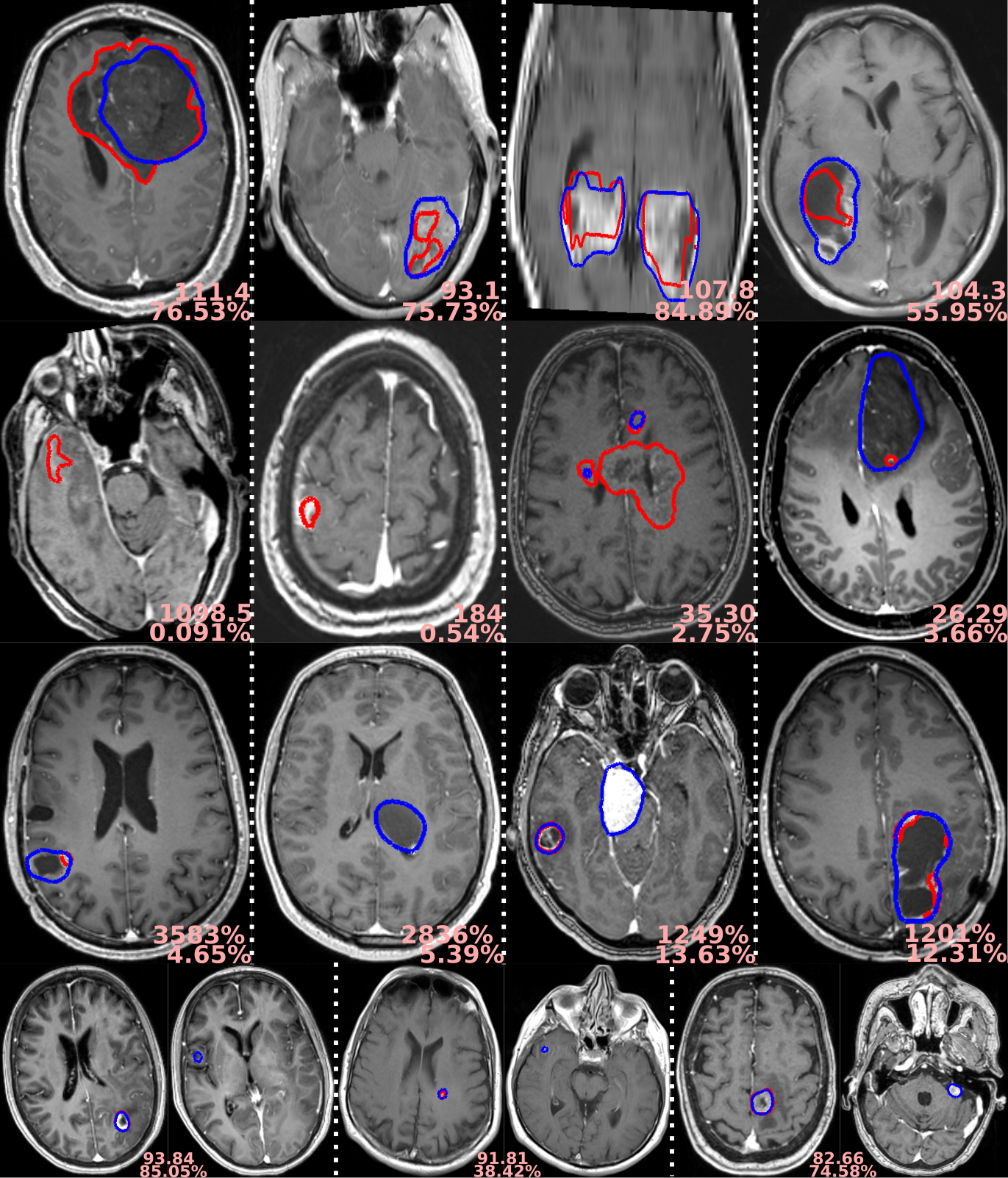}
\caption{Examples of glioblastoma outliers where the ground truth is shown in red and the automatic segmentation in blue. Each patient is separated by a dotted white line, and each row focuses on a different metric, from top to bottom: GCE, PBD, RAVD, and HD95. For each patient, both metric of interest and Dice scores are provided in light red.}
\label{fig:suppl-glioblastoma-outliers}
\end{figure}

The correlation matrix between some of the considered validation metrics is shown in \cref{fig:suppl-confusion-matrix-glioblastoma}. Visual segmentation results are illustrated in \cref{fig:suppl-glioblastoma-outliers} where the GCE, PBD, RAVD, and HD95 metrics scored orders of magnitude above their average, indicating clear outliers not identifiable by the Dice metric only.
In the first row, the two leftmost examples exhibit undersegmentation and oversegmentation of the tumor. The two rightmost examples are indicating subpar manual ground truth and where predictions from the model appears slightly more accurate.
In the second row, the two leftmost tumors where almost completely missed by the model, while only the contrast-enhancing part of the tumor was segmented in the two rightmost cases.
In the third row, the third patient features both a glioblastoma and a meningioma. Unfortunately, the trained model is not meant to perform tumor type classification and both tumor types are visually quite similar. As a result, both tumors have been segmented hence the low metric scores but for a $94.69$\% Dice over the sole glioblastoma. In the first and last patient, the cyst does not seem to have been included in the ground truth, focusing only on the contrast-enhanced parts.
Finally in the fourth row, all outlier cases featured a tumor with a very small volume whereby the HD95 scored low while all Dice scores reached above $80$\%.

\begin{table}[h]
\caption{Fold-wise overall segmentation performance for the glioblastoma tumor type.}
\label{tab:suppl-glioblastoma-foldwise}
\adjustbox{max width=\textwidth}{
\begin{tabular}{rr|cc|ccc|ccc}
 & & \multicolumn{2}{c|}{Pixel-wise} & \multicolumn{3}{c|}{Patient-wise} & \multicolumn{3}{c}{Object-wise} \tabularnewline
Fold & \# Samples & Dice & Dice-TP & F1-score & Recall & Precision & F1-score & Recall & Precision\tabularnewline
\hline
0 & 153 & $87.97\pm12.28$ & $88.55\pm10.03$ & $99.01$ & $99.35$ & $98.68$ & $93.27$ & $89.27$ & $97.64$\tabularnewline
1 & 103 & $88.54\pm13.95$ & $90.29\pm06.35$ & $98.19$ & $98.06$ & $98.32$ & $92.97$ & $90.58$ & $95.49$\tabularnewline
2 & 75 & $86.44\pm16.05$ & $86.44\pm16.05$ & $98.75$ & $100.00$ & $97.52$ & $94.44$ & $92.66$ & $96.30$\tabularnewline
3 & 72 & $86.17\pm15.70$ & $87.39\pm12.00$ & $97.17$ & $98.61$ & $95.77$ & $90.59$ & $89.80$ & $91.40$\tabularnewline
4 & 293 & $87.03\pm15.21$ & $89.16\pm06.87$ & $96.54$ & $97.61$ & $95.50$ & $90.28$ & $89.30$ & $91.28$\tabularnewline
5 & 83 & $88.22\pm10.79$ & $88.22\pm10.79$ & $98.57$ & $100.00$ & $97.19$ & $89.88$ & $85.50$ & $94.74$\tabularnewline
6 & 38 & $84.73\pm17.91$ & $84.73\pm17.91$ & $99.34$ & $100.00$ & $98.68$ & $80.34$ & $68.06$ & $98.04$\tabularnewline
7 & 74 & $83.49\pm19.42$ & $87.01\pm09.36$ & $96.99$ & $95.89$ & $98.12$ & $96.39$ & $96.39$ & $96.39$\tabularnewline
8 & 49 & $86.84\pm15.89$ & $88.65\pm09.87$ & $98.46$ & $97.96$ & $98.96$ & $83.67$ & $72.73$ & $98.48$\tabularnewline
9 & 457 & $82.82\pm21.85$ & $85.25\pm16.88$ & $96.10$ & $97.16$ & $95.07$ & $86.87$ & $82.44$ & $91.82$\tabularnewline
10 & 249 & $84.70\pm20.42$ & $87.88\pm12.38$ & $97.08$ & $96.39$ & $97.79$ & $88.36$ & $81.57$ & $96.37$\tabularnewline
11 & 134 & $86.17\pm15.45$ & $87.47\pm11.32$ & $98.87$ & $98.50$ & $99.24$ & $86.30$ & $76.61$ & $98.81$\tabularnewline
12 & 86 & $86.80\pm12.66$ & $87.82\pm08.51$ & $97.44$ & $98.84$ & $96.08$ & $92.05$ & $90.91$ & $93.22$\tabularnewline
13 & 171 & $85.12\pm11.19$ & $85.62\pm09.11$ & $97.17$ & $99.42$ & $95.02$ & $92.68$ & $94.67$ & $90.76$\tabularnewline
14 & 97 & $88.77\pm10.38$ & $88.77\pm10.38$ & $99.13$ & $100.00$ & $98.28$ & $88.15$ & $80.86$ & $96.88$\tabularnewline
\end{tabular}
}
\end{table}

\begin{table}[h]
\caption{Metrics correlation matrix for the glioblastoma segmentation. The color intensity of each cell represents the strength of the correlation, where blue denotes direct correlation and red denotes inverse correlation.}
\label{fig:suppl-confusion-matrix-glioblastoma}
\adjustbox{max width=\textwidth}{
\begin{tabular}{lrrrrrr|rr|rr|rrrrr|rrr|rr}
\toprule
 & \multicolumn{6}{c|}{Overlap} & \multicolumn{2}{c|}{Volume} & \multicolumn{2}{c|}{Information theory} & \multicolumn{5}{c|}{Probabilistic} & \multicolumn{3}{c|}{Spatial distance} & \multicolumn{2}{c}{Instance-wise}\tabularnewline
{} & Dice & TPR & TNR & PPV & IOU & GCE & VS & RAVD & MI & VOI & CKS & AUC & VC & MCC & PBD & HD95 & MHD & ASSD & ARI & OASSD\tabularnewline
Dice & \cellcolor{blue!100}1.0 & \cellcolor{blue!70}0.7 & \cellcolor{blue!28}0.29 & \cellcolor{blue!61}0.62 & \cellcolor{blue!98}0.98 & \cellcolor{red!22}-0.22 & \cellcolor{blue!94}0.94 & \cellcolor{red!34}-0.35 & \cellcolor{blue!98}0.99 & \cellcolor{red!23}-0.23 & \cellcolor{blue!99}1.0 & \cellcolor{blue!70}0.71 & \cellcolor{blue!78}0.78 & \cellcolor{blue!99}1.0 & \cellcolor{red!34}-0.34 & \cellcolor{red!54}-0.55 & \cellcolor{red!43}-0.43 & \cellcolor{red!70}-0.71 & \cellcolor{blue!99}1.0 & \cellcolor{red!30}-0.3\tabularnewline
TPR & \cellcolor{blue!70}0.7 & \cellcolor{blue!100}1.0 & \cellcolor{red!16}-0.17 & \cellcolor{red!6}-0.07 & \cellcolor{blue!70}0.71 & \cellcolor{red!8}-0.08 & \cellcolor{blue!62}0.62 & \cellcolor{blue!9}0.1 & \cellcolor{blue!70}0.7 & \cellcolor{red!8}-0.08 & \cellcolor{blue!70}0.7 & \cellcolor{blue!99}1.0 & \cellcolor{blue!51}0.51 & \cellcolor{blue!70}0.71 & \cellcolor{red!26}-0.26 & \cellcolor{red!37}-0.38 & \cellcolor{red!33}-0.34 & \cellcolor{red!47}-0.47 & \cellcolor{blue!70}0.7 & \cellcolor{red!20}-0.2\tabularnewline
TNR & \cellcolor{blue!28}0.29 & \cellcolor{red!16}-0.17 & \cellcolor{blue!100}1.0 & \cellcolor{blue!57}0.58 & \cellcolor{blue!28}0.28 & \cellcolor{red!75}-0.76 & \cellcolor{blue!28}0.29 & \cellcolor{red!36}-0.36 & \cellcolor{blue!33}0.33 & \cellcolor{red!75}-0.76 & \cellcolor{blue!28}0.29 & \cellcolor{red!16}-0.17 & \cellcolor{blue!22}0.23 & \cellcolor{blue!28}0.29 & \cellcolor{red!4}-0.04 & \cellcolor{red!16}-0.16 & \cellcolor{red!4}-0.04 & \cellcolor{red!27}-0.27 & \cellcolor{blue!29}0.29 & \cellcolor{red!22}-0.22\tabularnewline
PPV & \cellcolor{blue!61}0.62 & \cellcolor{red!6}-0.07 & \cellcolor{blue!57}0.58 & \cellcolor{blue!100}1.0 & \cellcolor{blue!64}0.64 & \cellcolor{red!24}-0.24 & \cellcolor{blue!54}0.55 & \cellcolor{red!49}-0.49 & \cellcolor{blue!64}0.64 & \cellcolor{red!24}-0.25 & \cellcolor{blue!61}0.62 & \cellcolor{red!6}-0.07 & \cellcolor{blue!47}0.47 & \cellcolor{blue!62}0.63 & \cellcolor{red!15}-0.16 & \cellcolor{red!37}-0.38 & \cellcolor{red!21}-0.21 & \cellcolor{red!47}-0.47 & \cellcolor{blue!61}0.62 & \cellcolor{red!21}-0.22\tabularnewline
IOU & \cellcolor{blue!98}0.98 & \cellcolor{blue!70}0.71 & \cellcolor{blue!28}0.28 & \cellcolor{blue!64}0.64 & \cellcolor{blue!100}1.0 & \cellcolor{red!23}-0.24 & \cellcolor{blue!89}0.9 & \cellcolor{red!28}-0.29 & \cellcolor{blue!99}0.99 & \cellcolor{red!23}-0.24 & \cellcolor{blue!98}0.98 & \cellcolor{blue!70}0.71 & \cellcolor{blue!71}0.71 & \cellcolor{blue!98}0.99 & \cellcolor{red!27}-0.28 & \cellcolor{red!55}-0.55 & \cellcolor{red!37}-0.37 & \cellcolor{red!69}-0.7 & \cellcolor{blue!98}0.98 & \cellcolor{red!30}-0.31\tabularnewline
GCE & \cellcolor{red!22}-0.22 & \cellcolor{red!8}-0.08 & \cellcolor{red!75}-0.76 & \cellcolor{red!24}-0.24 & \cellcolor{red!23}-0.24 & \cellcolor{blue!100}1.0 & \cellcolor{red!19}-0.19 & \cellcolor{blue!12}0.13 & \cellcolor{red!30}-0.3 & \cellcolor{blue!99}1.0 & \cellcolor{red!22}-0.23 & \cellcolor{red!8}-0.09 & \cellcolor{red!13}-0.14 & \cellcolor{red!22}-0.23 & \cellcolor{blue!2}0.02 & \cellcolor{blue!17}0.18 & \cellcolor{blue!2}0.03 & \cellcolor{blue!28}0.29 & \cellcolor{red!22}-0.23 & \cellcolor{blue!27}0.28\tabularnewline
\hline
VS & \cellcolor{blue!94}0.94 & \cellcolor{blue!62}0.62 & \cellcolor{blue!28}0.29 & \cellcolor{blue!54}0.55 & \cellcolor{blue!89}0.9 & \cellcolor{red!19}-0.19 & \cellcolor{blue!100}1.0 & \cellcolor{red!37}-0.37 & \cellcolor{blue!90}0.9 & \cellcolor{red!20}-0.2 & \cellcolor{blue!94}0.94 & \cellcolor{blue!62}0.62 & \cellcolor{blue!75}0.76 & \cellcolor{blue!92}0.92 & \cellcolor{red!35}-0.36 & \cellcolor{red!48}-0.48 & \cellcolor{red!43}-0.43 & \cellcolor{red!65}-0.65 & \cellcolor{blue!94}0.94 & \cellcolor{red!26}-0.26\tabularnewline
RAVD & \cellcolor{red!34}-0.35 & \cellcolor{blue!9}0.1 & \cellcolor{red!36}-0.36 & \cellcolor{red!49}-0.49 & \cellcolor{red!28}-0.29 & \cellcolor{blue!12}0.13 & \cellcolor{red!37}-0.37 & \cellcolor{blue!100}1.0 & \cellcolor{red!31}-0.31 & \cellcolor{blue!14}0.15 & \cellcolor{red!34}-0.35 & \cellcolor{blue!9}0.1 & \cellcolor{red!38}-0.39 & \cellcolor{red!34}-0.34 & \cellcolor{blue!17}0.18 & \cellcolor{blue!19}0.19 & \cellcolor{blue!14}0.14 & \cellcolor{blue!28}0.28 & \cellcolor{red!34}-0.35 & \cellcolor{blue!14}0.15\tabularnewline
\hline
MI & \cellcolor{blue!98}0.99 & \cellcolor{blue!70}0.7 & \cellcolor{blue!33}0.33 & \cellcolor{blue!64}0.64 & \cellcolor{blue!99}0.99 & \cellcolor{red!30}-0.3 & \cellcolor{blue!90}0.9 & \cellcolor{red!31}-0.31 & \cellcolor{blue!100}1.0 & \cellcolor{red!30}-0.31 & \cellcolor{blue!98}0.99 & \cellcolor{blue!70}0.7 & \cellcolor{blue!73}0.74 & \cellcolor{blue!99}0.99 & \cellcolor{red!30}-0.31 & \cellcolor{red!56}-0.56 & \cellcolor{red!39}-0.4 & \cellcolor{red!71}-0.71 & \cellcolor{blue!98}0.99 & \cellcolor{red!32}-0.32\tabularnewline
VOI & \cellcolor{red!23}-0.23 & \cellcolor{red!8}-0.08 & \cellcolor{red!75}-0.76 & \cellcolor{red!24}-0.25 & \cellcolor{red!23}-0.24 & \cellcolor{blue!99}1.0 & \cellcolor{red!20}-0.2 & \cellcolor{blue!14}0.15 & \cellcolor{red!30}-0.31 & \cellcolor{blue!100}1.0 & \cellcolor{red!23}-0.23 & \cellcolor{red!8}-0.08 & \cellcolor{red!15}-0.15 & \cellcolor{red!23}-0.24 & \cellcolor{blue!2}0.03 & \cellcolor{blue!18}0.18 & \cellcolor{blue!3}0.03 & \cellcolor{blue!29}0.3 & \cellcolor{red!23}-0.24 & \cellcolor{blue!28}0.28\tabularnewline
\hline
CKS & \cellcolor{blue!99}1.0 & \cellcolor{blue!70}0.7 & \cellcolor{blue!28}0.29 & \cellcolor{blue!61}0.62 & \cellcolor{blue!98}0.98 & \cellcolor{red!22}-0.23 & \cellcolor{blue!94}0.94 & \cellcolor{red!34}-0.35 & \cellcolor{blue!98}0.99 & \cellcolor{red!23}-0.23 & \cellcolor{blue!100}1.0 & \cellcolor{blue!70}0.71 & \cellcolor{blue!78}0.78 & \cellcolor{blue!99}1.0 & \cellcolor{red!34}-0.34 & \cellcolor{red!54}-0.55 & \cellcolor{red!43}-0.43 & \cellcolor{red!70}-0.71 & \cellcolor{blue!99}1.0 & \cellcolor{red!30}-0.3\tabularnewline
AUC & \cellcolor{blue!70}0.71 & \cellcolor{blue!99}1.0 & \cellcolor{red!16}-0.17 & \cellcolor{red!6}-0.07 & \cellcolor{blue!70}0.71 & \cellcolor{red!8}-0.09 & \cellcolor{blue!62}0.62 & \cellcolor{blue!9}0.1 & \cellcolor{blue!70}0.7 & \cellcolor{red!8}-0.08 & \cellcolor{blue!70}0.71 & \cellcolor{blue!100}1.0 & \cellcolor{blue!51}0.51 & \cellcolor{blue!71}0.71 & \cellcolor{red!26}-0.27 & \cellcolor{red!38}-0.38 & \cellcolor{red!33}-0.34 & \cellcolor{red!47}-0.47 & \cellcolor{blue!70}0.71 & \cellcolor{red!20}-0.2\tabularnewline
VC & \cellcolor{blue!78}0.78 & \cellcolor{blue!51}0.51 & \cellcolor{blue!22}0.23 & \cellcolor{blue!47}0.47 & \cellcolor{blue!71}0.71 & \cellcolor{red!13}-0.14 & \cellcolor{blue!75}0.76 & \cellcolor{red!38}-0.39 & \cellcolor{blue!73}0.74 & \cellcolor{red!15}-0.15 & \cellcolor{blue!78}0.78 & \cellcolor{blue!51}0.51 & \cellcolor{blue!100}1.0 & \cellcolor{blue!77}0.78 & \cellcolor{red!49}-0.49 & \cellcolor{red!51}-0.51 & \cellcolor{red!58}-0.58 & \cellcolor{red!71}-0.71 & \cellcolor{blue!78}0.78 & \cellcolor{red!22}-0.22\tabularnewline
MCC & \cellcolor{blue!99}1.0 & \cellcolor{blue!70}0.71 & \cellcolor{blue!28}0.29 & \cellcolor{blue!62}0.63 & \cellcolor{blue!98}0.99 & \cellcolor{red!22}-0.23 & \cellcolor{blue!92}0.92 & \cellcolor{red!34}-0.34 & \cellcolor{blue!99}0.99 & \cellcolor{red!23}-0.24 & \cellcolor{blue!99}1.0 & \cellcolor{blue!71}0.71 & \cellcolor{blue!77}0.78 & \cellcolor{blue!100}1.0 & \cellcolor{red!35}-0.36 & \cellcolor{red!55}-0.55 & \cellcolor{red!44}-0.44 & \cellcolor{red!71}-0.71 & \cellcolor{blue!99}1.0 & \cellcolor{red!30}-0.31\tabularnewline
PBD & \cellcolor{red!34}-0.34 & \cellcolor{red!26}-0.26 & \cellcolor{red!4}-0.04 & \cellcolor{red!15}-0.16 & \cellcolor{red!27}-0.28 & \cellcolor{blue!2}0.02 & \cellcolor{red!35}-0.36 & \cellcolor{blue!17}0.18 & \cellcolor{red!30}-0.31 & \cellcolor{blue!2}0.03 & \cellcolor{red!34}-0.34 & \cellcolor{red!26}-0.27 & \cellcolor{red!49}-0.49 & \cellcolor{red!35}-0.36 & \cellcolor{blue!100}1.0 & \cellcolor{blue!15}0.16 & \cellcolor{blue!97}0.97 & \cellcolor{blue!28}0.29 & \cellcolor{red!34}-0.34 & \cellcolor{blue!5}0.05\tabularnewline
\hline
HD95 & \cellcolor{red!54}-0.55 & \cellcolor{red!37}-0.38 & \cellcolor{red!16}-0.16 & \cellcolor{red!37}-0.38 & \cellcolor{red!55}-0.55 & \cellcolor{blue!17}0.18 & \cellcolor{red!48}-0.48 & \cellcolor{blue!19}0.19 & \cellcolor{red!56}-0.56 & \cellcolor{blue!18}0.18 & \cellcolor{red!54}-0.55 & \cellcolor{red!38}-0.38 & \cellcolor{red!51}-0.51 & \cellcolor{red!55}-0.55 & \cellcolor{blue!15}0.16 & \cellcolor{blue!100}1.0 & \cellcolor{blue!25}0.25 & \cellcolor{blue!89}0.89 & \cellcolor{red!55}-0.55 & \cellcolor{blue!13}0.14\tabularnewline
MHD & \cellcolor{red!43}-0.43 & \cellcolor{red!33}-0.34 & \cellcolor{red!4}-0.04 & \cellcolor{red!21}-0.21 & \cellcolor{red!37}-0.37 & \cellcolor{blue!2}0.03 & \cellcolor{red!43}-0.43 & \cellcolor{blue!14}0.14 & \cellcolor{red!39}-0.4 & \cellcolor{blue!3}0.03 & \cellcolor{red!43}-0.43 & \cellcolor{red!33}-0.34 & \cellcolor{red!58}-0.58 & \cellcolor{red!44}-0.44 & \cellcolor{blue!97}0.97 & \cellcolor{blue!25}0.25 & \cellcolor{blue!100}1.0 & \cellcolor{blue!39}0.4 & \cellcolor{red!43}-0.43 & \cellcolor{blue!5}0.06\tabularnewline
ASSD & \cellcolor{red!70}-0.71 & \cellcolor{red!47}-0.47 & \cellcolor{red!27}-0.27 & \cellcolor{red!47}-0.47 & \cellcolor{red!69}-0.7 & \cellcolor{blue!28}0.29 & \cellcolor{red!65}-0.65 & \cellcolor{blue!28}0.28 & \cellcolor{red!71}-0.71 & \cellcolor{blue!29}0.3 & \cellcolor{red!70}-0.71 & \cellcolor{red!47}-0.47 & \cellcolor{red!71}-0.71 & \cellcolor{red!71}-0.71 & \cellcolor{blue!28}0.29 & \cellcolor{blue!89}0.89 & \cellcolor{blue!39}0.4 & \cellcolor{blue!100}1.0 & \cellcolor{red!71}-0.71 & \cellcolor{blue!19}0.2\tabularnewline
\hline
ARI & \cellcolor{blue!99}1.0 & \cellcolor{blue!70}0.7 & \cellcolor{blue!29}0.29 & \cellcolor{blue!61}0.62 & \cellcolor{blue!98}0.98 & \cellcolor{red!22}-0.23 & \cellcolor{blue!94}0.94 & \cellcolor{red!34}-0.35 & \cellcolor{blue!98}0.99 & \cellcolor{red!23}-0.24 & \cellcolor{blue!99}1.0 & \cellcolor{blue!70}0.71 & \cellcolor{blue!78}0.78 & \cellcolor{blue!99}1.0 & \cellcolor{red!34}-0.34 & \cellcolor{red!55}-0.55 & \cellcolor{red!43}-0.43 & \cellcolor{red!71}-0.71 & \cellcolor{blue!100}1.0 & \cellcolor{red!30}-0.3\tabularnewline
OASSD & \cellcolor{red!30}-0.3 & \cellcolor{red!20}-0.2 & \cellcolor{red!22}-0.22 & \cellcolor{red!21}-0.22 & \cellcolor{red!30}-0.31 & \cellcolor{blue!27}0.28 & \cellcolor{red!26}-0.26 & \cellcolor{blue!14}0.15 & \cellcolor{red!32}-0.32 & \cellcolor{blue!28}0.28 & \cellcolor{red!30}-0.3 & \cellcolor{red!20}-0.2 & \cellcolor{red!22}-0.22 & \cellcolor{red!30}-0.31 & \cellcolor{blue!5}0.05 & \cellcolor{blue!13}0.14 & \cellcolor{blue!5}0.06 & \cellcolor{blue!19}0.2 & \cellcolor{red!30}-0.3 & \cellcolor{blue!100}1.0\tabularnewline
\bottomrule
\end{tabular}
}
\end{table}

\clearpage

\FloatBarrier
\newpage
\subsection{Lower grade glioma segmentation}

Overall segmentation performances for the lower grade glioma tumor type are reported fold-wise in \cref{tab:suppl-lgg-foldwise}. The average Dice score is quite stable across the different folds, at $75$\%, given the almost identical number of samples they hold. The same trend can be noticed for the patient-wise and object-wise metrics across the board.

\begin{figure}[!ht]
\centering
\includegraphics[scale=1.40]{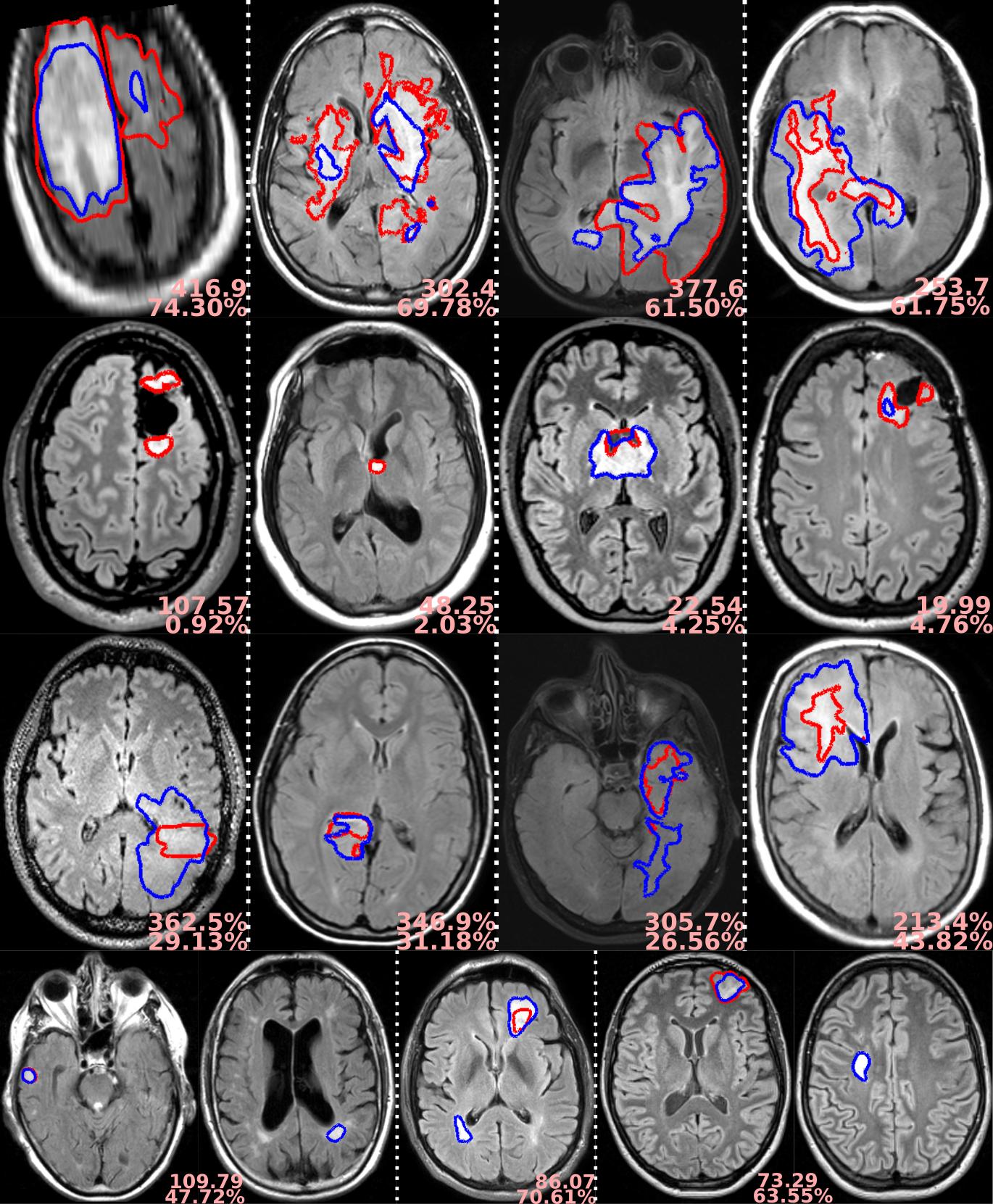}
\caption{Examples of lower grade glioma outliers where the ground truth is shown in red and the automatic segmentation in blue. Each patient is separated by a dotted white line, and each row focuses on a different metric, from top to bottom: GCE, PBD, RAVD, and HD95. For each patient, both metric of interest and Dice scores are provided in light red.}
\label{fig:suppl-lgg-outliers}
\end{figure}

The correlation matrix between some of the considered validation metrics is shown in \cref{fig:suppl-confusion-matrix-lgg}. Visual segmentation results are illustrated in \cref{fig:suppl-lgg-outliers} where the GCE, PBD, RAVD, and HD95 metrics scored orders of magnitude above their average, indicating clear outliers not identifiable by the Dice metric only.
In the first row, all examples exhibit extremely large and diffuse tumors whereby the manual ground truth is far from smooth and often parcelled. While Dice scores are overall above $60$\%, one can notice that only the brighter regions were segmented by the model.
In the second row, tumors with a small volume were not segmented while a bright non-tumor region has been segmented in the third case.
In the last row, small regions far from the main tumor location were incorrectly segmented, leading to high HD95 values for reasonable to good Dice scores.

\begin{table}[!h]
\caption{Fold-wise overall performance for the lower grade glioma segmentation.}
\label{tab:suppl-lgg-foldwise}
\adjustbox{max width=\textwidth}{
\begin{tabular}{rr|cc|ccc|ccc}
 & & \multicolumn{2}{c|}{Pixel-wise} & \multicolumn{3}{c|}{Patient-wise} & \multicolumn{3}{c}{Object-wise} \tabularnewline
Fold & \# Samples & Dice & Dice-TP & F1-score & Recall & Precision & F1-score & Recall & Precision\tabularnewline
\hline
0 & 132 & $74.97\pm28.56$ & $83.87\pm12.90$ & $91.09$ & $89.39$ & $92.85$ & $78.23$ & $72.78$ & $84.56$\tabularnewline
1 & 132 & $77.46\pm22.83$ & $80.50\pm17.21$ & $94.86$ & $96.21$ & $93.55$ & $80.59$ & $76.47$ & $85.19$\tabularnewline
2 & 132 & $74.98\pm25.20$ & $79.18\pm18.40$ & $94.68$ & $94.70$ & $94.67$ & $83.68$ & $77.78$ & $90.54$\tabularnewline
3 & 130 & $76.54\pm22.71$ & $80.24\pm15.61$ & $95.44$ & $95.38$ & $95.49$ & $80.94$ & $72.63$ & $91.39$\tabularnewline
4 & 132 & $73.01\pm29.77$ & $82.36\pm15.16$ & $91.98$ & $88.64$ & $95.58$ & $84.47$ & $78.18$ & $91.85$\tabularnewline
\end{tabular}
}
\end{table}

\begin{table}[h]
\caption{Metrics correlation matrix for the lower grade glioma segmentation. The color intensity of each cell represents the strength of the correlation, where blue denotes direct correlation and red denotes inverse correlation.}
\label{fig:suppl-confusion-matrix-lgg}
\adjustbox{max width=\textwidth}{
\begin{tabular}{lrrrrrr|rr|rr|rrrrr|rrr|rr}
\toprule
 & \multicolumn{6}{c|}{Overlap} & \multicolumn{2}{c|}{Volume} & \multicolumn{2}{c|}{Information theory} & \multicolumn{5}{c|}{Probabilistic} & \multicolumn{3}{c|}{Spatial distance} & \multicolumn{2}{c}{Instance-wise}\tabularnewline
{} & Dice & TPR & TNR & PPV & IOU & GCE & VS & RAVD & MI & VOI & CKS & AUC & VC & MCC & PBD & HD95 & MHD & ASSD & ARI & OASSD\tabularnewline
Dice & \cellcolor{blue!100}1.0 & \cellcolor{blue!82}0.83 & \cellcolor{blue!15}0.15 & \cellcolor{blue!48}0.48 & \cellcolor{blue!98}0.98 & \cellcolor{red!19}-0.2 & \cellcolor{blue!91}0.91 & \cellcolor{red!10}-0.11 & \cellcolor{blue!98}0.98 & \cellcolor{red!20}-0.2 & \cellcolor{blue!99}1.0 & \cellcolor{blue!82}0.83 & \cellcolor{blue!80}0.81 & \cellcolor{blue!99}1.0 & \cellcolor{red!45}-0.45 & \cellcolor{red!51}-0.51 & \cellcolor{red!65}-0.66 & \cellcolor{red!66}-0.67 & \cellcolor{blue!99}1.0 & \cellcolor{red!34}-0.34\tabularnewline
TPR & \cellcolor{blue!82}0.83 & \cellcolor{blue!100}1.0 & \cellcolor{red!15}-0.15 & \cellcolor{red!2}-0.02 & \cellcolor{blue!79}0.8 & \cellcolor{red!15}-0.16 & \cellcolor{blue!74}0.74 & \cellcolor{blue!28}0.28 & \cellcolor{blue!80}0.8 & \cellcolor{red!15}-0.15 & \cellcolor{blue!82}0.83 & \cellcolor{blue!99}1.0 & \cellcolor{blue!67}0.67 & \cellcolor{blue!82}0.82 & \cellcolor{red!40}-0.4 & \cellcolor{red!37}-0.37 & \cellcolor{red!55}-0.56 & \cellcolor{red!50}-0.51 & \cellcolor{blue!82}0.83 & \cellcolor{red!29}-0.29\tabularnewline
TNR & \cellcolor{blue!15}0.15 & \cellcolor{red!15}-0.15 & \cellcolor{blue!100}1.0 & \cellcolor{blue!49}0.49 & \cellcolor{blue!19}0.19 & \cellcolor{red!65}-0.66 & \cellcolor{blue!12}0.12 & \cellcolor{red!33}-0.34 & \cellcolor{blue!23}0.24 & \cellcolor{red!68}-0.68 & \cellcolor{blue!15}0.16 & \cellcolor{red!14}-0.14 & \cellcolor{blue!1}0.01 & \cellcolor{blue!16}0.16 & \cellcolor{blue!3}0.03 & \cellcolor{red!14}-0.14 & \cellcolor{red!0}-0.0 & \cellcolor{red!14}-0.15 & \cellcolor{blue!16}0.16 & \cellcolor{red!20}-0.21\tabularnewline
PPV & \cellcolor{blue!48}0.48 & \cellcolor{red!2}-0.02 & \cellcolor{blue!49}0.49 & \cellcolor{blue!100}1.0 & \cellcolor{blue!54}0.54 & \cellcolor{red!17}-0.17 & \cellcolor{blue!31}0.32 & \cellcolor{red!64}-0.65 & \cellcolor{blue!53}0.54 & \cellcolor{red!18}-0.19 & \cellcolor{blue!48}0.48 & \cellcolor{red!1}-0.02 & \cellcolor{blue!35}0.36 & \cellcolor{blue!51}0.51 & \cellcolor{red!1}-0.01 & \cellcolor{red!41}-0.41 & \cellcolor{red!18}-0.19 & \cellcolor{red!45}-0.45 & \cellcolor{blue!48}0.48 & \cellcolor{red!23}-0.23\tabularnewline
IOU & \cellcolor{blue!98}0.98 & \cellcolor{blue!79}0.8 & \cellcolor{blue!19}0.19 & \cellcolor{blue!54}0.54 & \cellcolor{blue!100}1.0 & \cellcolor{red!23}-0.24 & \cellcolor{blue!87}0.87 & \cellcolor{red!11}-0.11 & \cellcolor{blue!99}0.99 & \cellcolor{red!23}-0.24 & \cellcolor{blue!98}0.98 & \cellcolor{blue!79}0.8 & \cellcolor{blue!73}0.74 & \cellcolor{blue!98}0.98 & \cellcolor{red!36}-0.37 & \cellcolor{red!53}-0.53 & \cellcolor{red!59}-0.59 & \cellcolor{red!66}-0.66 & \cellcolor{blue!98}0.98 & \cellcolor{red!35}-0.36\tabularnewline
GCE & \cellcolor{red!19}-0.2 & \cellcolor{red!15}-0.16 & \cellcolor{red!65}-0.66 & \cellcolor{red!17}-0.17 & \cellcolor{red!23}-0.24 & \cellcolor{blue!100}1.0 & \cellcolor{red!15}-0.15 & \cellcolor{blue!2}0.02 & \cellcolor{red!30}-0.3 & \cellcolor{blue!99}0.99 & \cellcolor{red!20}-0.21 & \cellcolor{red!16}-0.16 & \cellcolor{red!3}-0.04 & \cellcolor{red!20}-0.21 & \cellcolor{red!2}-0.02 & \cellcolor{blue!18}0.18 & \cellcolor{blue!1}0.01 & \cellcolor{blue!24}0.24 & \cellcolor{red!21}-0.22 & \cellcolor{blue!31}0.32\tabularnewline
\hline
VS & \cellcolor{blue!91}0.91 & \cellcolor{blue!74}0.74 & \cellcolor{blue!12}0.12 & \cellcolor{blue!31}0.32 & \cellcolor{blue!87}0.87 & \cellcolor{red!15}-0.15 & \cellcolor{blue!100}1.0 & \cellcolor{red!10}-0.1 & \cellcolor{blue!86}0.86 & \cellcolor{red!15}-0.16 & \cellcolor{blue!91}0.91 & \cellcolor{blue!74}0.74 & \cellcolor{blue!74}0.75 & \cellcolor{blue!88}0.89 & \cellcolor{red!48}-0.49 & \cellcolor{red!38}-0.38 & \cellcolor{red!63}-0.63 & \cellcolor{red!57}-0.58 & \cellcolor{blue!91}0.91 & \cellcolor{red!25}-0.25\tabularnewline
RAVD & \cellcolor{red!10}-0.11 & \cellcolor{blue!28}0.28 & \cellcolor{red!33}-0.34 & \cellcolor{red!64}-0.65 & \cellcolor{red!11}-0.11 & \cellcolor{blue!2}0.02 & \cellcolor{red!10}-0.1 & \cellcolor{blue!100}1.0 & \cellcolor{red!11}-0.12 & \cellcolor{blue!4}0.04 & \cellcolor{red!10}-0.11 & \cellcolor{blue!27}0.28 & \cellcolor{red!10}-0.1 & \cellcolor{red!12}-0.12 & \cellcolor{blue!0}0.0 & \cellcolor{blue!11}0.11 & \cellcolor{blue!2}0.02 & \cellcolor{blue!12}0.12 & \cellcolor{red!10}-0.11 & \cellcolor{blue!2}0.02\tabularnewline
\hline
MI & \cellcolor{blue!98}0.98 & \cellcolor{blue!80}0.8 & \cellcolor{blue!23}0.24 & \cellcolor{blue!53}0.54 & \cellcolor{blue!99}0.99 & \cellcolor{red!30}-0.3 & \cellcolor{blue!86}0.86 & \cellcolor{red!11}-0.12 & \cellcolor{blue!100}1.0 & \cellcolor{red!31}-0.31 & \cellcolor{blue!98}0.99 & \cellcolor{blue!80}0.81 & \cellcolor{blue!75}0.76 & \cellcolor{blue!98}0.99 & \cellcolor{red!39}-0.4 & \cellcolor{red!53}-0.54 & \cellcolor{red!61}-0.61 & \cellcolor{red!67}-0.68 & \cellcolor{blue!98}0.99 & \cellcolor{red!38}-0.38\tabularnewline
VOI & \cellcolor{red!20}-0.2 & \cellcolor{red!15}-0.15 & \cellcolor{red!68}-0.68 & \cellcolor{red!18}-0.19 & \cellcolor{red!23}-0.24 & \cellcolor{blue!99}0.99 & \cellcolor{red!15}-0.16 & \cellcolor{blue!4}0.04 & \cellcolor{red!31}-0.31 & \cellcolor{blue!100}1.0 & \cellcolor{red!21}-0.21 & \cellcolor{red!15}-0.16 & \cellcolor{red!4}-0.04 & \cellcolor{red!21}-0.21 & \cellcolor{red!2}-0.02 & \cellcolor{blue!18}0.19 & \cellcolor{blue!1}0.01 & \cellcolor{blue!25}0.25 & \cellcolor{red!22}-0.22 & \cellcolor{blue!33}0.33\tabularnewline
\hline
CKS & \cellcolor{blue!99}1.0 & \cellcolor{blue!82}0.83 & \cellcolor{blue!15}0.16 & \cellcolor{blue!48}0.48 & \cellcolor{blue!98}0.98 & \cellcolor{red!20}-0.21 & \cellcolor{blue!91}0.91 & \cellcolor{red!10}-0.11 & \cellcolor{blue!98}0.99 & \cellcolor{red!21}-0.21 & \cellcolor{blue!100}1.0 & \cellcolor{blue!82}0.83 & \cellcolor{blue!80}0.81 & \cellcolor{blue!99}1.0 & \cellcolor{red!45}-0.45 & \cellcolor{red!51}-0.51 & \cellcolor{red!65}-0.66 & \cellcolor{red!67}-0.67 & \cellcolor{blue!99}1.0 & \cellcolor{red!34}-0.35\tabularnewline
AUC & \cellcolor{blue!82}0.83 & \cellcolor{blue!99}1.0 & \cellcolor{red!14}-0.14 & \cellcolor{red!1}-0.02 & \cellcolor{blue!79}0.8 & \cellcolor{red!16}-0.16 & \cellcolor{blue!74}0.74 & \cellcolor{blue!27}0.28 & \cellcolor{blue!80}0.81 & \cellcolor{red!15}-0.16 & \cellcolor{blue!82}0.83 & \cellcolor{blue!100}1.0 & \cellcolor{blue!67}0.67 & \cellcolor{blue!82}0.83 & \cellcolor{red!40}-0.4 & \cellcolor{red!37}-0.37 & \cellcolor{red!56}-0.56 & \cellcolor{red!50}-0.51 & \cellcolor{blue!82}0.83 & \cellcolor{red!29}-0.29\tabularnewline
VC & \cellcolor{blue!80}0.81 & \cellcolor{blue!67}0.67 & \cellcolor{blue!1}0.01 & \cellcolor{blue!35}0.36 & \cellcolor{blue!73}0.74 & \cellcolor{red!3}-0.04 & \cellcolor{blue!74}0.75 & \cellcolor{red!10}-0.1 & \cellcolor{blue!75}0.76 & \cellcolor{red!4}-0.04 & \cellcolor{blue!80}0.81 & \cellcolor{blue!67}0.67 & \cellcolor{blue!100}1.0 & \cellcolor{blue!81}0.81 & \cellcolor{red!51}-0.51 & \cellcolor{red!51}-0.52 & \cellcolor{red!73}-0.73 & \cellcolor{red!72}-0.72 & \cellcolor{blue!80}0.81 & \cellcolor{red!23}-0.24\tabularnewline
MCC & \cellcolor{blue!99}1.0 & \cellcolor{blue!82}0.82 & \cellcolor{blue!16}0.16 & \cellcolor{blue!51}0.51 & \cellcolor{blue!98}0.98 & \cellcolor{red!20}-0.21 & \cellcolor{blue!88}0.89 & \cellcolor{red!12}-0.12 & \cellcolor{blue!98}0.99 & \cellcolor{red!21}-0.21 & \cellcolor{blue!99}1.0 & \cellcolor{blue!82}0.83 & \cellcolor{blue!81}0.81 & \cellcolor{blue!100}1.0 & \cellcolor{red!45}-0.46 & \cellcolor{red!52}-0.52 & \cellcolor{red!66}-0.67 & \cellcolor{red!67}-0.68 & \cellcolor{blue!99}1.0 & \cellcolor{red!35}-0.36\tabularnewline
PBD & \cellcolor{red!45}-0.45 & \cellcolor{red!40}-0.4 & \cellcolor{blue!3}0.03 & \cellcolor{red!1}-0.01 & \cellcolor{red!36}-0.37 & \cellcolor{red!2}-0.02 & \cellcolor{red!48}-0.49 & \cellcolor{blue!0}0.0 & \cellcolor{red!39}-0.4 & \cellcolor{red!2}-0.02 & \cellcolor{red!45}-0.45 & \cellcolor{red!40}-0.4 & \cellcolor{red!51}-0.51 & \cellcolor{red!45}-0.46 & \cellcolor{blue!100}1.0 & \cellcolor{blue!16}0.16 & \cellcolor{blue!87}0.87 & \cellcolor{blue!25}0.26 & \cellcolor{red!45}-0.45 & \cellcolor{blue!10}0.1\tabularnewline
\hline
HD95 & \cellcolor{red!51}-0.51 & \cellcolor{red!37}-0.37 & \cellcolor{red!14}-0.14 & \cellcolor{red!41}-0.41 & \cellcolor{red!53}-0.53 & \cellcolor{blue!18}0.18 & \cellcolor{red!38}-0.38 & \cellcolor{blue!11}0.11 & \cellcolor{red!53}-0.54 & \cellcolor{blue!18}0.19 & \cellcolor{red!51}-0.51 & \cellcolor{red!37}-0.37 & \cellcolor{red!51}-0.52 & \cellcolor{red!52}-0.52 & \cellcolor{blue!16}0.16 & \cellcolor{blue!100}1.0 & \cellcolor{blue!37}0.37 & \cellcolor{blue!84}0.84 & \cellcolor{red!51}-0.51 & \cellcolor{blue!16}0.16\tabularnewline
MHD & \cellcolor{red!65}-0.66 & \cellcolor{red!55}-0.56 & \cellcolor{red!0}-0.0 & \cellcolor{red!18}-0.19 & \cellcolor{red!59}-0.59 & \cellcolor{blue!1}0.01 & \cellcolor{red!63}-0.63 & \cellcolor{blue!2}0.02 & \cellcolor{red!61}-0.61 & \cellcolor{blue!1}0.01 & \cellcolor{red!65}-0.66 & \cellcolor{red!56}-0.56 & \cellcolor{red!73}-0.73 & \cellcolor{red!66}-0.67 & \cellcolor{blue!87}0.87 & \cellcolor{blue!37}0.37 & \cellcolor{blue!100}1.0 & \cellcolor{blue!45}0.45 & \cellcolor{red!65}-0.66 & \cellcolor{blue!16}0.16\tabularnewline
ASSD & \cellcolor{red!66}-0.67 & \cellcolor{red!50}-0.51 & \cellcolor{red!14}-0.15 & \cellcolor{red!45}-0.45 & \cellcolor{red!66}-0.66 & \cellcolor{blue!24}0.24 & \cellcolor{red!57}-0.58 & \cellcolor{blue!12}0.12 & \cellcolor{red!67}-0.68 & \cellcolor{blue!25}0.25 & \cellcolor{red!67}-0.67 & \cellcolor{red!50}-0.51 & \cellcolor{red!72}-0.72 & \cellcolor{red!67}-0.68 & \cellcolor{blue!25}0.26 & \cellcolor{blue!84}0.84 & \cellcolor{blue!45}0.45 & \cellcolor{blue!100}1.0 & \cellcolor{red!67}-0.67 & \cellcolor{blue!22}0.23\tabularnewline
\hline
ARI & \cellcolor{blue!99}1.0 & \cellcolor{blue!82}0.83 & \cellcolor{blue!16}0.16 & \cellcolor{blue!48}0.48 & \cellcolor{blue!98}0.98 & \cellcolor{red!21}-0.22 & \cellcolor{blue!91}0.91 & \cellcolor{red!10}-0.11 & \cellcolor{blue!98}0.99 & \cellcolor{red!22}-0.22 & \cellcolor{blue!99}1.0 & \cellcolor{blue!82}0.83 & \cellcolor{blue!80}0.81 & \cellcolor{blue!99}1.0 & \cellcolor{red!45}-0.45 & \cellcolor{red!51}-0.51 & \cellcolor{red!65}-0.66 & \cellcolor{red!67}-0.67 & \cellcolor{blue!100}1.0 & \cellcolor{red!34}-0.35\tabularnewline
OASSD & \cellcolor{red!34}-0.34 & \cellcolor{red!29}-0.29 & \cellcolor{red!20}-0.21 & \cellcolor{red!23}-0.23 & \cellcolor{red!35}-0.36 & \cellcolor{blue!31}0.32 & \cellcolor{red!25}-0.25 & \cellcolor{blue!2}0.02 & \cellcolor{red!38}-0.38 & \cellcolor{blue!33}0.33 & \cellcolor{red!34}-0.35 & \cellcolor{red!29}-0.29 & \cellcolor{red!23}-0.24 & \cellcolor{red!35}-0.36 & \cellcolor{blue!10}0.1 & \cellcolor{blue!16}0.16 & \cellcolor{blue!16}0.16 & \cellcolor{blue!22}0.23 & \cellcolor{red!34}-0.35 & \cellcolor{blue!100}1.0\tabularnewline
\bottomrule
\end{tabular}
}
\end{table}

\FloatBarrier
\newpage
\subsection{Meningioma segmentation}

Overall segmentation performances for the meningioma tumor type are reported fold-wise in \cref{tab:suppl-meningioma-foldwise}. The average Dice score is quite stable across the different folds, at $75$\%, given the almost identical number of samples they hold. The same trend can be noticed for the patient-wise and object-wise metrics across the board.

\begin{figure}[!ht]
\centering
\includegraphics[scale=1.40]{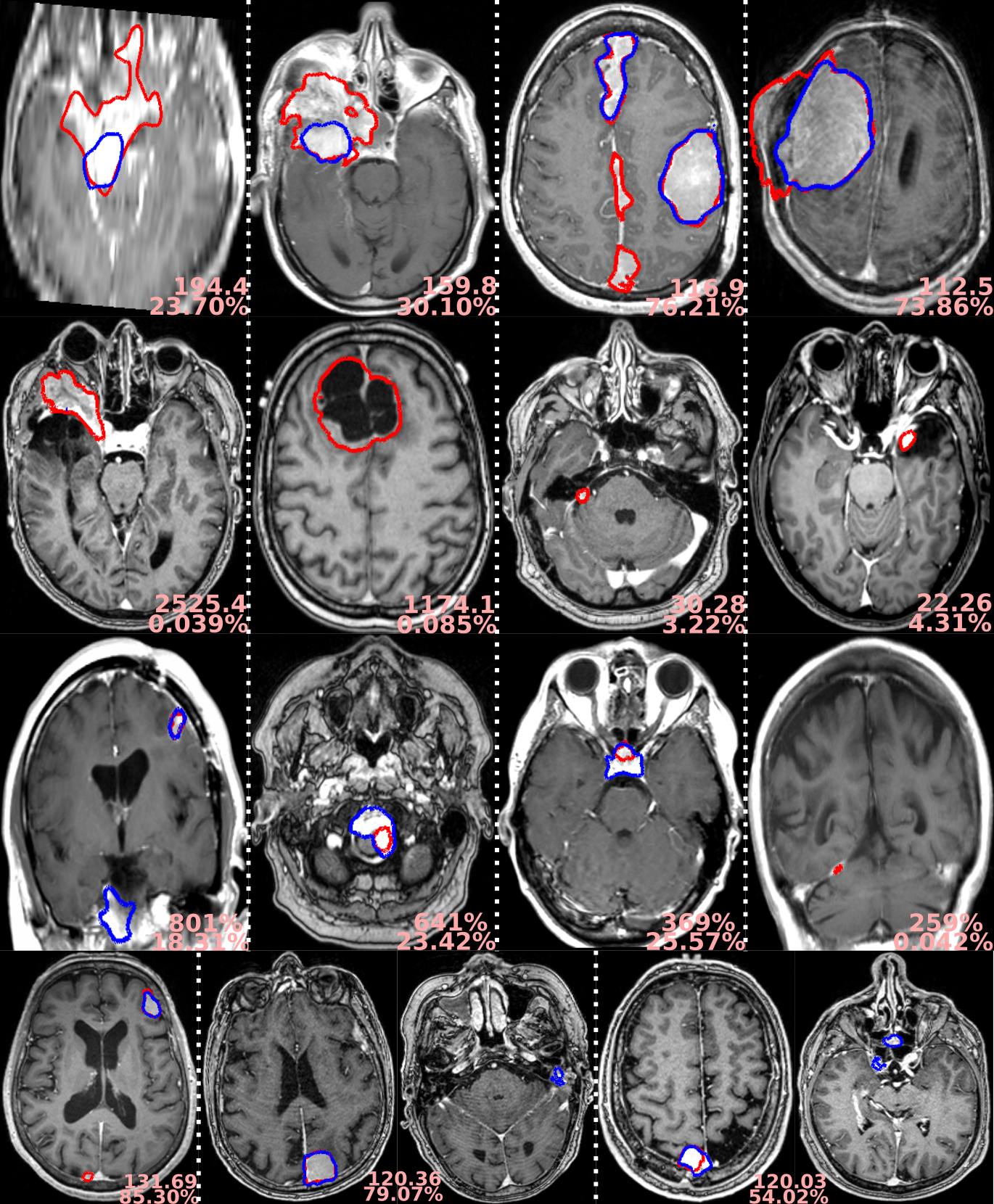}
\caption{Examples of meningioma outliers where the ground truth is shown in red and the automatic segmentation in blue. Each patient is separated by a dotted white line, and each row focuses on a different metric, from top to bottom: GCE, PBD, RAVD, and HD95. For each patient, both metric of interest and Dice scores are provided in light red.}
\label{fig:suppl-meningioma-outliers}
\end{figure}

The correlation matrix between some of the considered validation metrics is shown in \cref{fig:suppl-confusion-matrix-menin}. Visual segmentation results are illustrated in \cref{fig:suppl-meningioma-outliers} where the GCE, PBD, RAVD, and HD95 metrics scored orders of magnitude above their average, indicating clear outliers not identifiable by the Dice metric only. Coronal slices are provided in case of a large slice thickness.

\begin{table}[h!]
\caption{Fold-wise overall performance for the meningioma segmentation.}
\label{tab:suppl-meningioma-foldwise}
\adjustbox{max width=\textwidth}{
\begin{tabular}{rr|cc|ccc|ccc}
 & & \multicolumn{2}{c|}{Pixel-wise} & \multicolumn{3}{c|}{Patient-wise} & \multicolumn{3}{c}{Object-wise} \tabularnewline
Fold & \# Samples & Dice & Dice-TP & F1-score & Recall & Precision & F1-score & Recall & Precision\tabularnewline
\hline
0 & 144 & $77.30\pm28.18$ & $84.32\pm16.55$ & $87.88$ & $91.67$ & $84.39$ & $77.60$ & $83.33$ & $72.60$\tabularnewline
1 & 144 & $74.93\pm31.61$ & $85.63\pm15.02$ & $91.83$ & $87.50$ & $96.60$ & $83.08$ & $75.00$ & $93.10$\tabularnewline
2 & 144 & $74.94\pm29.94$ & $83.65\pm16.50$ & $91.23$ & $89.58$ & $92.94$ & $87.58$ & $87.82$ & $87.34$\tabularnewline
3 & 143 & $74.76\pm32.53$ & $87.62\pm10.66$ & $91.20$ & $85.31$ & $97.97$ & $87.20$ & $79.25$ & $96.92$\tabularnewline
4 & 144 & $73.07\pm30.46$ & $82.85\pm15.57$ & $91.19$ & $88.19$ & $94.40$ & $83.82$ & $79.23$ & $88.96$\tabularnewline
\end{tabular}
}
\end{table}

\begin{table}[h!]
\caption{Metrics correlation matrix for the meningioma segmentation. The color intensity of each cell represents the strength of the correlation, where blue denotes direct correlation and red denotes inverse correlation.}
\label{fig:suppl-confusion-matrix-menin}
\adjustbox{max width=\textwidth}{
\begin{tabular}{lrrrrrr|rr|rr|rrrrr|rrr|rr}
\toprule
 & \multicolumn{6}{c|}{Overlap} & \multicolumn{2}{c|}{Volume} & \multicolumn{2}{c|}{Information theory} & \multicolumn{5}{c|}{Probabilistic} & \multicolumn{3}{c|}{Spatial distance} & \multicolumn{2}{c}{Instance-wise}\tabularnewline
{} & Dice & TPR & TNR & PPV & IOU & GCE & VS & RAVD & MI & VOI & CKS & AUC & VC & MCC & PBD & HD95 & MHD & ASSD & ARI & OASSD\tabularnewline
Dice & \cellcolor{blue!100}1.0 & \cellcolor{blue!85}0.85 & \cellcolor{blue!11}0.11 & \cellcolor{blue!39}0.4 & \cellcolor{blue!98}0.98 & \cellcolor{red!33}-0.34 & \cellcolor{blue!93}0.94 & \cellcolor{red!4}-0.04 & \cellcolor{blue!99}0.99 & \cellcolor{red!34}-0.35 & \cellcolor{blue!99}1.0 & \cellcolor{blue!85}0.85 & \cellcolor{blue!81}0.82 & \cellcolor{blue!99}1.0 & \cellcolor{red!28}-0.29 & \cellcolor{red!43}-0.43 & \cellcolor{red!67}-0.68 & \cellcolor{red!56}-0.56 & \cellcolor{blue!99}1.0 & \cellcolor{red!27}-0.28\tabularnewline
TPR & \cellcolor{blue!85}0.85 & \cellcolor{blue!100}1.0 & \cellcolor{red!12}-0.13 & \cellcolor{red!7}-0.07 & \cellcolor{blue!83}0.84 & \cellcolor{red!34}-0.34 & \cellcolor{blue!77}0.77 & \cellcolor{blue!40}0.4 & \cellcolor{blue!84}0.85 & \cellcolor{red!34}-0.34 & \cellcolor{blue!85}0.85 & \cellcolor{blue!99}1.0 & \cellcolor{blue!61}0.62 & \cellcolor{blue!85}0.85 & \cellcolor{red!25}-0.26 & \cellcolor{red!26}-0.26 & \cellcolor{red!53}-0.54 & \cellcolor{red!32}-0.33 & \cellcolor{blue!85}0.85 & \cellcolor{red!29}-0.29\tabularnewline
TNR & \cellcolor{blue!11}0.11 & \cellcolor{red!12}-0.13 & \cellcolor{blue!100}1.0 & \cellcolor{blue!41}0.41 & \cellcolor{blue!11}0.11 & \cellcolor{red!48}-0.49 & \cellcolor{blue!12}0.12 & \cellcolor{red!42}-0.42 & \cellcolor{blue!15}0.15 & \cellcolor{red!52}-0.52 & \cellcolor{blue!11}0.12 & \cellcolor{red!12}-0.13 & \cellcolor{blue!17}0.18 & \cellcolor{blue!11}0.11 & \cellcolor{blue!3}0.03 & \cellcolor{red!18}-0.18 & \cellcolor{red!6}-0.06 & \cellcolor{red!23}-0.24 & \cellcolor{blue!11}0.12 & \cellcolor{red!24}-0.24\tabularnewline
PPV & \cellcolor{blue!39}0.4 & \cellcolor{red!7}-0.07 & \cellcolor{blue!41}0.41 & \cellcolor{blue!100}1.0 & \cellcolor{blue!45}0.46 & \cellcolor{red!5}-0.06 & \cellcolor{blue!28}0.28 & \cellcolor{red!70}-0.7 & \cellcolor{blue!43}0.44 & \cellcolor{red!6}-0.07 & \cellcolor{blue!39}0.4 & \cellcolor{red!7}-0.07 & \cellcolor{blue!37}0.37 & \cellcolor{blue!41}0.41 & \cellcolor{blue!5}0.06 & \cellcolor{red!43}-0.44 & \cellcolor{red!16}-0.17 & \cellcolor{red!52}-0.53 & \cellcolor{blue!39}0.4 & \cellcolor{red!3}-0.03\tabularnewline
IOU & \cellcolor{blue!98}0.98 & \cellcolor{blue!83}0.84 & \cellcolor{blue!11}0.11 & \cellcolor{blue!45}0.46 & \cellcolor{blue!100}1.0 & \cellcolor{red!33}-0.33 & \cellcolor{blue!89}0.89 & \cellcolor{red!4}-0.05 & \cellcolor{blue!99}0.99 & \cellcolor{red!33}-0.34 & \cellcolor{blue!98}0.98 & \cellcolor{blue!83}0.84 & \cellcolor{blue!75}0.76 & \cellcolor{blue!98}0.98 & \cellcolor{red!22}-0.23 & \cellcolor{red!46}-0.47 & \cellcolor{red!60}-0.61 & \cellcolor{red!57}-0.58 & \cellcolor{blue!98}0.98 & \cellcolor{red!27}-0.27\tabularnewline
GCE & \cellcolor{red!33}-0.34 & \cellcolor{red!34}-0.34 & \cellcolor{red!48}-0.49 & \cellcolor{red!5}-0.06 & \cellcolor{red!33}-0.33 & \cellcolor{blue!100}1.0 & \cellcolor{red!32}-0.33 & \cellcolor{red!6}-0.06 & \cellcolor{red!38}-0.38 & \cellcolor{blue!99}1.0 & \cellcolor{red!33}-0.34 & \cellcolor{red!34}-0.34 & \cellcolor{red!17}-0.17 & \cellcolor{red!33}-0.33 & \cellcolor{blue!5}0.06 & \cellcolor{blue!19}0.2 & \cellcolor{blue!18}0.18 & \cellcolor{blue!27}0.27 & \cellcolor{red!34}-0.34 & \cellcolor{blue!46}0.47\tabularnewline
\hline
VS & \cellcolor{blue!93}0.94 & \cellcolor{blue!77}0.77 & \cellcolor{blue!12}0.12 & \cellcolor{blue!28}0.28 & \cellcolor{blue!89}0.89 & \cellcolor{red!32}-0.33 & \cellcolor{blue!100}1.0 & \cellcolor{red!5}-0.06 & \cellcolor{blue!89}0.9 & \cellcolor{red!34}-0.34 & \cellcolor{blue!93}0.94 & \cellcolor{blue!77}0.77 & \cellcolor{blue!78}0.78 & \cellcolor{blue!91}0.92 & \cellcolor{red!31}-0.31 & \cellcolor{red!36}-0.37 & \cellcolor{red!68}-0.68 & \cellcolor{red!51}-0.51 & \cellcolor{blue!93}0.94 & \cellcolor{red!24}-0.25\tabularnewline
RAVD & \cellcolor{red!4}-0.04 & \cellcolor{blue!40}0.4 & \cellcolor{red!42}-0.42 & \cellcolor{red!70}-0.7 & \cellcolor{red!4}-0.05 & \cellcolor{red!6}-0.06 & \cellcolor{red!5}-0.06 & \cellcolor{blue!100}1.0 & \cellcolor{red!4}-0.04 & \cellcolor{red!4}-0.04 & \cellcolor{red!4}-0.04 & \cellcolor{blue!40}0.4 & \cellcolor{red!20}-0.2 & \cellcolor{red!3}-0.03 & \cellcolor{red!10}-0.11 & \cellcolor{blue!22}0.22 & \cellcolor{blue!5}0.06 & \cellcolor{blue!34}0.34 & \cellcolor{red!4}-0.04 & \cellcolor{red!9}-0.09\tabularnewline
\hline
MI & \cellcolor{blue!99}0.99 & \cellcolor{blue!84}0.85 & \cellcolor{blue!15}0.15 & \cellcolor{blue!43}0.44 & \cellcolor{blue!99}0.99 & \cellcolor{red!38}-0.38 & \cellcolor{blue!89}0.9 & \cellcolor{red!4}-0.04 & \cellcolor{blue!100}1.0 & \cellcolor{red!39}-0.39 & \cellcolor{blue!99}0.99 & \cellcolor{blue!84}0.85 & \cellcolor{blue!78}0.78 & \cellcolor{blue!99}0.99 & \cellcolor{red!25}-0.26 & \cellcolor{red!46}-0.46 & \cellcolor{red!64}-0.64 & \cellcolor{red!58}-0.58 & \cellcolor{blue!99}0.99 & \cellcolor{red!30}-0.3\tabularnewline
VOI & \cellcolor{red!34}-0.35 & \cellcolor{red!34}-0.34 & \cellcolor{red!52}-0.52 & \cellcolor{red!6}-0.07 & \cellcolor{red!33}-0.34 & \cellcolor{blue!99}1.0 & \cellcolor{red!34}-0.34 & \cellcolor{red!4}-0.04 & \cellcolor{red!39}-0.39 & \cellcolor{blue!100}1.0 & \cellcolor{red!35}-0.35 & \cellcolor{red!34}-0.35 & \cellcolor{red!19}-0.2 & \cellcolor{red!34}-0.34 & \cellcolor{blue!7}0.08 & \cellcolor{blue!20}0.21 & \cellcolor{blue!20}0.21 & \cellcolor{blue!28}0.29 & \cellcolor{red!35}-0.35 & \cellcolor{blue!47}0.47\tabularnewline
\hline
CKS & \cellcolor{blue!99}1.0 & \cellcolor{blue!85}0.85 & \cellcolor{blue!11}0.12 & \cellcolor{blue!39}0.4 & \cellcolor{blue!98}0.98 & \cellcolor{red!33}-0.34 & \cellcolor{blue!93}0.94 & \cellcolor{red!4}-0.04 & \cellcolor{blue!99}0.99 & \cellcolor{red!35}-0.35 & \cellcolor{blue!100}1.0 & \cellcolor{blue!85}0.85 & \cellcolor{blue!81}0.81 & \cellcolor{blue!99}1.0 & \cellcolor{red!28}-0.29 & \cellcolor{red!43}-0.43 & \cellcolor{red!67}-0.68 & \cellcolor{red!56}-0.56 & \cellcolor{blue!99}1.0 & \cellcolor{red!27}-0.28\tabularnewline
AUC & \cellcolor{blue!85}0.85 & \cellcolor{blue!99}1.0 & \cellcolor{red!12}-0.13 & \cellcolor{red!7}-0.07 & \cellcolor{blue!83}0.84 & \cellcolor{red!34}-0.34 & \cellcolor{blue!77}0.77 & \cellcolor{blue!40}0.4 & \cellcolor{blue!84}0.85 & \cellcolor{red!34}-0.35 & \cellcolor{blue!85}0.85 & \cellcolor{blue!100}1.0 & \cellcolor{blue!61}0.62 & \cellcolor{blue!85}0.85 & \cellcolor{red!25}-0.26 & \cellcolor{red!26}-0.26 & \cellcolor{red!53}-0.54 & \cellcolor{red!32}-0.33 & \cellcolor{blue!85}0.85 & \cellcolor{red!29}-0.29\tabularnewline
VC & \cellcolor{blue!81}0.82 & \cellcolor{blue!61}0.62 & \cellcolor{blue!17}0.18 & \cellcolor{blue!37}0.37 & \cellcolor{blue!75}0.76 & \cellcolor{red!17}-0.17 & \cellcolor{blue!78}0.78 & \cellcolor{red!20}-0.2 & \cellcolor{blue!78}0.78 & \cellcolor{red!19}-0.2 & \cellcolor{blue!81}0.81 & \cellcolor{blue!61}0.62 & \cellcolor{blue!100}1.0 & \cellcolor{blue!81}0.81 & \cellcolor{red!40}-0.41 & \cellcolor{red!43}-0.43 & \cellcolor{red!76}-0.77 & \cellcolor{red!61}-0.62 & \cellcolor{blue!81}0.81 & \cellcolor{red!15}-0.15\tabularnewline
MCC & \cellcolor{blue!99}1.0 & \cellcolor{blue!85}0.85 & \cellcolor{blue!11}0.11 & \cellcolor{blue!41}0.41 & \cellcolor{blue!98}0.98 & \cellcolor{red!33}-0.33 & \cellcolor{blue!91}0.92 & \cellcolor{red!3}-0.03 & \cellcolor{blue!99}0.99 & \cellcolor{red!34}-0.34 & \cellcolor{blue!99}1.0 & \cellcolor{blue!85}0.85 & \cellcolor{blue!81}0.81 & \cellcolor{blue!100}1.0 & \cellcolor{red!32}-0.32 & \cellcolor{red!43}-0.44 & \cellcolor{red!69}-0.7 & \cellcolor{red!56}-0.56 & \cellcolor{blue!99}1.0 & \cellcolor{red!28}-0.28\tabularnewline
PBD & \cellcolor{red!28}-0.29 & \cellcolor{red!25}-0.26 & \cellcolor{blue!3}0.03 & \cellcolor{blue!5}0.06 & \cellcolor{red!22}-0.23 & \cellcolor{blue!5}0.06 & \cellcolor{red!31}-0.31 & \cellcolor{red!10}-0.11 & \cellcolor{red!25}-0.26 & \cellcolor{blue!7}0.08 & \cellcolor{red!28}-0.29 & \cellcolor{red!25}-0.26 & \cellcolor{red!40}-0.41 & \cellcolor{red!32}-0.32 & \cellcolor{blue!100}1.0 & \cellcolor{blue!6}0.07 & \cellcolor{blue!83}0.84 & \cellcolor{blue!12}0.13 & \cellcolor{red!28}-0.29 & \cellcolor{blue!14}0.14\tabularnewline
\hline
HD95 & \cellcolor{red!43}-0.43 & \cellcolor{red!26}-0.26 & \cellcolor{red!18}-0.18 & \cellcolor{red!43}-0.44 & \cellcolor{red!46}-0.47 & \cellcolor{blue!19}0.2 & \cellcolor{red!36}-0.37 & \cellcolor{blue!22}0.22 & \cellcolor{red!46}-0.46 & \cellcolor{blue!20}0.21 & \cellcolor{red!43}-0.43 & \cellcolor{red!26}-0.26 & \cellcolor{red!43}-0.43 & \cellcolor{red!43}-0.44 & \cellcolor{blue!6}0.07 & \cellcolor{blue!100}1.0 & \cellcolor{blue!30}0.31 & \cellcolor{blue!90}0.9 & \cellcolor{red!43}-0.43 & \cellcolor{blue!11}0.12\tabularnewline
MHD & \cellcolor{red!67}-0.68 & \cellcolor{red!53}-0.54 & \cellcolor{red!6}-0.06 & \cellcolor{red!16}-0.17 & \cellcolor{red!60}-0.61 & \cellcolor{blue!18}0.18 & \cellcolor{red!68}-0.68 & \cellcolor{blue!5}0.06 & \cellcolor{red!64}-0.64 & \cellcolor{blue!20}0.21 & \cellcolor{red!67}-0.68 & \cellcolor{red!53}-0.54 & \cellcolor{red!76}-0.77 & \cellcolor{red!69}-0.7 & \cellcolor{blue!83}0.84 & \cellcolor{blue!30}0.31 & \cellcolor{blue!100}1.0 & \cellcolor{blue!44}0.45 & \cellcolor{red!67}-0.68 & \cellcolor{blue!20}0.21\tabularnewline
ASSD & \cellcolor{red!56}-0.56 & \cellcolor{red!32}-0.33 & \cellcolor{red!23}-0.24 & \cellcolor{red!52}-0.53 & \cellcolor{red!57}-0.58 & \cellcolor{blue!27}0.27 & \cellcolor{red!51}-0.51 & \cellcolor{blue!34}0.34 & \cellcolor{red!58}-0.58 & \cellcolor{blue!28}0.29 & \cellcolor{red!56}-0.56 & \cellcolor{red!32}-0.33 & \cellcolor{red!61}-0.62 & \cellcolor{red!56}-0.56 & \cellcolor{blue!12}0.13 & \cellcolor{blue!90}0.9 & \cellcolor{blue!44}0.45 & \cellcolor{blue!100}1.0 & \cellcolor{red!56}-0.57 & \cellcolor{blue!14}0.14\tabularnewline
\hline
ARI & \cellcolor{blue!99}1.0 & \cellcolor{blue!85}0.85 & \cellcolor{blue!11}0.12 & \cellcolor{blue!39}0.4 & \cellcolor{blue!98}0.98 & \cellcolor{red!34}-0.34 & \cellcolor{blue!93}0.94 & \cellcolor{red!4}-0.04 & \cellcolor{blue!99}0.99 & \cellcolor{red!35}-0.35 & \cellcolor{blue!99}1.0 & \cellcolor{blue!85}0.85 & \cellcolor{blue!81}0.81 & \cellcolor{blue!99}1.0 & \cellcolor{red!28}-0.29 & \cellcolor{red!43}-0.43 & \cellcolor{red!67}-0.68 & \cellcolor{red!56}-0.57 & \cellcolor{blue!100}1.0 & \cellcolor{red!28}-0.28\tabularnewline
OASSD & \cellcolor{red!27}-0.28 & \cellcolor{red!29}-0.29 & \cellcolor{red!24}-0.24 & \cellcolor{red!3}-0.03 & \cellcolor{red!27}-0.27 & \cellcolor{blue!46}0.47 & \cellcolor{red!24}-0.25 & \cellcolor{red!9}-0.09 & \cellcolor{red!30}-0.3 & \cellcolor{blue!47}0.47 & \cellcolor{red!27}-0.28 & \cellcolor{red!29}-0.29 & \cellcolor{red!15}-0.15 & \cellcolor{red!28}-0.28 & \cellcolor{blue!14}0.14 & \cellcolor{blue!11}0.12 & \cellcolor{blue!20}0.21 & \cellcolor{blue!14}0.14 & \cellcolor{red!28}-0.28 & \cellcolor{blue!100}1.0\tabularnewline
\bottomrule
\end{tabular}
}
\end{table}

\FloatBarrier
\newpage
\subsection{Metastases segmentation}

Overall segmentation performances for the meningioma tumor type are reported fold-wise in \cref{tab:suppl-metastasis-foldwise}. The average Dice score is quite stable across the different folds, at $87$\%, given the almost identical number of samples they hold. The same trend can be noticed for the patient-wise and object-wise metrics across the board.

\begin{figure}[!ht]
\centering
\includegraphics[scale=1.40]{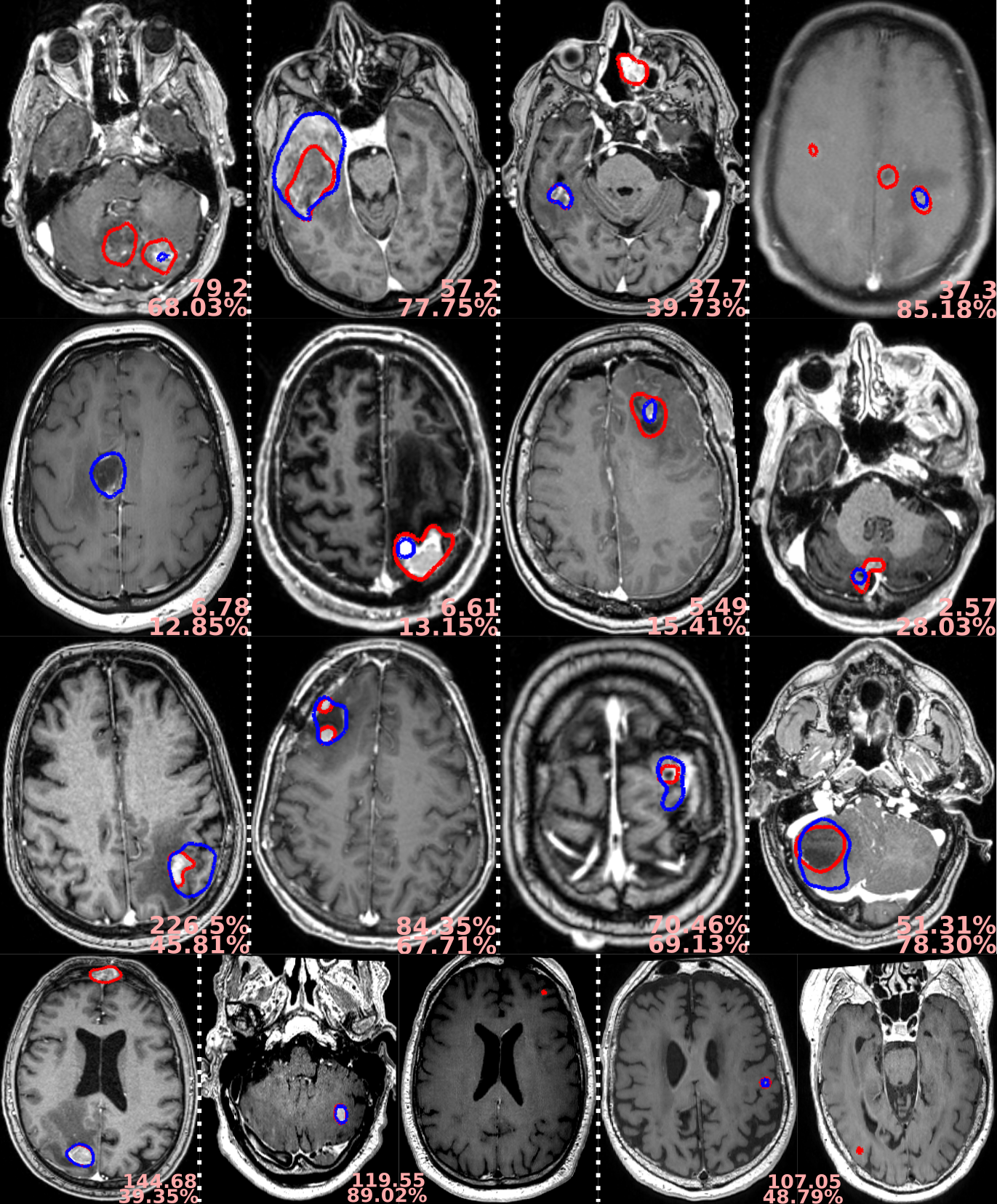}
\caption{Examples of metastasis outliers where the ground truth is shown in red and the automatic segmentation in blue. Each patient is separated by a dotted white line, and each row focuses on a different metric, from top to bottom: GCE, PBD, RAVD, and HD95. For each patient, both metric of interest and Dice scores are provided in light red.}
\label{fig:suppl-metastase-outliers}
\end{figure}

The correlation matrix between some of the considered validation metrics is shown in \cref{fig:suppl-confusion-matrix-meta}. Visual segmentation results are illustrated in \cref{fig:suppl-metastase-outliers} where the GCE, PBD, RAVD, and HD95 metrics scored orders of magnitude above their average, indicating clear outliers not identifiable by the Dice metric only.
Overall, the metastasis model has subpar precision because of the smallest amount of training data, suffering a bit from transfer learning from the HGG model, which can be addressed with more training data.

As can be seen in the second row, the model segmented a cavity in the leftmost example. A large tumor portion has been missed at the top of the head in the second column, maybe as a resulting limitation from performing skull stripping.

\begin{table}[!h]
\caption{Fold-wise overall performance for the metastase segmentation.}
\label{tab:suppl-metastasis-foldwise}
\adjustbox{max width=\textwidth}{
\begin{tabular}{rr|cc|ccc|ccc}
 & & \multicolumn{2}{c|}{Pixel-wise} & \multicolumn{3}{c|}{Patient-wise} & \multicolumn{3}{c}{Object-wise} \tabularnewline
Fold & \# Samples & Dice & Dice-TP & F1-score & Recall & Precision & F1-score & Recall & Precision\tabularnewline
\hline
0 & 79 & $87.51\pm19.98$ & $90.96\pm10.05$ & $96.74$ & $96.20$ & $97.28$ & $87.87$ & $82.73$ & $93.68$\tabularnewline
1 & 80 & $89.03\pm16.82$ & $91.31\pm09.05$ & $98.05$ & $97.50$ & $98.61$ & $89.08$ & $82.81$ & $96.36$\tabularnewline
2 & 80 & $89.63\pm15.16$ & $89.63\pm15.16$ & $98.73$ & $100.00$ & $97.50$ & $91.16$ & $86.73$ & $96.08$\tabularnewline
3 & 75 & $84.13\pm22.26$ & $86.43\pm17.60$ & $96.87$ & $97.33$ & $96.41$ & $87.94$ & $82.48$ & $94.17$\tabularnewline
4 & 80 & $88.15\pm19.88$ & $91.58\pm09.81$ & $97.25$ & $96.25$ & $98.27$ & $87.46$ & $79.22$ & $97.60$\tabularnewline
\end{tabular}
}
\end{table}

\begin{table}[h]
\caption{Metrics correlation matrix for the metastasis segmentation. The color intensity of each cell represents the strength of the correlation, where blue denotes direct correlation and red denotes inverse correlation.}
\label{fig:suppl-confusion-matrix-meta}
\adjustbox{max width=\textwidth}{
\begin{tabular}{lrrrrrr|rr|rr|rrrrr|rrr|rr}
\toprule
 & \multicolumn{6}{c|}{Overlap} & \multicolumn{2}{c|}{Volume} & \multicolumn{2}{c|}{Information theory} & \multicolumn{5}{c|}{Probabilistic} & \multicolumn{3}{c|}{Spatial distance} & \multicolumn{2}{c}{Instance-wise}\tabularnewline
{} & Dice & TPR & TNR & PPV & IOU & GCE & VS & RAVD & MI & VOI & CKS & AUC & VC & MCC & PBD & HD95 & MHD & ASSD & ARI & OASSD\tabularnewline
Dice & \cellcolor{blue!100}1.0 & \cellcolor{blue!88}0.89 & \cellcolor{blue!11}0.12 & \cellcolor{blue!41}0.41 & \cellcolor{blue!98}0.99 & \cellcolor{red!27}-0.27 & \cellcolor{blue!95}0.96 & \cellcolor{red!10}-0.1 & \cellcolor{blue!99}0.99 & \cellcolor{red!29}-0.3 & \cellcolor{blue!99}1.0 & \cellcolor{blue!88}0.89 & \cellcolor{blue!72}0.73 & \cellcolor{blue!99}1.0 & \cellcolor{red!82}-0.82 & \cellcolor{red!43}-0.44 & \cellcolor{red!74}-0.74 & \cellcolor{red!57}-0.57 & \cellcolor{blue!99}1.0 & \cellcolor{red!72}-0.73\tabularnewline
TPR & \cellcolor{blue!88}0.89 & \cellcolor{blue!100}1.0 & \cellcolor{red!9}-0.09 & \cellcolor{red!0}-0.01 & \cellcolor{blue!88}0.88 & \cellcolor{red!15}-0.16 & \cellcolor{blue!84}0.84 & \cellcolor{blue!26}0.26 & \cellcolor{blue!87}0.87 & \cellcolor{red!17}-0.18 & \cellcolor{blue!88}0.89 & \cellcolor{blue!99}1.0 & \cellcolor{blue!56}0.57 & \cellcolor{blue!88}0.88 & \cellcolor{red!64}-0.65 & \cellcolor{red!42}-0.43 & \cellcolor{red!50}-0.51 & \cellcolor{red!48}-0.48 & \cellcolor{blue!88}0.89 & \cellcolor{red!61}-0.61\tabularnewline
TNR & \cellcolor{blue!11}0.12 & \cellcolor{red!9}-0.09 & \cellcolor{blue!100}1.0 & \cellcolor{blue!48}0.48 & \cellcolor{blue!14}0.15 & \cellcolor{red!87}-0.88 & \cellcolor{blue!8}0.09 & \cellcolor{red!19}-0.2 & \cellcolor{blue!18}0.18 & \cellcolor{red!84}-0.85 & \cellcolor{blue!11}0.12 & \cellcolor{red!9}-0.09 & \cellcolor{blue!11}0.12 & \cellcolor{blue!13}0.13 & \cellcolor{red!3}-0.04 & \cellcolor{red!8}-0.08 & \cellcolor{red!12}-0.12 & \cellcolor{red!16}-0.16 & \cellcolor{blue!11}0.12 & \cellcolor{red!25}-0.25\tabularnewline
PPV & \cellcolor{blue!41}0.41 & \cellcolor{red!0}-0.01 & \cellcolor{blue!48}0.48 & \cellcolor{blue!100}1.0 & \cellcolor{blue!44}0.45 & \cellcolor{red!35}-0.36 & \cellcolor{blue!31}0.32 & \cellcolor{red!65}-0.66 & \cellcolor{blue!46}0.46 & \cellcolor{red!35}-0.35 & \cellcolor{blue!41}0.41 & \cellcolor{red!0}-0.01 & \cellcolor{blue!41}0.41 & \cellcolor{blue!44}0.44 & \cellcolor{red!36}-0.37 & \cellcolor{red!20}-0.2 & \cellcolor{red!53}-0.54 & \cellcolor{red!35}-0.35 & \cellcolor{blue!41}0.41 & \cellcolor{red!34}-0.35\tabularnewline
IOU & \cellcolor{blue!98}0.99 & \cellcolor{blue!88}0.88 & \cellcolor{blue!14}0.15 & \cellcolor{blue!44}0.45 & \cellcolor{blue!100}1.0 & \cellcolor{red!30}-0.3 & \cellcolor{blue!92}0.92 & \cellcolor{red!6}-0.06 & \cellcolor{blue!99}1.0 & \cellcolor{red!32}-0.32 & \cellcolor{blue!98}0.99 & \cellcolor{blue!88}0.88 & \cellcolor{blue!67}0.68 & \cellcolor{blue!99}0.99 & \cellcolor{red!73}-0.73 & \cellcolor{red!45}-0.46 & \cellcolor{red!69}-0.69 & \cellcolor{red!56}-0.57 & \cellcolor{blue!98}0.99 & \cellcolor{red!72}-0.73\tabularnewline
GCE & \cellcolor{red!27}-0.27 & \cellcolor{red!15}-0.16 & \cellcolor{red!87}-0.88 & \cellcolor{red!35}-0.36 & \cellcolor{red!30}-0.3 & \cellcolor{blue!100}1.0 & \cellcolor{red!22}-0.23 & \cellcolor{blue!4}0.05 & \cellcolor{red!34}-0.35 & \cellcolor{blue!99}1.0 & \cellcolor{red!27}-0.28 & \cellcolor{red!15}-0.16 & \cellcolor{red!21}-0.21 & \cellcolor{red!28}-0.29 & \cellcolor{blue!14}0.14 & \cellcolor{blue!16}0.17 & \cellcolor{blue!19}0.19 & \cellcolor{blue!29}0.29 & \cellcolor{red!27}-0.28 & \cellcolor{blue!44}0.44\tabularnewline
\hline
VS & \cellcolor{blue!95}0.96 & \cellcolor{blue!84}0.84 & \cellcolor{blue!8}0.09 & \cellcolor{blue!31}0.32 & \cellcolor{blue!92}0.92 & \cellcolor{red!22}-0.23 & \cellcolor{blue!100}1.0 & \cellcolor{red!10}-0.11 & \cellcolor{blue!92}0.92 & \cellcolor{red!25}-0.25 & \cellcolor{blue!95}0.96 & \cellcolor{blue!84}0.84 & \cellcolor{blue!71}0.71 & \cellcolor{blue!94}0.94 & \cellcolor{red!82}-0.83 & \cellcolor{red!38}-0.39 & \cellcolor{red!71}-0.72 & \cellcolor{red!52}-0.53 & \cellcolor{blue!95}0.96 & \cellcolor{red!71}-0.72\tabularnewline
RAVD & \cellcolor{red!10}-0.1 & \cellcolor{blue!26}0.26 & \cellcolor{red!19}-0.2 & \cellcolor{red!65}-0.66 & \cellcolor{red!6}-0.06 & \cellcolor{blue!4}0.05 & \cellcolor{red!10}-0.11 & \cellcolor{blue!100}1.0 & \cellcolor{red!8}-0.09 & \cellcolor{blue!5}0.05 & \cellcolor{red!10}-0.1 & \cellcolor{blue!26}0.26 & \cellcolor{red!43}-0.44 & \cellcolor{red!10}-0.11 & \cellcolor{blue!35}0.35 & \cellcolor{blue!3}0.03 & \cellcolor{blue!62}0.63 & \cellcolor{blue!21}0.21 & \cellcolor{red!10}-0.1 & \cellcolor{red!7}-0.08\tabularnewline
\hline
MI & \cellcolor{blue!99}0.99 & \cellcolor{blue!87}0.87 & \cellcolor{blue!18}0.18 & \cellcolor{blue!46}0.46 & \cellcolor{blue!99}1.0 & \cellcolor{red!34}-0.35 & \cellcolor{blue!92}0.92 & \cellcolor{red!8}-0.09 & \cellcolor{blue!100}1.0 & \cellcolor{red!37}-0.37 & \cellcolor{blue!99}0.99 & \cellcolor{blue!87}0.88 & \cellcolor{blue!70}0.7 & \cellcolor{blue!99}1.0 & \cellcolor{red!77}-0.77 & \cellcolor{red!45}-0.45 & \cellcolor{red!71}-0.72 & \cellcolor{red!58}-0.58 & \cellcolor{blue!99}0.99 & \cellcolor{red!74}-0.74\tabularnewline
VOI & \cellcolor{red!29}-0.3 & \cellcolor{red!17}-0.18 & \cellcolor{red!84}-0.85 & \cellcolor{red!35}-0.35 & \cellcolor{red!32}-0.32 & \cellcolor{blue!99}1.0 & \cellcolor{red!25}-0.25 & \cellcolor{blue!5}0.05 & \cellcolor{red!37}-0.37 & \cellcolor{blue!100}1.0 & \cellcolor{red!30}-0.3 & \cellcolor{red!18}-0.18 & \cellcolor{red!23}-0.23 & \cellcolor{red!31}-0.31 & \cellcolor{blue!17}0.17 & \cellcolor{blue!17}0.17 & \cellcolor{blue!21}0.22 & \cellcolor{blue!31}0.31 & \cellcolor{red!30}-0.3 & \cellcolor{blue!47}0.47\tabularnewline
\hline
CKS & \cellcolor{blue!99}1.0 & \cellcolor{blue!88}0.89 & \cellcolor{blue!11}0.12 & \cellcolor{blue!41}0.41 & \cellcolor{blue!98}0.99 & \cellcolor{red!27}-0.28 & \cellcolor{blue!95}0.96 & \cellcolor{red!10}-0.1 & \cellcolor{blue!99}0.99 & \cellcolor{red!30}-0.3 & \cellcolor{blue!100}1.0 & \cellcolor{blue!88}0.89 & \cellcolor{blue!72}0.73 & \cellcolor{blue!99}1.0 & \cellcolor{red!82}-0.82 & \cellcolor{red!43}-0.44 & \cellcolor{red!74}-0.74 & \cellcolor{red!57}-0.57 & \cellcolor{blue!99}1.0 & \cellcolor{red!73}-0.73\tabularnewline
AUC & \cellcolor{blue!88}0.89 & \cellcolor{blue!99}1.0 & \cellcolor{red!9}-0.09 & \cellcolor{red!0}-0.01 & \cellcolor{blue!88}0.88 & \cellcolor{red!15}-0.16 & \cellcolor{blue!84}0.84 & \cellcolor{blue!26}0.26 & \cellcolor{blue!87}0.88 & \cellcolor{red!18}-0.18 & \cellcolor{blue!88}0.89 & \cellcolor{blue!100}1.0 & \cellcolor{blue!56}0.57 & \cellcolor{blue!88}0.88 & \cellcolor{red!64}-0.65 & \cellcolor{red!42}-0.43 & \cellcolor{red!50}-0.51 & \cellcolor{red!48}-0.48 & \cellcolor{blue!88}0.89 & \cellcolor{red!61}-0.61\tabularnewline
VC & \cellcolor{blue!72}0.73 & \cellcolor{blue!56}0.57 & \cellcolor{blue!11}0.12 & \cellcolor{blue!41}0.41 & \cellcolor{blue!67}0.68 & \cellcolor{red!21}-0.21 & \cellcolor{blue!71}0.71 & \cellcolor{red!43}-0.44 & \cellcolor{blue!70}0.7 & \cellcolor{red!23}-0.23 & \cellcolor{blue!72}0.73 & \cellcolor{blue!56}0.57 & \cellcolor{blue!100}1.0 & \cellcolor{blue!72}0.73 & \cellcolor{red!72}-0.73 & \cellcolor{red!46}-0.47 & \cellcolor{red!88}-0.88 & \cellcolor{red!78}-0.79 & \cellcolor{blue!72}0.73 & \cellcolor{red!32}-0.33\tabularnewline
MCC & \cellcolor{blue!99}1.0 & \cellcolor{blue!88}0.88 & \cellcolor{blue!13}0.13 & \cellcolor{blue!44}0.44 & \cellcolor{blue!99}0.99 & \cellcolor{red!28}-0.29 & \cellcolor{blue!94}0.94 & \cellcolor{red!10}-0.11 & \cellcolor{blue!99}1.0 & \cellcolor{red!31}-0.31 & \cellcolor{blue!99}1.0 & \cellcolor{blue!88}0.88 & \cellcolor{blue!72}0.73 & \cellcolor{blue!100}1.0 & \cellcolor{red!81}-0.81 & \cellcolor{red!45}-0.45 & \cellcolor{red!74}-0.74 & \cellcolor{red!58}-0.58 & \cellcolor{blue!99}1.0 & \cellcolor{red!72}-0.72\tabularnewline
PBD & \cellcolor{red!82}-0.82 & \cellcolor{red!64}-0.65 & \cellcolor{red!3}-0.04 & \cellcolor{red!36}-0.37 & \cellcolor{red!73}-0.73 & \cellcolor{blue!14}0.14 & \cellcolor{red!82}-0.83 & \cellcolor{blue!35}0.35 & \cellcolor{red!77}-0.77 & \cellcolor{blue!17}0.17 & \cellcolor{red!82}-0.82 & \cellcolor{red!64}-0.65 & \cellcolor{red!72}-0.73 & \cellcolor{red!81}-0.81 & \cellcolor{blue!100}1.0 & \cellcolor{blue!30}0.3 & \cellcolor{blue!80}0.81 & \cellcolor{blue!47}0.48 & \cellcolor{red!82}-0.82 & \cellcolor{blue!54}0.55\tabularnewline
\hline
HD95 & \cellcolor{red!43}-0.44 & \cellcolor{red!42}-0.43 & \cellcolor{red!8}-0.08 & \cellcolor{red!20}-0.2 & \cellcolor{red!45}-0.46 & \cellcolor{blue!16}0.17 & \cellcolor{red!38}-0.39 & \cellcolor{blue!3}0.03 & \cellcolor{red!45}-0.45 & \cellcolor{blue!17}0.17 & \cellcolor{red!43}-0.44 & \cellcolor{red!42}-0.43 & \cellcolor{red!46}-0.47 & \cellcolor{red!45}-0.45 & \cellcolor{blue!30}0.3 & \cellcolor{blue!100}1.0 & \cellcolor{blue!43}0.43 & \cellcolor{blue!78}0.79 & \cellcolor{red!43}-0.44 & \cellcolor{blue!14}0.15\tabularnewline
MHD & \cellcolor{red!74}-0.74 & \cellcolor{red!50}-0.51 & \cellcolor{red!12}-0.12 & \cellcolor{red!53}-0.54 & \cellcolor{red!69}-0.69 & \cellcolor{blue!19}0.19 & \cellcolor{red!71}-0.72 & \cellcolor{blue!62}0.63 & \cellcolor{red!71}-0.72 & \cellcolor{blue!21}0.22 & \cellcolor{red!74}-0.74 & \cellcolor{red!50}-0.51 & \cellcolor{red!88}-0.88 & \cellcolor{red!74}-0.74 & \cellcolor{blue!80}0.81 & \cellcolor{blue!43}0.43 & \cellcolor{blue!100}1.0 & \cellcolor{blue!68}0.68 & \cellcolor{red!74}-0.74 & \cellcolor{blue!38}0.39\tabularnewline
ASSD & \cellcolor{red!57}-0.57 & \cellcolor{red!48}-0.48 & \cellcolor{red!16}-0.16 & \cellcolor{red!35}-0.35 & \cellcolor{red!56}-0.57 & \cellcolor{blue!29}0.29 & \cellcolor{red!52}-0.53 & \cellcolor{blue!21}0.21 & \cellcolor{red!58}-0.58 & \cellcolor{blue!31}0.31 & \cellcolor{red!57}-0.57 & \cellcolor{red!48}-0.48 & \cellcolor{red!78}-0.79 & \cellcolor{red!58}-0.58 & \cellcolor{blue!47}0.48 & \cellcolor{blue!78}0.79 & \cellcolor{blue!68}0.68 & \cellcolor{blue!100}1.0 & \cellcolor{red!57}-0.57 & \cellcolor{blue!21}0.22\tabularnewline
\hline
ARI & \cellcolor{blue!99}1.0 & \cellcolor{blue!88}0.89 & \cellcolor{blue!11}0.12 & \cellcolor{blue!41}0.41 & \cellcolor{blue!98}0.99 & \cellcolor{red!27}-0.28 & \cellcolor{blue!95}0.96 & \cellcolor{red!10}-0.1 & \cellcolor{blue!99}0.99 & \cellcolor{red!30}-0.3 & \cellcolor{blue!99}1.0 & \cellcolor{blue!88}0.89 & \cellcolor{blue!72}0.73 & \cellcolor{blue!99}1.0 & \cellcolor{red!82}-0.82 & \cellcolor{red!43}-0.44 & \cellcolor{red!74}-0.74 & \cellcolor{red!57}-0.57 & \cellcolor{blue!100}1.0 & \cellcolor{red!73}-0.73\tabularnewline
OASSD & \cellcolor{red!72}-0.73 & \cellcolor{red!61}-0.61 & \cellcolor{red!25}-0.25 & \cellcolor{red!34}-0.35 & \cellcolor{red!72}-0.73 & \cellcolor{blue!44}0.44 & \cellcolor{red!71}-0.72 & \cellcolor{red!7}-0.08 & \cellcolor{red!74}-0.74 & \cellcolor{blue!47}0.47 & \cellcolor{red!73}-0.73 & \cellcolor{red!61}-0.61 & \cellcolor{red!32}-0.33 & \cellcolor{red!72}-0.72 & \cellcolor{blue!54}0.55 & \cellcolor{blue!14}0.15 & \cellcolor{blue!38}0.39 & \cellcolor{blue!21}0.22 & \cellcolor{red!73}-0.73 & \cellcolor{blue!100}1.0\tabularnewline
\bottomrule
\end{tabular}
}
\end{table}

\newpage
\FloatBarrier

\clearpage
\subsection{Runtime}
Using a representative MRI volume of each tumor type considered, the following four tables indicate runtime for the different steps performed during the process to generate the standardized clinical reports. Results are reported when using the high-end computer, either using the stand-alone software solution or the 3D Slicer plugin. 

The most noticeable difference comes from the penultimate step (i.e., Apply reg.) going from $183$\,seconds with the lower grade glioma volume to $792$\,seconds with the metastasis volume. The size of the MRI volume has a huge impact on the runtime necessary to apply the inverse registration transform on all cortical and subcortical structures atlases, in order to express them in the patient space. For the metastasis tumor type, the other steps performing volume-wise operations (e.g., tumor segmentation or registration) also have a longer runtime given the dimensions of the MRI volume, when compared to a typical glioblastoma MR scan.

\begin{table}[!ht]
\caption{Runtime distribution per step in the standardized report pipeline, computed and averaged using an high-end computer for the glioblastoma tumor type, each expressed in seconds.}
\label{fig:suppl-runtime-glioblastoma-highend}
\adjustbox{max width=\textwidth}{
\begin{tabular}{l|cc|ccc|c}
 & Brain segm. & Tumor segm. & Reg. preproc & Registration & Apply reg. & Features comp.\tabularnewline
\hline
Raidionics & $25.72\pm0.123$ & $21.99\pm0.177$ & $1.394\pm0.016$ & $37.40\pm1.41$ & $154.56\pm0.91$ & $112.55\pm1.25$\tabularnewline
Raidionics-Slicer & $23.04\pm00.64$ & $20.63\pm00.07$ & $1.539\pm0.109$ & $95.18\pm0.88$ & $240.13\pm0.20$ & $120.63\pm4.05$\tabularnewline
\end{tabular}
}
\end{table}

\begin{table}[!ht]
\caption{Runtime distribution per step in the standardized report pipeline, computed and averaged using an high-end computer for the lower grade glioma tumor type, each expressed in seconds.}
\label{fig:suppl-runtime-lower grade glioma-highend}
\adjustbox{max width=\textwidth}{
\begin{tabular}{l|cc|ccc|c}
 & Brain segm. & Tumor segm. & Reg. preproc & Registration & Apply reg. & Features comp.\tabularnewline
\hline
Raidionics & $18.29\pm0.402$ & $16.69\pm0.427$ & $0.778\pm0.013$ & $32.10\pm1.32$ & $112.12\pm3.34$ & $90.28\pm1.03$\tabularnewline
Raidionics-Slicer & $16.58\pm0.201$ & $16.24\pm0.286$ & $0.858\pm0.032$ & $74.76\pm1.97$ & $183.41\pm0.88$ &$89.91\pm1.92$ \tabularnewline
\end{tabular}
}
\end{table}

\begin{table}[!ht]
\caption{Runtime distribution per step in the standardized report pipeline, computed and averaged using an high-end computer for the meningioma tumor type, each expressed in seconds.}
\label{fig:suppl-runtime-meningioma-highend}
\adjustbox{max width=\textwidth}{
\begin{tabular}{l|cc|ccc|c}
 & Brain segm. & Tumor segm. & Reg. preproc & Registration & Apply reg. & Features comp.\tabularnewline
\hline
Raidionics & $19.08\pm0.445$ & $17.21\pm0.425$ & $0.865\pm0.023$ & $37.79\pm2.63$ & $107.77\pm1.98$ & $146.08\pm4.12$\tabularnewline
Raidionics-Slicer & $17.70\pm0.624$ & $17.07\pm0.571$ & $0.924\pm0.016$ & $90.58\pm1.03$ & $178.92\pm2.71$ & $144.35\pm2.49$ \tabularnewline
\end{tabular}
}
\end{table}

\begin{table}[!ht]
\caption{Runtime distribution per step in the standardized report pipeline, computed and averaged using an high-end computer for the metastasis tumor type, each expressed in seconds.}
\label{fig:suppl-runtime-metastasis-highend}
\adjustbox{max width=\textwidth}{
\begin{tabular}{l|cc|ccc|c}
 & Brain segm. & Tumor segm. & Reg. preproc & Registration & Apply reg. & Features comp.\tabularnewline
\hline
Raidionics & $68.58\pm1.454$ & $63.32\pm1.310$ & $6.906\pm0.222$ & $72.09\pm1.12$ & $573.33\pm9.76$ & $136.96\pm0.98$\tabularnewline
Raidionics-Slicer & $59.37\pm1.266$ & $53.85\pm0.358$ & $7.097\pm0.174$ & $139.56\pm3.70$ & $792.76\pm17.13$ & $135.71\pm3.71$ \tabularnewline
\end{tabular}
}
\end{table}